\documentclass[aps,rmp,reprint,noeprint,amsmath,amssymb,graphicx,longbibliography]{revtex4-1}

\usepackage{bm}
\usepackage[hidelinks]{hyperref}
\usepackage{graphicx}

\begin{document}

\title{Quantum optical coherence theory based on Feynman\rq{}s path integral}

\author{Jianbin Liu$^*$, Hui Chen,  Huaibin Zheng, Yuchen He, Zhuo Xu}
\affiliation{Electronic Materials Research Laboratory, Key Laboratory of the Ministry of Education, International Center for Dielectric Research, School of Electronic Science and Engineering, Xi'an Jiaotong University, Xi'an 710049, China}
\author{Yu Zhou, Fuli Li} 
\affiliation{MOE Key Laboratory for Nonequilibrium Synthesis and Modulation of Condensed Matter, School of Physics, Xi'an Jiaotong University, Xi'an 710049, China}

\email[]{liujianbin@xjtu.edu.cn}

\date{\today{}}

\begin{abstract}
 
Compared to classical optical coherence theory based on Maxwell\rq{}s electromagnetic theory and Glauber\rq{}s quantum optical coherence theory based on matrix mechanics formulation of quantum mechanics, quantum optical coherence theory based on Feynman\rq{}s path integral formulation of quantum mechanics provides a novel tool to study optical coherence. It has the advantage of understanding the connection between mathematical calculations and physical interpretations better. Quantum optical coherence theory based on Feynman\rq{}s path integral is introduced and reviewed in this paper. Based on the results of transient first-order interference of two independent light beams, it is predicted that the classical model for electric field of thermal light introduced  by classical optical textbooks may not be accurate. The physics of two-photon bunching of thermal light and Hong-Ou-Mandel dip of entangled photon pairs is the same, which can be interpreted by constructive and destructive two-photon interference, respectively. Quantum optical coherence theory based on Feynman\rq{}s path integral is helpful to understand the coherence properties of light, which may eventually lead us to the answer of the question: what is a photon? 
   
\end{abstract}

\maketitle

\tableofcontents{}

\section{Introduction}\label{sec-intro}

Ever since the interference of light was introduced by Young in 1801 \cite{young1802ii}, optical coherence has been one of the most important topics in the development of physics. With contributions from generations of scientists, two main optical coherence theories have been established \cite{born2013principles,mandel1995optical}. One is classical optical coherence theory based on Maxwell\rq{}s electromagnetic theory \cite{born2013principles}. The other one is Glauber\rq{}s quantum optical coherence theory based on matrix mechanics formulation of quantum mechanics \cite{glauber1963coherent,glauber1963quantum}. It was proved by Glauber and Sudarshan independently that classical and quantum optical coherence theories are mathematically equivalent for classical light \cite{sudarshan1963equivalence,glauber1963photon}. However, classical optical coherence theory can not interpret non-classical phenomena such as photoelectric effect \cite{hertz1887ueber,einstein1905heuristic}, photon antibuching \cite{miller1967anti,stoler1974photon,kimble1977photon}, Bell inequality violation \cite{bell1964einstein,bell1966problem} and so on. On the other hand, Glauber\rq{}s quantum optical coherence theory can interpret both classical and non-classical phenomena. 

As Glauber\rq{}s quantum optical coherence theory can explain all the optical coherence phenomena \cite{glauber2006nobel}, why introduce another quantum optical coherence theory based on Feynman\rq{}s path integral? This is the question we asked ourselves when we employed Feynman\rq{}s path integral to explain the subwavelength interference of light \cite{liu2010unified}. Similar question bothered Young \cite{young1802ii} and Feynman \cite{feynman1948space}, too. At the beginning of Young\rq{}s paper on the theory of light and colours \cite{young1802ii}, he explained why he employed wave theory: \textit{Although the invention of plausible hypotheses, independent of any connection with experimental observations, can be of very little use in the promotion of natural knowledge; yet the discovery of simple and uniform principles, by which a great number of apparently heterogeneous phenomena are reduced to coherent and universal laws, must ever be allowed to be of considerable importance towards the improvement of the human intellect.}

When Feynman introduced path integral formulation of quantum mechanics, he also faced similar question. There were wave mechanics and matrix mechanics formulations of quantum mechanics. Why introduce a third and equivalent formulation of quantum mechanics? Feynman\rq{}s answer to the question is \cite{feynman1948space}: \textit{The formulation is mathematically equivalent to the more usual formulations. There are, therefore, no fundamentally new results. However, there is a pleasure in recognizing old things from a new point of view. Also, there are problems for which the new point of view offers a distinct advantage.}

The answers given by Young and Feynman are also the reasons why we introduced quantum optical coherence theory based on Feynman\rq{}s path integral. As introduced in Sect. III, quantum optical coherence theory based on Feynman\rq{}s path integral can indeed offer deeper insight about the physics of the first- and second-order interference of two independent light beams, which is essential to understand the nature of light.

The remaining parts of the paper are organized as follows. A short history of quantum optical coherence theory based on Feynman\rq{}s path integral and some basic concepts are introduced in Sect. \ref{sec-basic}.  Applying path integral to interpret the first- and second-order interference of light are in Sects. \ref{sec-first} and \ref{sec-second}, respectively. Generalizing the method developed in Sects. \ref{sec-first} and \ref{sec-second} to the third- and higher-order interference of light can be found in Sect. \ref{sec-third}. Section \ref{sec-atom} introduces quantum atomic coherence theory based on Feynman\rq{}s path integral by analogy of quantum optical coherence theory based on Feynman\rq{}s path integral. Our conclusions are summarized in Sect. \ref{sec-summary}.

\section{Basic concepts in quantum optical coherence theory based on Feynman\rq{}s path integral}\label{sec-basic}

\subsection{History of quantum optical coherence theory based on Feynman\rq{}s path integral}

The history of classical optical coherence theory began in the 17th century when Grimaldi discovered diffraction of light \cite{hecht2012optics}. With the contributions from Young \cite{young1802ii,young1804bakerian}, Fresnel \cite{fresnel1821memoire}, and many other scientists \cite{hecht2012optics,born2013principles}, classical optical coherence theory has been established based on Maxwell\rq{}s electromagnetic theory \cite{maxwell1865viii}. The detail history of classical optical coherence can be found in Refs. \cite{born2013principles,hecht2012optics,mandel1965coherence}.

The history of quantum optical coherence theory dates back to Dirac and Feynman \cite{dirac1930principles,feynman2011feynman}. In his famous book on quantum mechanics,  Dirac discussed the first-order interference of light with the concept of photon and concluded  that \textit{Each photon then interferes only with itself. Interferences between two different photons never occurs} \cite{dirac1930principles}. Feynman employed path integral to discuss the first-order interference of electrons in a Young\rq{}s double-slit interferometer in his lectures \cite{feynman2011feynman}. All the discussions about optical coherence were limited to the first-order before 1956. Both quantum and classical optical coherence theories can explain the first-order interference of light. The need for quantum optical coherence theory was not so urgent then. In 1956, Hanbury Brown and Twiss found that photons emitted by thermal light source are not independent \cite{brown1956correlation,hanbury1956test}, which is known as two-photon bunching of thermal light. People began to realize that there are second- and higher-order interference of light besides the first-order interference. In 1963, Glauber published a series of papers explaining optical coherence in quantum theory \cite{glauber1963coherent,glauber1963quantum,glauber1963photon}, which laid the foundation of quantum optical coherence theory. Glauber was awarded 2005 Nobel physics prize for his contribution to quantum theory of optical coherence \cite{glauber2006nobel}. Glauber\rq{}s quantum optical coherence theory was employed by most scientists in quantum optics and introduced in many quantum optics textbooks \cite{mandel1995optical,scully1997quantum,loudon2000quantum,shih2020introduction}.

Glauber\rq{}s quantum optical coherence theory is based on matrix mechanics formulation of quantum mechanics \cite{heisenberg1925uber,heisenberg1949physical}. There are two more equivalent formulations of quantum mechanics called wave mechanics \cite{schrodinger1926undulatory} and Feynman\rq{}s path integral \cite{feynman1948space,feynman2010quantum}.  In principle, one can establish quantum optical coherence theory based on wave mechanics and Feynman\rq{}s path integral formulations, too. In the following parts, we will introduce quantum optical coherence theory based on Feynman\rq{}s path integral.

\begin{figure}[htb]
\centering
\includegraphics[width=80mm]{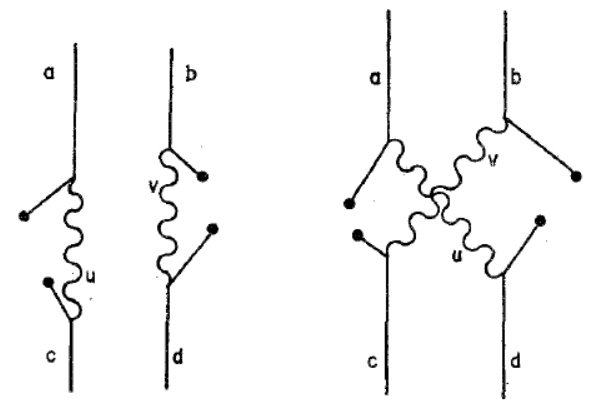}
\caption{Interpretation of two-photon bunching of thermal light in Feynman\rq{}s path integral \cite{fano1961quantum}. Atoms a and b are two atoms that can emit a photon each. Atoms c and d are two atom that can absorb a photon each. $u$ and $v$ are two paths. }\label{1-fano-1961}
\end{figure}

Quantum optical coherence theory based on Feynman\rq{}s path integral was first introduced by Feynman himself. In his lectures on physics, Feynman examined Young\rq{}s double-slit interference of electrons, which is similar as the Young\rq{}s double-slit interference of photons \cite{feynman2011feynman}. Shortly after Hanbury Brown and Twiss\rq{}s experiments \cite{brown1956correlation,hanbury1956test}, Fano employed Feynman\rq{}s path integral to interpret two-photon bunching of thermal light \cite{fano1961quantum}. As shown in Fig. \ref{1-fano-1961}, there are two different and indistinguishable paths for two photons emitted by atoms a and b to be absorbed by atoms c and d.  The first path is the photon emitted by atom a is absorbed by atom c and the photon emitted by atom b is absorbed by atom d. The second path is the photon emitted by atom a is absorbed by atom d and the photon emitted by atom b is absorbed by atom c. Two-photon bunching of thermal light is interpreted as the superposition of two different two-photon probability amplitudes corresponding to these two different paths, which is later named as two-photon interference \cite{shih2020introduction}. In his lecture on quantum electrodynamics, Feynman also employed similar idea to interpret two-photon bunching of thermal light \cite{feynman2006qed}. 

As mentioned before, Glauber\rq{}s quantum optical coherence theory was employed to study optical coherence in most researches in quantum optics. There are limited number of researches employing Feynman\rq{}s path integral to study optical coherence. Hillery and Zubairy employed Feynman\rq{}s path integral to discuss some problems in nonlinear optics and calculated the path integral representation for coherent state propagator \cite{hillery1982path}. Yabuki employed Feynman\rq{}s path integral to discuss the nonlinear contribution in the first-order interference of light in a Young\rq{}s double-slit interferometer \cite{yabuki1986feynman}. Barut and Basri discussed single-photon interference in the single- and double-slit interferometers with Feynman\rq{}s path integral \cite{barut1992path}. Braidotti \textit{et. al.} employed Feynman\rq{}s path integral to discuss some non-paraxial optics problems \cite{braidotti2018path}. St{\"o}hr showed that Feynman\rq{}s path integral reduces to the conventional wave formalism of light diffraction in the lowest order \cite{stohr2019overcoming}. Wallner employed Feynman\rq{}s path integral to discuss the interference of entangled photon pairs \cite{wallner2021feynman}. Wen \textit{et al.} experimentally measured Feynman\rq{}s propagator for single photons \cite{wen2023demonstration}.

We have employed Feynman\rq{}s path integral to discuss the first- and second-order interference of light since 2010 and obtained some interesting conclusions \cite{liu2010unified,liu2011observation,liu2013spatial,liu2014second,liu2014changing,liu2015two,liu2015first,bai2016transient,zhou2017superbunching,zhou2019experimental,luo2021two,luo2021observing}. For instance, it is predicted that there is no transient first-order interference pattern by superposing two independent thermal light beams \cite{bai2016transient}, which is different from the one of superposing two independent laser light beams \cite{javan1962frequency,magyar1963interference,lipsett1963coherence}. The difference leads to an interesting conclusion that classical electric field models for laser and thermal light within the coherence time should be different, which is contrary to the well-accepted electric model introduced in classical optics textbooks \cite{born2013principles}. Detail discussions about this problem can be found in Sect. \ref{sec-first}.

The researches on optical coherence with Feynman\rq{}s path integral performed by other groups \cite{fano1961quantum,feynman2011feynman,feynman2006qed,hillery1982path,yabuki1986feynman,barut1992path,braidotti2018path,wallner2021feynman} and our group \cite{liu2010unified,liu2011observation,liu2013spatial,liu2014second,liu2014changing,liu2015two,liu2015first,bai2016transient,zhou2017superbunching,zhou2019experimental,luo2021two,luo2021observing}  are all about certain specified interference phenomena of light. No systematic study about quantum optical coherence theory based on Feynman\rq{}s path integral is reported.  This paper presents a systematic introduction and review about quantum optical coherence theory based on Feynman\rq{}s path integral.

\subsection{Principles of quantum optical coherence theory based on Feynman\rq{}s path integral}

The key to discuss optical coherence in Feynman\rq{}s path integral is superposition principle. As Feynman pointed out in his lectures, \textit{it contains the only mystery (of quantum mechanics) } \cite{feynman2011feynman}. Feynman summarized how to calculate the probability of an event with superposition principle \cite{feynman2011feynman}:

(I) \textit{The probability of an event in an ideal experiment is given by the square of the absolute value of a complex number $\phi$ which is called the probability amplitude:
\center{$P$=probability,\\
$\phi$=probability amplitude,\\
$P=|\phi|^2$\\}}

(II) \textit{When an event can occur in several alternative ways, the probability amplitudes for the event is the sum of the probability amplitudes for each way considered separately. There is interference:
\center{$\phi=\phi_1+\phi_2$,\\
$P=|\phi_1+\phi_2|^2$\\}}

(III) \textit{If an experiment is performed which is capable of determining whether one or another alternative is actually taken, the probability of the event is the sum of the probabilities for each alternative. The interference is lost:
\center{$P=P_1+P_2$\\}}

There are two premises for calculating the probability of an event with the above assumptions. The first premise is how to calculate the probability amplitude, $\phi$, for each path. The second premise is how to decide two different alternatives for an event are distinguishable or not. 

\subsubsection{Probability amplitude for a path}

The probability amplitude for a path is first given by Dirac by analogy of action in classical dynamics theory \cite{dirac1930principles}.  Dirac assumed that the solution of schr\"{o}dinger wave equation is
\begin{equation}\label{pa-dirac}
\langle q_{t_b}|q_{t_a}\rangle=e^{iS/\hbar},
\end{equation}
where $q_{t_a}$ and $q_{t_b}$ are canonical coordinates of particle at time $t_a$ and $t_b$, respectively, $\hbar=h/(2\pi)$ and $h$ is Planck constant, $S$ is classical action given by \cite{taylor2005classical}
$$
S_{x(t)}=\int_{t_a}^{t_b} dt L(x,\dot{x},t),
$$
$x(t)$ represents the path, $L(x,\dot{x},t)$ is the Lagrangian of the particle taking the path $x(t)$, $\dot{x}$ is short for $\partial  x/ \partial t$. Equation (\ref{pa-dirac}) means that the probability amplitude for a particle starting from $q_{t_a}$ at time $t_a$ and ending at $q_{t_b}$ at time $t_b$ equals $e^{iS/\hbar}$. 

Based on the above assumption, Dirac concluded that the probability amplitude of the path can be expressed as the integral of all the possible paths 
\begin{eqnarray}\label{pa-dirac-2}
\langle q_{t_b}|q_{t_a}\rangle& =& \int ... \int \langle q_{t_b}|q_{t_m}\rangle dq_{t_m} \langle q_{t_m}| q_{t_{m-1}} \rangle dq_{t_{m-1}} ... \nonumber\\
&& \times \langle q_{t_2}| q_{t_{1}} \rangle dq_{t_{1}} \langle q_{t_1}| q_{t_{a}} \rangle,
\end{eqnarray}
where $t_j=t_a+(j-1)\times (t_b-t_a)/m$  ($j=1, 2, ..., m$) and $q_{t_j}$ is the corresponding canonical coordinate at time $t_j$.  Equation (\ref{pa-dirac-2}) is in fact the key result for path integral. The exactly same equation can be found in Feynman\rq{}s book \cite{feynman2010quantum}. The difference between Dirac and Feynma\rq{}s work on path integral is that Dirac gave the probability amplitude by analogy of classical dynamics and he used the word, \textit{analogous to}, while Feynman proved that the probability amplitude in quantum mechanics is \textit{equal to} Eq. (\ref{pa-dirac}) \cite{feynman2010quantum}.

The probability amplitude for photons, usually called photon\rq{}s Feynman propagator, has a simple form in momentum space in Feynman\rq{}s Gauge \cite{greiner2013field}
\begin{equation}\label{propagator-p}
D_F^{\mu\nu}(k)=\frac{-g_{\mu\nu}}{k^2+i\epsilon},
\end{equation}
where $g^{\mu\nu}$ is the metric tensor in four-dimensional Minkowski spacetime,  $k$ is a four-dimension wave vector, $(k_0,k_1,k_2,k_3)$, $k_0=\omega$, $(k_1,k_2,k_3)$ are three components of usual wave vector, $k^2 = k^\mu k_\mu = k_0^2 - k_1^2 - k_2^2 - k_3^2$, $\epsilon$ is an infinitesimal quantity introduced to ensure convergence of the integral.

Photon\rq{}s Feynman propagator in coordinate space can be obtained via Fourier transform of Eq. (\ref{propagator-p})
\begin{equation}\label{propagator-x}
D_F^{\mu\nu}(x;y)=\int \frac{d^4k}{(2\pi)^4}e^{-ik\cdot(y-x)}D_F^{\mu\nu}(k),
\end{equation}
where $x$ and $y$ are four-dimensional Minkowski coordinate vectors, $x=(x_0,x_1,x_2,x_3)$, $x_0$ is time $t$, $(x_1,x_2,x_3)$ is the three components of the usual coordinate vector $\vec{x}$. Equation (\ref{propagator-x}) can be further simplified as  \cite{greiner2013field}
\begin{equation}\label{propagator-x1}
D_F(x;y)= \int \frac{d^3k}{(2\pi)^3}\frac{i}{(2\omega)^2} e^{-i[\omega (t_y-t_x)-\vec{k}\cdot(\vec{y}-\vec{x})]},
\end{equation}
in which $t_y>t_x$ is assumed.

Comparing photon\rq{}s Feynman propagator, Eq. (\ref{propagator-x1}), with Green function in classical optics \cite{born2013principles}
\begin{equation}\label{green}
G(\vec{x},t_x;\vec{y},t_y)= \int d\omega \frac{-i}{|\vec{y}-\vec{x}|\lambda} e^{-i[\omega (t_y-t_x)-\vec{k}\cdot (\vec{y}-\vec{x})]},
\end{equation}
it is easy to find out that these two functions have similar expressions with different contributions for each momentum component. The reason is that all the paths are taken into account in Eq. (\ref{propagator-x1}) with equal amplitude, while all the paths are taken into account in Eq. (\ref{green})  with unequal amplitudes. 

It has been proved by Feynman that if the Lagrangian of a particle has the form of \cite{feynman2010quantum}
\begin{eqnarray}\label{larg-gauss}
&&L(\dot{x},x,t)\nonumber\\
&=&a(t)\dot{x}^2+b(t)\dot{x} x+c(t)x^2+d(t)\dot{x}+e(t)x+f(t),
\end{eqnarray}
the particle\rq{}s Feynman propagator can be expressed as
\begin{equation}\label{larg-kernel}
K(x_a,x_b)=e^{iS_{cl}(x_a,x_b)/\hbar}F(t_a,t_b),
\end{equation}
where one dimension case is assumed for simplicity, $x_a$ and $x_b$ are the starting and ending positions of the particle, respectively, $S_{cl}(a,b)$ is the classical action of the particle, $F(t_a,t_b)$ is a function of the starting and ending time. Equation (\ref{larg-kernel}) means that the particle\rq{}s Feynman propagator going from $x_a$ to $x_b$ can be related to the classical propagation function with an extra function which is only dependent on the starting and ending time. The Lagrangian of free photons has the form of Eq. (\ref{larg-gauss}) \cite{srednicki2007quantum}. Equation (\ref{larg-kernel}) should be valid for photons. Classical Green function can be used as photon\rq{}s Feynman propagator in certain cases.

\subsubsection{Distinguishable and indistinguishable paths}\label{B2}

Indistinguishable paths are a unique concept in quantum theory. All the paths are distinguishable in classical theory. One reason is that measurements are different in quantum and classical theories \cite{wheeler2014quantum}. In classical theory, measurement precision can be arbitrarily high (at least in principle) without introducing unpredictable influence to the measured system. In quantum theory, Measurement precision cannot be arbitrarily high because measurements will introduce unpredictable influence and the uncertainty of the measurement is limited by Heisenberg uncertainty principle. Another reason is that there are no identical particles in classical theory. As Griffiths and Schroeter mention in their book that \textit{all electrons are utterly identical, in a way that no two classical objects can never be}, identical particle is a unique concept in quantum theory \cite{griffiths2018introduction}.

\begin{figure}[htb]
\centering
\includegraphics[width=55mm]{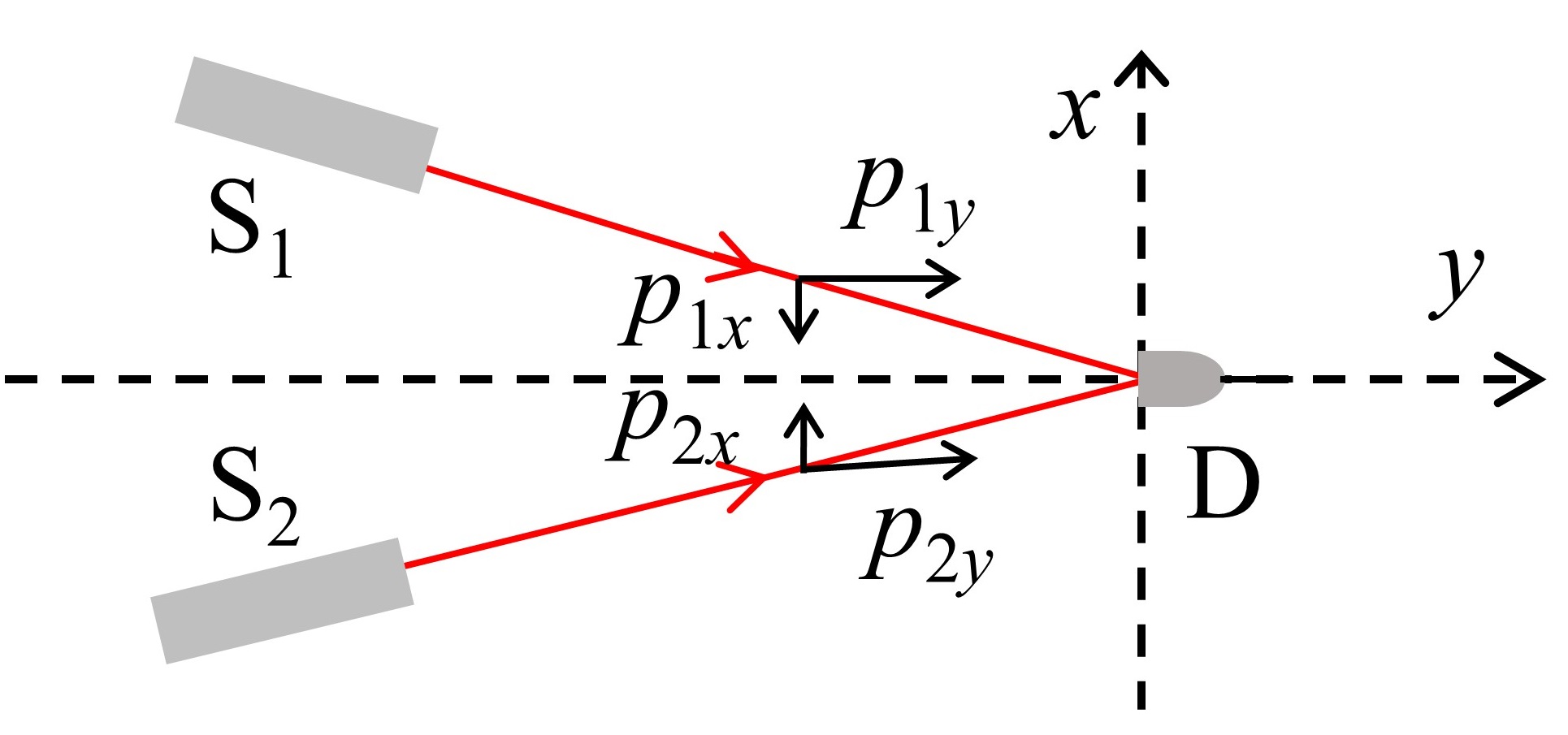}
\caption{Distinguishable and indistinguishable paths. S$_1$ and S$_2$ are two sources which can emit quantum or classical particles. D is a single-particle detector.}\label{2-different-alternatives}
\end{figure}

Let us take the scheme shown in Fig. \ref{2-different-alternatives} as an example to illustrate distinguishable and indistinguishable paths. S$_1$ and S$_2$ are two sources that can emit either quantum or classical particles. D is a single-particle detector. If the detector D detects a particle. The particle comes from either S$_1$ or S$_2$. In classical theory, the particles emitted by these two sources are distinguishable. Even if all the properties such as mass, volume, color and so on, are identical\footnote{This assumption is impossible in classical theory. For instance, two particles having the same weight in classical theory is only valid in some precision. Classical particles are infinitely divisible. One can always find higher precision in which these two particles having different weights.}, the momentum difference between these two cases can be employed to distinguish these two paths. Different paths are always distinguishable in classical theory. Things become different in quantum theory.  Assuming the particles emitted by S$_1$ and S$_2$ are identical, the only difference between these two paths is the momentum difference. The measurement precision of momentum in quantum theory is limited by Heisenberg uncertainty principle \cite{heisenberg1949physical}
\begin{equation}
\Delta p_x \Delta x \geq h,
\end{equation}
where $\Delta p_x$ and $\Delta x$ are the uncertainties of momentum and position of the measured particle, respectively. The uncertainty of position is usually determined by the detection mechanism of the employed detector. Photon is usually detected by photoelectric effect. The position uncertainty of photon detection equals the size of the atom absorbing the photon, which is usually at $10^{-10}$ m range for metal atoms \cite{shackelford2016introduction}. The paths with momentum difference less than $h/10^{-10}$ are indistinguishable. Different paths can be indistinguishable in quantum theory. Heisenberg uncertainty principle is employed to analyze the path difference when detail measuring apparatus is taken into account. 

\subsection{What is a photon?}\label{II-C}

The concept of photon can trace back to Planck \cite{plank1901gesetz} and Einstein \cite{einstein1905heuristic}. The answer to the question of what is a photon is still not satisfactory. However, it does not stop photons \textit{appear everywhere even when there is no need for them} \cite{schleich2003photon}. As stated by Einstein in 1951: \textit{All these fifty years of conscious brooding have brought me no nearer to the answer to the question, -What are light quanta? Nowadays every Tom, Dick and Harry thinks he knows it, but he is mistaken} \cite{afshar2007paradox}.  \textit{Despite significant progress in our understanding, it remains an open question} \cite{roychoudhuri2003nature}. In quantum optical coherence theory based on Feynman\rq{}s path integral, the concept of photon is unavoidable. It is impossible for us to answer the question of what is a  photon right now. However, it is helpful to clarify what we are talking about when the concept of photon is employed. 

In short, we picture photon as a particle, which is described by photon wave function $\psi(\vec{r},t)$. The probability to find a photon at $(\vec{r},t)$ in a volume $dV$ is given by $|\psi(\vec{r},t)|^2dV$. The initial phase of a photon is determined by the emission time.

\subsubsection{Photon is a particle} 

Wave-particle duality is well-accepted for photon. When we talk about photon in this paper, we picture it as a particle like electron, proton and so on. The particle is described by a photon wave function $\psi(\vec{r},t)$, which is different from the concept of particle in classical theory.

The evidence that photon is a particle includes photoelectric effect \cite{hertz1887ueber} and photon antibunching effect \cite{paul1982photon}. Hertz observed the following phenomena when shining a light beam on the surface of a metal:\\
(I) The kinetic energy $E$ of the ejected electron is equal to $\hbar \nu - \Phi$, where $\nu$ is the frequency of incidental light, and $\Phi$ is the work function of the metal. \\
(II) The rate of electron ejection is proportional to the square of the incident electric filed.\\
(III) The time delay $T$ between the light shining on the metal and the ejected photo-electron is very short.\\
It is experimentally proved by Lawrence and Beams that $T$ is shorter than 3 ns \cite{lawrence1928element}. Forrester \textit{et al.} further proved that $T$ is much shorter than 0.1 ns \cite{forrester1955photoelectric}. The first two phenomena can be interpreted by wave theory. The last phenomenon can only be interpreted by treating photon is a particle \cite{muthukrishnan2017concept}.

The second evidence for that photon is particle is photon antibuching effect \cite{paul1982photon}. The physics of photon antibunching is that one photon cannot trigger two separated single-photon detectors simultaneously. The key to observe photon antibunching is to generate single-photon state. Kimble \textit{et al.} \cite{kimble1977photon} and Grangier \textit{et al.} \cite{grangier1986experimental} observed photon antibunching effect with different ways to generate single-photon state.  Classical electromagnetic wave theory can not interpret photon antibunching effect \cite{paul1982photon, klyshko1996nonclassical}. 

There are also other evidence for photon is a particle. For instance, when a light beam with low intensity is incident into an interferometer, the first-order interference pattern builds up point by point on the observation plane \cite{dimitrova2008wave}.

\subsubsection{Photon\rq{}s wave function}

Strictly speaking, there is no wave function for photon since there is no position operator for photon \cite{bohm2012quantum}. However, one can define photon\rq{}s position operator from the point of view of detection. Photon is usually detected by photoelectric effect. The position resolution is limited by the size of atom. As long as the position accuracy of photon\rq{}s wave function is lower than the size of atom, one can define and use photon\rq{}s wave function by analogy of the wave function of a massive particle.

The wave function of a single photon emitted by a two-level atom is \cite{scully1997quantum}
\begin{equation}\label{wave-function}
\psi(\vec{r},t)=K\frac{\sin \eta}{r}\theta(t-\frac{r}{c})\exp{[-i(\omega+i\frac{\Gamma}{2})(t-\frac{r}{c})]},
\end{equation}
where $K$ is a normalization constant, $r=|\vec{r}-\vec{r}_0|$ is the distance between the position of the atom, $\vec{r}_0$, and the position of photon, $\vec{r}$, $\eta$ is the angle of $\vec{r}$ related to the electric dipole of atom, $\theta(t-\frac{r}{c})$ is a step function, which indicates that a photon can be detected at  $\vec{r}$ only when it propagates to position $\vec{r}$, $\omega$ is the frequency of photon, $\Gamma$ is the width of the up-level energy of two-level system, $\exp{[-i(\omega+i\frac{\Gamma}{2})(t-\frac{r}{c})]}$ indicates that the probability of detecting the photon decays by $e^{-\Gamma t}$.

Equation (\ref{wave-function}) gives the temporal part of photon\rq{}s wave function. The spatial part of photon\rq{}s wave function can be treated the same as the spatial structure of classical electric field \cite{scully1997quantum}. For instance, the electric field emitted by a point source is spherical wave in classical theory. The spatial wave function of single photon emitted by a point source is also spherical.
 
\subsubsection{Photon\rq{}s phase}

Photon\rq{}s phase is defined by analogy of the phase of electric field in classical electromagnetic theory \cite{dirac1930principles}. Classical electric field can be expressed as \cite{born2013principles}
\begin{equation}\label{2-classical-field}
\vec{E}(\vec{r},t)=\vec{E}_0 \exp{[-i(\omega t -\vec{k}\cdot\vec{r})+i\varphi_0]},
\end{equation}
where $E_0$ is the amplitude of the field, $\omega t$ is the phase related to field propagating in time, $\vec{k}\cdot\vec{r}$ is the phase related to the field propagating in space, $\varphi_0$ is the initial phase of the field, $\varphi=-\omega t +\vec{k}\cdot\vec{r}+\varphi_0$ is the phase of electric field in classical theory. 

Dirac gave a definition of quantum phase \cite{dirac1927quantum}. However, the phase operator defined by Dirac encounters difficulties when it is expanded in terms of number states. Several other definitions of quantum phase were also given \cite{susskind1964quantum,barnett1986phase,pegg1997quantum}. The usually employed phase operator is defined by phase shifts \cite{pegg1997quantum,loudon2000quantum}
\begin{equation}\label{2-phase-state}
|\varphi_m\rangle=\frac{1}{\sqrt{1+r}}\sum_{n=0}^r e^{in\varphi_m}|n\rangle,
\end{equation}
where $\varphi_m=2\pi m/(r+1)$ is a discrete phase to simplify the calculation, $m=0,1,2...r$, and $r$ is a very large positive integer. When $r$ goes to infinity, the discrete phase becomes continuous. 

The uncertainties of phase ($\Delta \varphi$) and number of photons ($\Delta N$) satisfies the following uncertainty principle \cite{pegg1997quantum},
\begin{equation}
\Delta N \Delta \varphi  \geq \frac{1}{2}.
\end{equation}
For number state, $|N\rangle$, the uncertainty of photon number is zero, which means that the phases of photons in a number state are completely uncertain. It can be proved that the phase distribution function,
\begin{equation}\label{2-phase-density}
P(\varphi)=\frac{1}{2\pi}|\sum_0^\infty c_n \exp{(-in\varphi)}|^2,
\end{equation}
of single-photon state equals $1/2\pi$, which means that the phase of the photon in a single-photon state is equally distributed between 0 and $2 \pi$ \cite{pegg1997quantum}.

Feynman assumed that the initial phase of a photon is determined by the emission time \cite{feynman2006qed}. Based on the phase relationship between different photons in a light beam, light can be categorized into two groups. The first group is that the initial phases of photons in a light beam are random. The photons in this type of light are usually emitted by spontaneous emission. Chaotic light, thermal light, number state are examples of this type of light. The second group is that the initial phases of photons in a light beam are identical. The photons in this type of light are usually emitted by stimulated emission. Single-mode laser, amplified spontaneous emission belongs to the second type of light.

\subsection{Coherence volume and degeneracy factor of light}

The key to applying superposition principle in Feynman\rq{}s path integral is to determine whether different paths for an event are distinguishable or not.  Heisenberg uncertainty principle is employed to judge whether different paths are distinguishable or not. Indistinguishable paths are close related to indistinguishable particles. Photons within the same coherence volume are indistinguishable \cite{martienssen1964coherence}.  The uncertainties of coordinates ($\Delta x$, $\Delta y$, and $\Delta z$) and momentum  ($\Delta p_x$, $\Delta p_y$, and $\Delta p_z$) of photons within the same coherence volume satisfies the following equations,
\begin{eqnarray}
&&\Delta x \Delta p_x = h,\nonumber \\
&&\Delta y \Delta p_y = h,\\
&&\Delta z \Delta p_z = h.\nonumber
\end{eqnarray}
Assuming the light beam propagating in $z$ direction and the frequency bandwidth of light is $\Delta v$, it can be proved that the coherence volume equals the transverse coherence area times the longitudinal coherence length, $c/\Delta v$  \cite{martienssen1964coherence}. 

The average number of photons within one coherence volume is called degeneracy factor of the light. For blackbody radiation, the degeneracy factor is given by 
\begin{equation}
\delta=\frac{1}{e^{h\nu/k_BT}-1},
\end{equation}
where $h$ is the Planck constant, $\nu$ is the frequency of radiation, $k_B$ is the Boltzmann constant, $T$ is the temperature of blackbody. Thermal light can be treated as blackbody radiation. The degeneracy factor of usual thermal light is much less than 1. For instance, the degeneracy factor of sun light at 532 nm equals 0.0073. On the other hand, the degeneracy factor of laser light can be much larger than 1. For instance, the degeneracy factor a 1-mW single-mode continuous-wave He-Ne laser light beam equals $3\times 10^9$ \cite{mandel1995optical}.

\section{First-order interference of light}\label{sec-first}

The first-order interference of light can be well interpreted in classical optical coherence theory based on Maxwell\rq{}s electromagnetic theory \cite{born2013principles}. Dirac firstly interpreted the first-order interference of light in the language of photons and gave the well-known statement about the first-order interference of light: ``\textit{Each photon then interferes only with itself. Interferences between two different photons never occurs} \cite{dirac1930principles}.\rq{}\rq{} The contents in this section are mainly based on Dirac\rq{}s statement. For a comprehensive review about the first-order interference of light, please refer to Refs. \cite{born2013principles,hecht2012optics,hariharan2003optical}.

\subsection{First-order interference of one light beam}

\begin{figure}[htb]
\centering
\includegraphics[width=55mm]{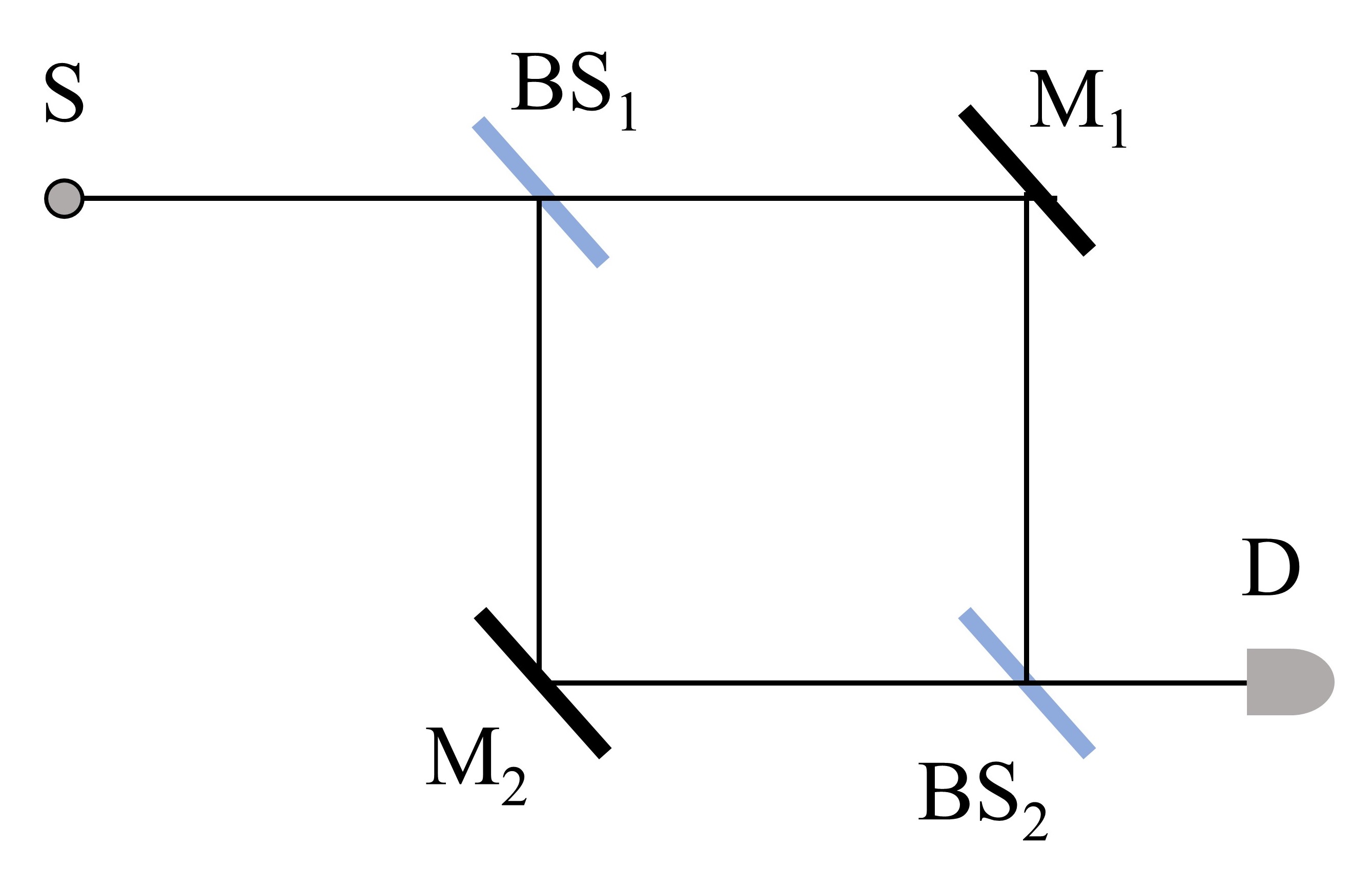}
\caption{Mach-Zehnder interferometer. S is a light source. BS$_1$ and BS$_2$ are two non-polarizing 1:1 beam splitter. M$_1$ and M$_2$ are two mirrors. D is a single-photon detector.}\label{3-MZ-interferometer}
\end{figure}

The first-order interference of one light beam means that light emitted by one source is split into two or more components and then combined together to observe the first-order interference pattern. It corresponds to each photon interferes only with itself in Dirac\rq{}s statement. Take Mach-Zehnder interferometer shown in Fig. \ref{3-MZ-interferometer} as an example to show how to employ Feynman\rq{}s path integral to discuss the first-order interference of one light beam.

There are two different paths for a single-photon detector, D, detecting a photon in Fig. \ref{3-MZ-interferometer}. The first path is the detected photon taking the upper path (transmitting through beam splitter BS$_1$,  reflected by mirror M$_1$, and reflected by beam splitter BS$_2$) and recorded as path 1 for simplicity. The probability for the photon taking the path is $1/2$ and the probability amplitude for the photon taking the path is $A_1(S,D)$. The second path is the photon taking the lower path (reflected by BS$_1$ and M$_2$, transmitting through BS$_2$) and recorded as path 2. The probability for the photon taking the path is $1/2$ and the probability amplitude for photon taking the path is $A_2(S,D)$. If these two different paths to trigger a photon detection event are indistinguishable, the probability function for the $j$th detected photon is 
\begin{eqnarray}\label{Pj}
P_j(t)&=&|\frac{1}{\sqrt{2}}A_1(S,D)+\frac{1}{\sqrt{2}} A_2(S,D)|^2\nonumber\\
&\propto&|e^{i\varphi_j}K_1(S,D)+e^{i\varphi_j}K_2(S,D)|^2,
\end{eqnarray}
where $j$ is a positive integer, $\varphi_j$ is the initial phase of the $j$th detected photon, $K_1(S,D)$ is photon\rq{}s Feynman propagator from the source to the detector via path 1, $A_1(S,D)=e^{i\varphi_j}K_1(S,D)$. Similar definition holds for $K_2(S,D)$. The first-order interference of one light beam is due to photon interferes with itself. The initial phase is identical for the same photon taking two different paths.

The first-order interference pattern is proportional to the sum of the probability functions of all the detected photons,
\begin{eqnarray}\label{I}
I(t) \propto \sum_{j=1}^N P_j(t),
\end{eqnarray}
where $N$ is the number of the detected photons. For a large enough number of photons, 
\begin{eqnarray}\label{I2}
I(t) \propto \langle P_j(t) \rangle,
\end{eqnarray}
where $\langle ... \rangle$ is ensemble average by taking all possible realizations \cite{mandel1995optical,shih2020introduction}.
Substituting Eq. (\ref{Pj}) into Eq. (\ref{I2}), the first-order interference pattern in Fig. \ref{3-MZ-interferometer} is
\begin{eqnarray}\label{I3}
I(t) \propto \langle |K_1(S,D)+K_2(S,D)|^2 \rangle.
\end{eqnarray}
The initial phase of the photon, $\varphi_j$, does not contribute to the first-order interference pattern in Fig. \ref{3-MZ-interferometer}. Regardless of whether thermal or laser light is used, as long as the frequency bandwidth and spatial distribution are identical, the observed first-order interference patterns will be the same. The reason is that the first-order interference is a result of photon interferes with itself. 

For a point light source, the photon\rq{}s Feynman propagator can be approximated as
\begin{equation}\label{kernel}
K(\vec{r}_0,t_0;\vec{r},t) \propto \frac{1}{|\vec{r}-\vec{r}_0|}e^{-i[\omega_0 (t-t_0)-\vec{k}\cdot (\vec{r}-\vec{r}_0)]},
\end{equation}
where $(\vec{r}_0,t_0)$ is the space-time coordinate of the photon being emitted, $(\vec{r},t)$ is the  space-time coordinate of the photon being detected, $\omega_0$ is angular frequency, $\vec{k}$ is wave vector. For simplicity, single-frequency light is assumed firstly. Substituting Eq. (\ref{kernel}) into Eq. (\ref{I3}) and considering the temporal part, the first-order interference pattern in Fig. \ref{3-MZ-interferometer} is
\begin{eqnarray}\label{I-temporal}
I(t_1-t_2) &\propto& P_j(t_1-t_2)\nonumber\\
 &\propto& 1+\cos[\omega_0(t_1-t_2)],
\end{eqnarray}
where $t_1$ and $t_2$ are the travailing time of the photon from S to D via path 1 and 2, respectively. The first-order interference pattern is proportional to the probability function of the $j$th detected photon in Fig. \ref{3-MZ-interferometer}, which is the same as the first-order interference pattern obtained in classical optical coherence theory \cite{born2013principles}.

The first-order interference pattern of light with finite frequency bandwidth can be calculated in the same way. Assuming S is a point single-photon source, the photon wave function can be expressed as \cite{bialynicki1996v,davis2018measuring}
\begin{equation}\label{single-photon}
|1\rangle=\int d\omega f(\omega) \hat{a}^\dag(\omega) |0\rangle,
\end{equation}
where $f(\omega)$ is the frequency distribution function, $ \hat{a}^\dag(\omega)$ is the creation operator, $ |0\rangle$ is vacuum state. When the time difference between the photon taking two paths, $|t_1-t_2|$, is longer than the coherence time of the employed single-photon state, the two different paths to trigger a photon detection event at D are distinguishable. The probability function for the $j$th detected photon equals
\begin{eqnarray}\label{wp-distinguishable}
&&P_j(t)\nonumber\\
&=& |\frac{1}{\sqrt{2}}e^{i\varphi_j}K_1(S,D)|^2+|\frac{1}{\sqrt{2}}e^{i\varphi_j}K_2(S,D)|^2  \\
&\propto& |\int d\omega f(\omega)e^{-i\omega (t-t_{01})}|^2+|\int d\omega\rq{} f(\omega\rq{})e^{-i\omega\rq{} (t-t_{02})}|^2, \nonumber
\end{eqnarray}
where $t$ is the time of photon being detected, $t_1=t-t_{01}$, $t_{01}$ is the emitting time of the photon detected by D via path 1, $t_2=t-t_{02}$ and $t_{02}$ is the emitting time of the photon detected by D via path 2. The probability function of the $j$th detected photon is the sum of the probability functions of the photon taking two different paths. There is no first-order interference.

When the time difference, $|t_1-t_2|$, is shorter than the coherence time of the employed single-photon state, these two different paths to trigger a photon detection event at D are indistinguishable. The probability function of the $j$th detected photon is
\begin{eqnarray}\label{wp-indistinguishable}
P_j(t)&=& |\frac{1}{\sqrt{2}}e^{i\varphi_j}K_1(S,D)+\frac{1}{\sqrt{2}}e^{i\varphi_j}K_2(S,D)|^2\\
&\propto& |\int d\omega f(\omega)e^{-i\omega (t-t_{01})}+ \int d\omega' f(\omega')e^{-i\omega' (t-t_{02})}|^2 \nonumber
\end{eqnarray}
Assume the frequency distribution function is Gaussian,
\begin{equation}\label{guassian}
f(\omega)=A \exp{[-\frac{(\omega-\omega_0)^2}{2\sigma^2}]},
\end{equation} 
where $A$ is a normalization constant,  $\sigma^2$ is variance of Gaussian function, $\omega_0$ is the central frequency. Equation (\ref{wp-indistinguishable}) can be simplified as \cite{shih2020introduction}
\begin{equation}
P_j(t_1-t_2)=1+ \frac{e^{-\sigma^2(t_1^2+t_2^2)/2}}{e^{-\sigma^2t_1^2}+e^{-\sigma^2t_2^2}} \cos{\omega_0 (t_1-t_2)}.
\end{equation}
The first-order interference pattern of light can not be observed with only one photon. The first-order interference pattern of a single-photon state can be observed by preparing many identical single-photon states and put these single-photon states into the interferometer one by one. By collecting large enough number of photons, the first-order interference pattern with single-photon states in Fig. \ref{3-MZ-interferometer} is
\begin{eqnarray}
I(t_1-t_2) &\propto& \langle P_j(t_1-t_2)  \rangle \nonumber\\
&=& 1+ \frac{e^{-\sigma^2(t_1^2+t_2^2)/2}}{e^{-\sigma^2t_1^2}+e^{-\sigma^2t_2^2}} \cos{\omega_0 (t_1-t_2)},
\end{eqnarray}
which is the same the probability function of the $j$th detected photon.

\begin{figure}[htb]
\centering
\includegraphics[width=80mm]{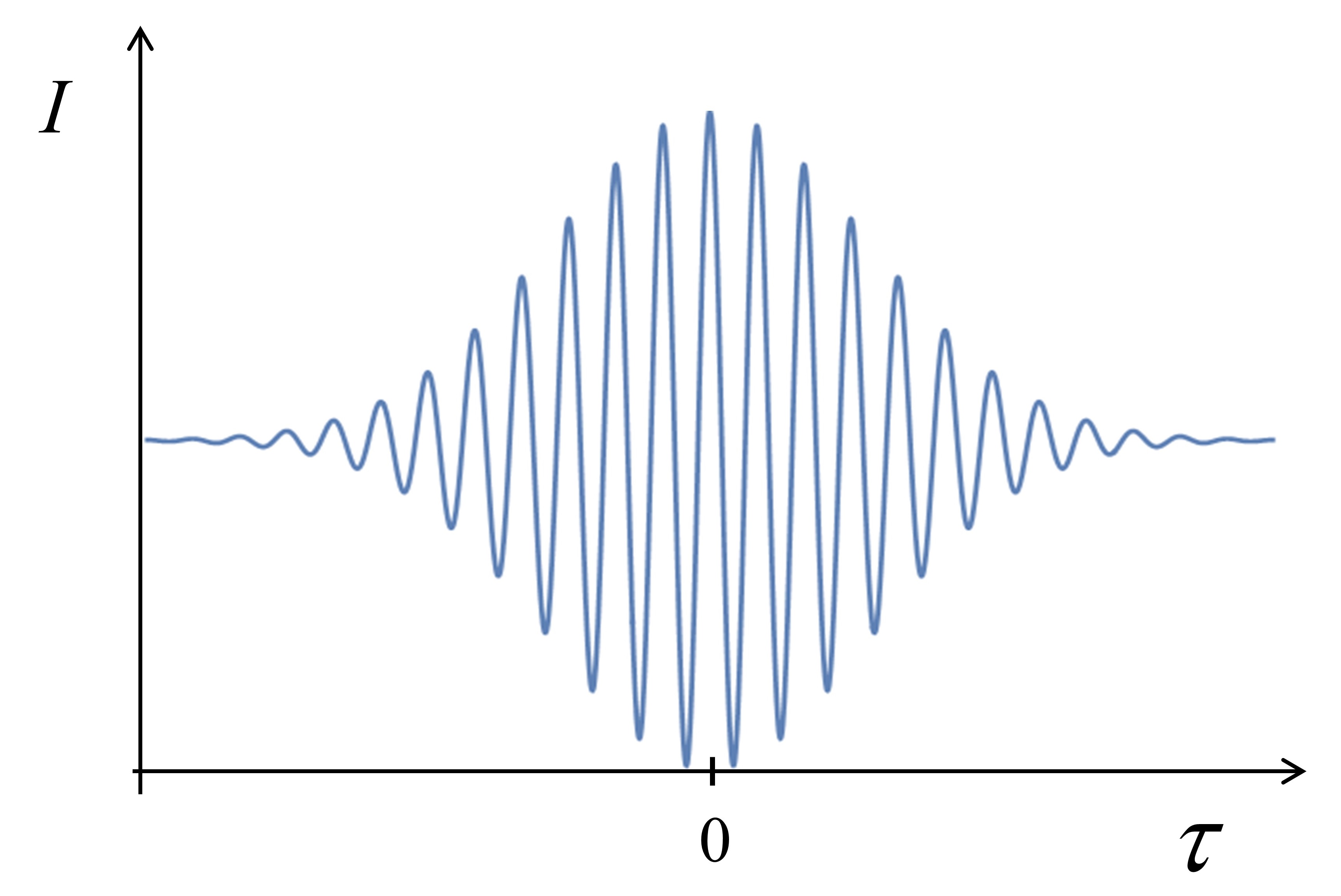}
\caption{First-order temporal interference pattern of single-photon states in a Mach-Zehnder interferometer. I is intensity of light and $\tau$ is time difference between photon taking these two paths.}\label{4-MZ-interference}
\end{figure}

Figure \ref{4-MZ-interference} shows the simulated first-order interference pattern of single-photon states with Gaussian spectrum in a Mach-Zehnder interferometer. $\tau=t_1-t_2$ is the time difference between paths 1 and 2, I is the intensity of light. When the time difference equals 0, the visibility of the first-order interference pattern can reach 100\%. As the time difference becomes larger, the visibility of the first-order interference pattern decreases. When the time difference is larger than the coherence time of light, the visibility goes to zero, which is consistent with Eq. (\ref{wp-indistinguishable}). The simulated result in Fig. \ref{4-MZ-interference} is consistent with the experimental results with single photons \cite{braig2003experimental}

Similar method can be employed to calculate the first-order spatial interference pattern of one light beam in Feynman\rq{}s path integral. 

\subsection{First-order interference of two independent light beams}\label{B-two}

The first-order interference of one light beam confirms Dirac\rq{}s statement that photon only interferes with itself. Based on Popper\rq{}s philosophy that a scientific conclusion can not be proved to be true \cite{popper2005logic}. The first-order interference of one light beam does not prove Dirac\rq{}s statement is correct. On the contrary, any scientific conclusion should be possible to be disproved. Dirac\rq{}s statement may be disproved if one can observe the first-order interference pattern with two independent light beams. 

When Young first discussed the interference of light, he pointed out that there is first-order interference pattern of two light beams from different origins by analogy of sound waves \cite{young1802ii}: \textit{When two Undulations, from different Origins, coincide either perfectly or very nearly in Direction, their joint effect is a Combination of the Motions belonging to each.} Later, Young realized that it is impossible to observe first-order interference pattern of light from two different origins. He observed the first-order interference pattern of sunlight by splitting one beam into two and then combining these two beams together \cite{young1804bakerian}. It is now well-accepted that in order to observe the first-order interference pattern with two light beams,  these two light beams must have the same polarization and frequency, and fixed relative phase \cite{hecht2012optics}. The easiest way to meet these conditions is that these two light beams are splitting from the same one. Most of the first-order interferometers are designed in this way \cite{born2013principles}.

Thing becomes different when laser was invented \cite{maiman1960stimulated,hecht2010short}.  Several groups superposed two light beams emitted by two independent lasers and observed the first-order interference pattern within the coherence time \cite{javan1962frequency,magyar1963interference,lipsett1963coherence}, which is usually called transient first-order interference pattern. Does the observed transient first-order interference pattern with two independent laser light beams means that the second part of Dirac\rq{}s statement is disproved? Further experiments shows that the first-order interference pattern exists when the intensities of these two laser light beams are so low that there is only one photon in the interferometer in most time \cite{pfleegor1968further,radloff1968interference,hariharan1993interference}. It is interpreted by Pfleegor and Mandel that the observed transient first-order interference pattern does not conflict with Dirac\rq{}s statement \cite{pfleegor1967interference}. It is the detection of a photon that puts the photon into a superposition state and causes interference \cite{pfleegor1967interference}. What Glauber said about these experiments is easier to understand \cite{glauber1995dirac}: \textit{The things that interfere in quantum mechanics are not particles. They are probability amplitudes for certain events. It is the fact that probability amplitudes add up like a complex numbers that is responsible for all quantum mechanics interferences.}

\subsubsection{First-order interference of two independent laser light beams}\label{B1}

\begin{figure}[htb]
\centering
\includegraphics[width=55mm]{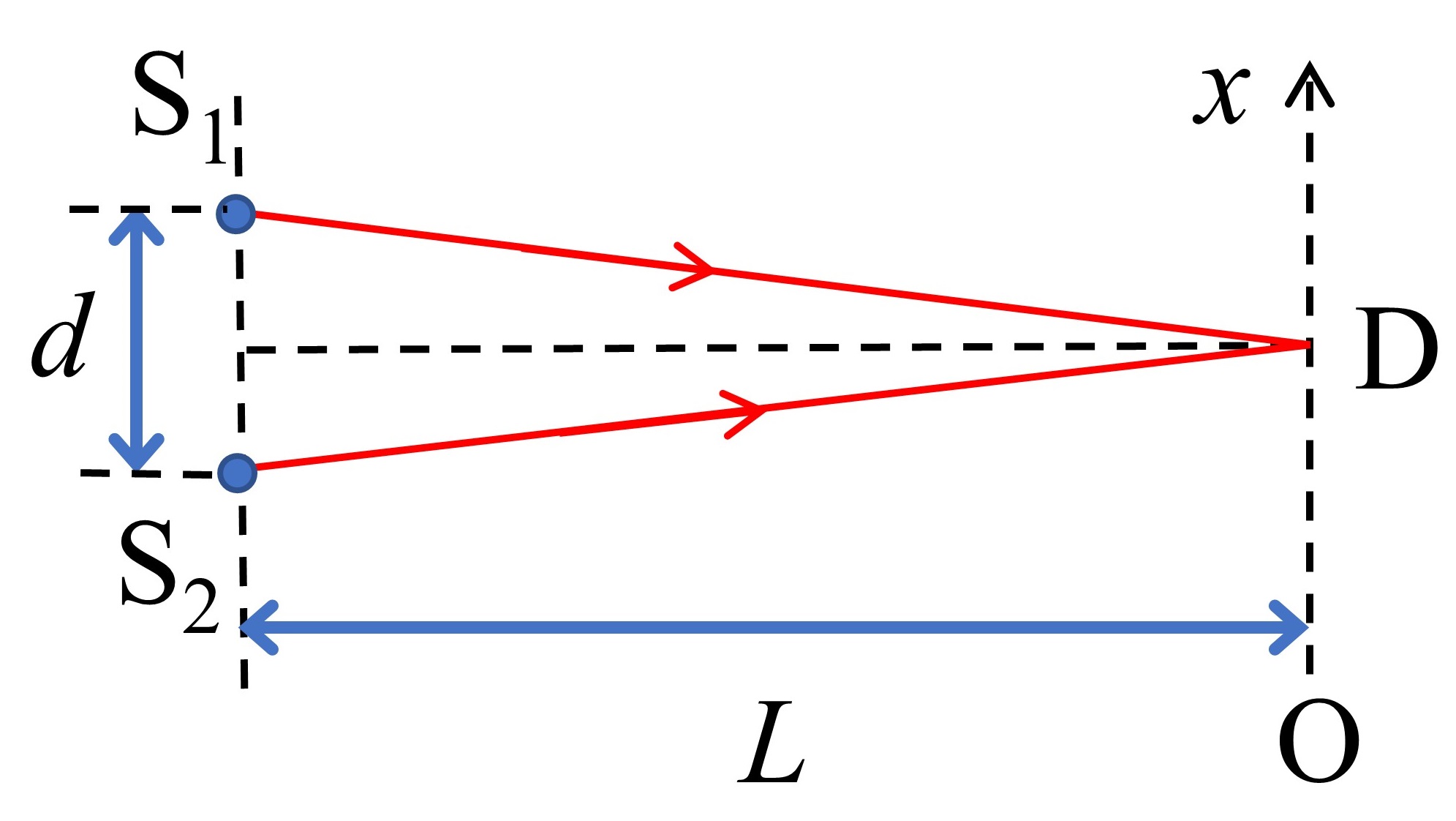}
\caption{The first-order interference of two independent light beams. S$_1$ and S$_2$ are two independent light sources. The distance between two sources is $d$. The distance between the source and observation planes equals $L$. O is the observation plane. D is a single-photon detector.}\label{5-two-lasers}
\end{figure}

In the scheme of the first-order interference of two independent laser light beams shown in Fig. \ref{5-two-lasers}, there are two different paths for the single-photon detector, D, detecting a photon. The first path is that the detected photon comes from laser S$_1$. The probability for the detected photon comes from S$_1$ is $I_1/(I_1+I_2)$, where $I_1$ and $I_2$ are the intensities of light emitted by S$_1$ and S$_2$, respectively. The corresponding probability amplitude is $A_{1}(x)$. The second one is that the detected photon comes from laser S$_2$. The probability for the detected photon comes from S$_2$ is $I_2/(I_1+I_2)$ and the corresponding probability amplitude is $A_{2}(x)$. If these two laser light beams have the same polarization and frequency bandwidth, these two different paths can only be distinguished by momentum difference. As introduced in Sect. \ref{B2}, when the distance between two lasers satisfies
\begin{equation}
d\leq \frac{\lambda L}{\Delta x},
\end{equation}
these two different paths are indistinguishable, where $L$ is the distance between source and observation planes, $\lambda$ is the wavelength of the photon, and $\Delta x$ is the position uncertainty of photon detection. The probability function of the $j$th detected photon is
\begin{eqnarray}\label{Pj-two-lasers}
P_j(x)&=& |\sqrt{\frac{I_1}{I_1+I_2}}A_{j1}(x)+\sqrt{\frac{I_2}{I_1+I_2}}A_{j2}(x)|^2\nonumber\\
&\propto&| \sqrt{I_1}e^{i\varphi_{j1}}K_{j1}(x)+ \sqrt{I_2}e^{i\varphi_{j2}}K_{j2}(x)|^2,
\end{eqnarray}
where $\varphi_{j1}$ and $\varphi_{j2}$ are the initial phases of the $j$th detected photon emitted by S$_1$ and S$_2$, respectively, $K_{j1}(x)$ and $K_{j2}(x)$ are the corresponding photon\rq{}s Feynman propagators.

For two point light sources with equal intensity, Eq. (\ref{Pj-two-lasers}) can be simplified as 
\begin{equation}\label{Pj-two-lasers2}
P_j(x)\propto 1+\cos[\frac{2\pi d}{\lambda L}x+(\varphi_{j1}-\varphi_{j2})],
\end{equation}
where paraxial approximation and one-dimension are employed to simplify the calculation \cite{born2013principles}. The initial phases of photons in a single-mode continuous-wave laser light beam are identical within one coherence volume. The initial phases of photons in different coherence volumes are random. The initial phases of two independent lasers, $\varphi_{j1}$ and $\varphi_{j2}$, are independent. However, the relative phase, $\varphi_{j1}-\varphi_{j2}$, is fixed within one coherence volume for two independent lasers. The observed intensity is proportional to the sum of the probability functions of all the detected photons
\begin{eqnarray}
I(x) &\propto& \sum_j^N P_j(x)\nonumber\\
& \propto & \langle 1+\cos[\frac{2\pi d}{\lambda L}x+(\varphi_{j1}-\varphi_{j2})] \rangle.
\end{eqnarray}
When the observation time is shorter than the coherence time, $\varphi_{j1}-\varphi_{j2}$ is a random but fixed number. The observed interference pattern of two independent lasers is
 \begin{eqnarray}
I(x) &\propto& 1+\cos[\frac{2\pi d}{\lambda L}x+(\varphi_{j1}-\varphi_{j2})], 
\end{eqnarray}
where the first-order interference pattern can be observed.

When the observation time is much longer than the coherence time, the sum of random phases averages to a constant. The observed intensity
 \begin{eqnarray}
I(x) &\propto& 1, 
\end{eqnarray}
where the first-order interference pattern of two independent laser light beams disappears for a long observation time.

The results above are consistent with Magyar and Mandel\rq{}s experiments \cite{magyar1963interference}. Even through the physics of the interference of two independent lasers is in question for a long time \cite{mandel1964quantum,louradour1993interference,wallace1994comment,davis1994comment,paul1986interference}, the experimental results are certain and the predictions in quantum optical coherence theory is consistent with the ones obtained in classical optical coherence theory.

\subsubsection{First-order interference of two independent thermal light beams}\label{B-laser}

Things become different for the first-order interference of two independent thermal light beams. There are two major differences between photons in laser and thermal light. The first difference is that the initial phases of photons in thermal light are random within one coherence volume and the initial phases of photons in laser light are identical within one coherence volume. The second difference is that the degeneracy factor of thermal light is usually much less than one and the degeneracy factor of laser light can be much larger than one. The latter difference indicates that it may be impossible to detect large enough number of photons to observe whether there is transient first-order interference pattern of two independent thermal light beams within the coherence time or not. No experimental observation of the transient first-order interference pattern of two independent thermal light beams was reported as the one for two independent lasers \cite{magyar1963interference}.

\begin{figure}[htb]
\centering
\includegraphics[width=70mm]{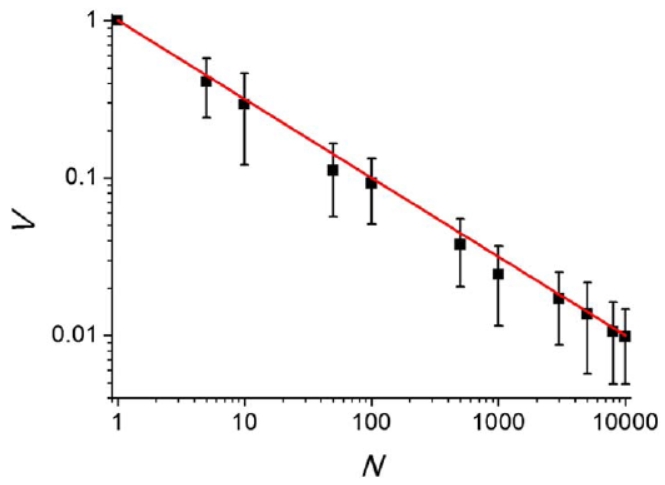}
\caption{The visibility of first-order interference pattern of two independent thermal light beams vs. the number of detected photons \cite{bai2016transient}. $V$ is the visibility of first-order interference pattern and $N$ is the number of detected photons. }\label{6-thermal-visibility}
\end{figure}

Theoretically, one can assume that the degeneracy factor of thermal light is much larger than one and discuss whether there exists transient first-order interference pattern or not. The scheme for the first-order interference of two independent thermal light beams is shown in Fig. \ref{5-two-lasers}, where S$_1$ and S$_2$ are two identical and independent thermal light sources. There are two different paths for the single-photon detector D detecting a photon. Assume the intensities of these two light beams are equal and the same assumption is always employed in the following situations unless specified otherwise. With the same process as the one in Sect. \ref{B1}, the probability function for the $j$th detected photon is
\begin{equation}\label{Pj-two-thermal}
P_j(x)\propto 1+\cos[\frac{2\pi d}{\lambda L}x+(\varphi_{j1}-\varphi_{j2})],
\end{equation}
where paraxial approximation and one-dimension are assumed. The probability function for $N$ detected photons is
\begin{eqnarray}\label{Pn-two-thermal}
P_N(x)&=& \sum _j^N 1+\cos[\frac{2\pi d}{\lambda L}x+(\varphi_{j1}-\varphi_{j2})].
\end{eqnarray}
Since the initial phases of photons in thermal light are random, $\varphi_{j1}-\varphi_{j2}$ is random for each detected photon. For a large enough number of photons, Eq. (\ref{Pn-two-thermal}) can be simplified as \cite{bai2016transient}
\begin{equation}\label{3-30}
P_N(x)\propto 1+\frac{1}{\sqrt{N}}\cos(\frac{2\pi d}{\lambda L}x+\varphi),
\end{equation}
where $\varphi$ is the result of the sum of large number of random phases. The visibility of the first-order interference pattern after detecting $N$ photon is
\begin{equation}\label{Vn}
V_N=\frac{P_{Nmax}-P_{Nmin}}{P_{Nmax}+P_{Nmin}}=\frac{1}{\sqrt{N}},
\end{equation}
which decreases as the number of detected photons increases. Figure \ref{6-thermal-visibility} shows the relationship between the visibility of the simulated first-order interference pattern of two independent thermal light beams and the number of detected photons. $V$ is the visibility and $N$ is the number of detected photons. The red solid line is theoretical value given by Eq. (\ref{Vn}) and square dots are numerical results. Theoretical and numerical results are consistent \cite{bai2016transient}.

The first-order interference pattern of light can not be observed with only one or several photons due to one photon only triggers one photo-electron \cite{dimitrova2008wave}. There must be large enough number of photons to retrieve the first-order interference pattern. However, the visibility decreases as the number of detected photons increases via Eq. (\ref{Vn}). The first-order interference pattern of two independent thermal light beams may not be observable. The reason is similar as the one for the first-order interference pattern of two independent laser light beams can not be observed for a long observation time. 

\subsubsection{First-order interference of two independent non-classical light beams}
 
Feynman\rq{}s path integral can also be employed to calculate the first-order interference of two independent non-classical light beams. Assume S$_1$ and S$_2$ in Fig.  \ref{5-two-lasers} are two identical and independent single-photon sources. There are two different paths for single-photon detector detecting a photon. The detected photon is emitted by either S$_1$ or S$_2$. The probability amplitudes for these two paths are $A_{1}(x)$ and $A_{2}(x)$, respectively. When the distance between these two sources is less than $\lambda L/\Delta x$, these two difference paths can not be distinguished by momentum difference. However, these conditions do not guarantee that these two paths are indistinguishable as the one in laser and thermal light cases.

When there is only one photon in the interferometer, these two difference paths are always distinguishable. One can measure the status of the single-photon sources to ensure the detected photon comes from S$_1$ or S$_2$ without changing the detection results. The probability function for the $j$th detected photon is 
\begin{eqnarray}
P_j(x)&=&|\frac{1}{\sqrt{2}}A_{j1}(x)|^2+|\frac{1}{\sqrt{2}}A_{j2}(x)|^2\nonumber\\
&\propto& |e^{i\varphi_{j1}}K_{j1}(x)|^2+|e^{i\varphi_{j2}}K_{j2}(x)|^2.
\end{eqnarray}
The probability function equals the sum of the probability functions for these two paths. There is no interference.

When these two single-photon sources emit photons simultaneously, there are two photons in the interferometer and these two different paths can be indistinguishable when all the other conditions are satisfied. The probability function for the $j$th detected photon is 
 \begin{eqnarray}
P_j\rq{}(x)= |\frac{1}{\sqrt{2}}e^{i\varphi_{j1}}K_{j1}(x)+\frac{1}{\sqrt{2}}e^{i\varphi_{j2}}K_{j2}(x)|^2.
\end{eqnarray}
 
The probability function after detecting $N$ photons is the sum of the probability function corresponding to these two cases,
\begin{eqnarray}\label{two-single}
P_N(x)=\sum_{j=1}^{N_1}P_j(x)+\sum_{j=1}^{N_2}P_j\rq{}(x),
\end{eqnarray}
where $N_1$ is the number of photons when only one source emitted photons, $N_2$ is the number of photons when two sources emitted photons simultaneously, $N_1+N_2=N$. With the same method as the one for the first-order interference of two independent thermal light beams,  Eq. (\ref{two-single}) can be simplified as
\begin{eqnarray}\label{two-single-2}
P_N(x)\propto 1+ \frac{N_2}{N}\frac{1}{\sqrt{N_2}}\cos(\frac{2\pi d}{\lambda L}x+\varphi).
\end{eqnarray}
 $N_2$ is much less than $N$ for the probability of two independent single-photon sources simultaneously emitting one photon each is much less than the probability of these two sources emitting one photon only. The visibility of the first-order interference pattern of two independent single-photon states is much less than the one of two-independent thermal light beams. It is impossible to observe the first-order interference pattern of two independent single-photon states for either short or long time comparing to the coherence time.
 
 With the same method above, one can calculate the first-order interference of two independent number states with different numbers of photons, one thermal and one laser light beams, and so on. No transient first-order interference pattern can be observed except the one of two independent laser light beams. No first-order interference pattern exists for two independent light beams with long observing time.

\subsubsection{Remakes about the classical models of laser and thermal light field}

With the results above, it is interesting to re-visit the electric field models for laser and thermal light in classical electromagnetic theory. For simplicity, quasi-monochromatic light and plane wave are assumed. The electric field for single-mode continuous-wave laser light at space-time coordinate $(\vec{r},t)$ is \cite{loudon2000quantum,born2013principles}
\begin{equation}\label{classical-e}
\vec{E}(\vec{r},t)=\vec{E}_j e^{-i(\omega_0 t - \vec{k}\cdot \vec{r}-\varphi_j)},
\end{equation}
where $\vec{E}_j$ is the amplitude, $\omega_0$ is the central angular frequency, $\vec{k}$ is the wave vector, $\varphi_j$ is the initial phase of the field. $\vec{E}_j$ and $\varphi_j$ are constant within one coherence time and change randomly between different coherence times.

The same model as Eq. (\ref{classical-e}) is employed to describe electric field of thermal light within one coherence time. For instance, in Born and Wolf\rq{}s book \cite{born2013principles}, they wrote: \textit{In a monochromatic wave field the amplitude of the vibrations at any point \text{P} is constant, while the phase varies linearly with time. This is no longer the case in a wave field produced by a real source: the amplitude and phase undergo irregular fluctuation, the rapidity of which depends essentially on the effective width $\Delta \nu$ of the spectrum. The complex amplitude remains substantially constant only during a time interval $\Delta t$ which is small compared to the reciprocal of the effective spectral width $\Delta \nu$; in such a time interval the change of the relative phase of any two Fourier components is much less than $2\pi$ and the addition of such components represents a disturbance which in this time interval behaves like a monochromatic wave with the mean frequency. However, this is not true for a longer time interval. The characteristic time $\Delta t=1/\Delta v$ is of the order of the coherence time introduced in \S7.5.8.} It means that the amplitude and phase of electric field produced by a real source undergo irregular fluctuations for a long time interval. The amplitude and phase of electric field can be seen as constant within one coherence time. In other words, the electric field produced by real source, \textit{i.e.}, laser or thermal light sources, can be described by Eq. (\ref{classical-e}). The same electric field model is employed to describe laser and thermal light field within the coherence time. This model was employed by Forrester to calculate the first-order interference of two independent thermal light beams, too \cite{forrester1956coherence}.

In classical theory, it is straightforward to obtain the conclusion that there is transient first-order interference pattern by superposing two independent thermal light beams as the one of two independent laser light beams. This conclusion is in conflict with the conclusion obtained in Feynman\rq{}s path integral in Sect. \ref{B-laser}.  Experiments should be employed to verify these two different conclusions. However, due to the very low degeneracy factor of thermal light, it is difficult to observe transient first-order interference pattern of two independent thermal light beams directly.

Forrester \textit{et al.} claimed that they observed \textit{interference between independently generated light waves} \cite{forrester1955photoelectric}. Detail analysis of their experiments indicates that their experiment belongs to one beam interference including nonlinear effect as the slits in a Young\rq{}s double-slit interferometer, which is similar as the ones with single atom \cite{grangier1985quantum,eichmann1993young} and atom ensembles \cite{afzelius2007interference} as the slits in a Young\rq{}s double-slit interferometer. Forrester \textit{et al.}\rq{}s experiments can not be interpreted as the results of first-order interference of two independent thermal light beams. 

If there is no transient first-order interference pattern with two independent thermal light beams, the classical model for electric field of thermal light within the coherence time may not be the same as the one of laser light. This result is helpful to understand the physics of thermal light and be worthy of further experimental studies.

\subsection{First-order interference of multiple independent light beams}\label{B-three}

\begin{figure}[htb]
\centering
\includegraphics[width=55mm]{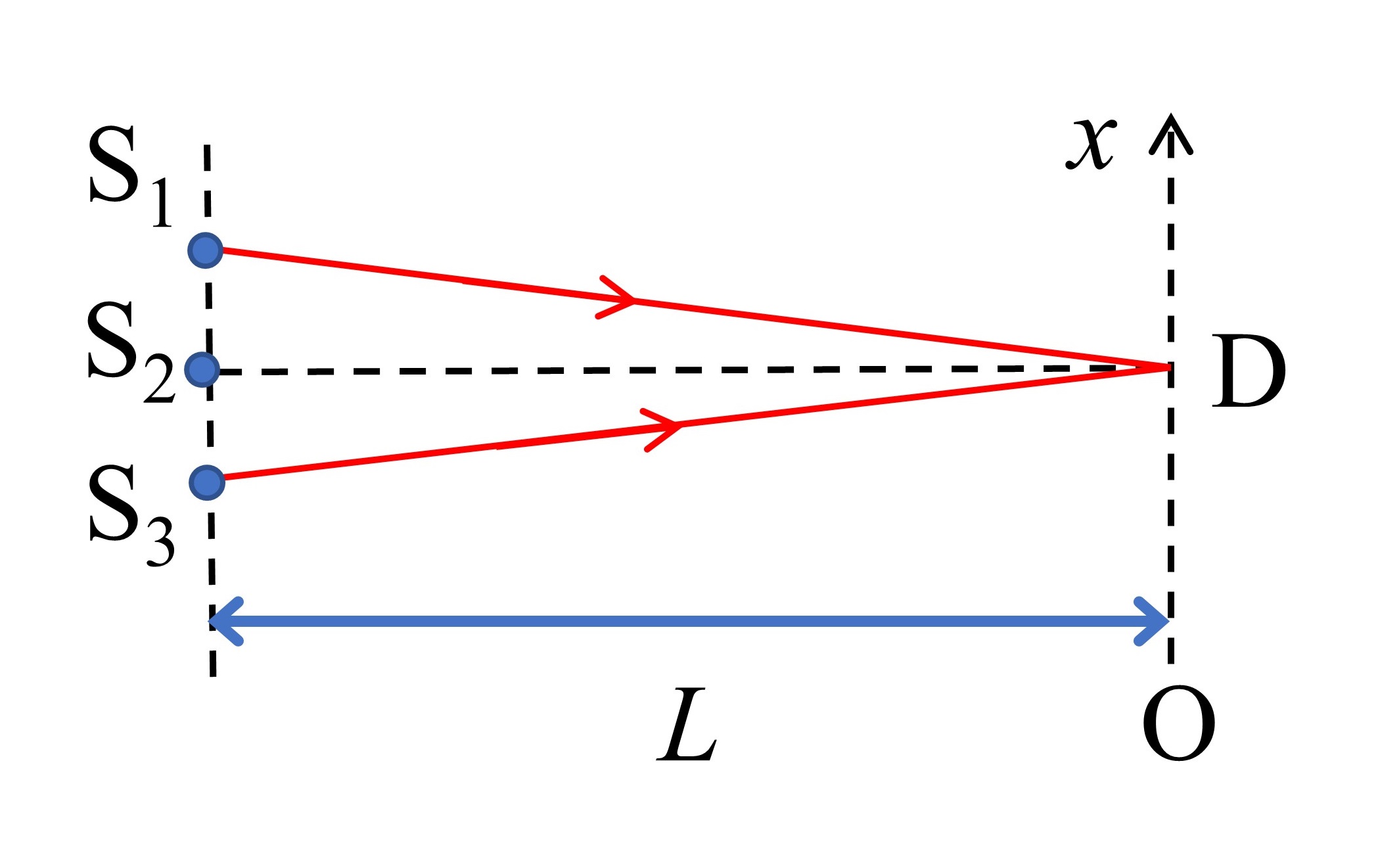}
\caption{First-order interference of three independent light beams. S$_1$, S$_2$, and S$_3$ are three independent light sources. D is a single-photon detector in the observation plane, O. $L$ is the distance between the source and observation planes.} \label{7-three-source}
\end{figure}

The same method can be employed to calculate the first-order interference of multiple independent light beams. Figure \ref{7-three-source} shows the scheme for the first-order interference of three light beams emitted by three independent light sources, S$_1$, S$_2$, and S$_3$.

There are three different paths for the single-photon detector, D, detecting a photon. The first path is that the detected photon is emitted by S$_1$ and the corresponding probability amplitude is $A_1(S,D)$. The second and third paths are defined similarly and the corresponding probability amplitudes are $A_2(S,D)$ and $A_3(S,D)$, respectively. For simplicity, these three light sources are assumed to be identical single-mode continuous-wave lasers with equal intensities. The polarizations and frequency bandwidths of the photons emitted by these three lasers are the same. When the distance between $S_1$ and S$_3$ is less than $\lambda L/\Delta x$, these three different paths to trigger a photon detection event are indistinguishable. The probability function for the $j$th detected photon is
\begin{eqnarray}\label{three-44}
&&P_j(x)\nonumber\\
&=& |\frac{1}{\sqrt{3}}[A_1(S,D)+A_2(S,D)+A_3(S,D)]|^2\\
&\propto&|e^{i\varphi_{j1}}K_1(S,D)+e^{i\varphi_{j2}}K_2(S,D)+e^{i\varphi_{j3}}K_3(S,D)|^2\nonumber,
\end{eqnarray}
where $\varphi_{j1}$, $\varphi_{j2}$ and $\varphi_{j3}$ are the initial phases of the $j$th detected photon emitted by S$_1$, S$_2$ and S$_3$, respectively. $K_1(S,D)$, $K_2(S,D)$ and $K_3(S,D)$ are the corresponding photon\rq{}s Feynman propagators for the three paths.

\begin{figure}[htb]
\centering
\includegraphics[width=55mm]{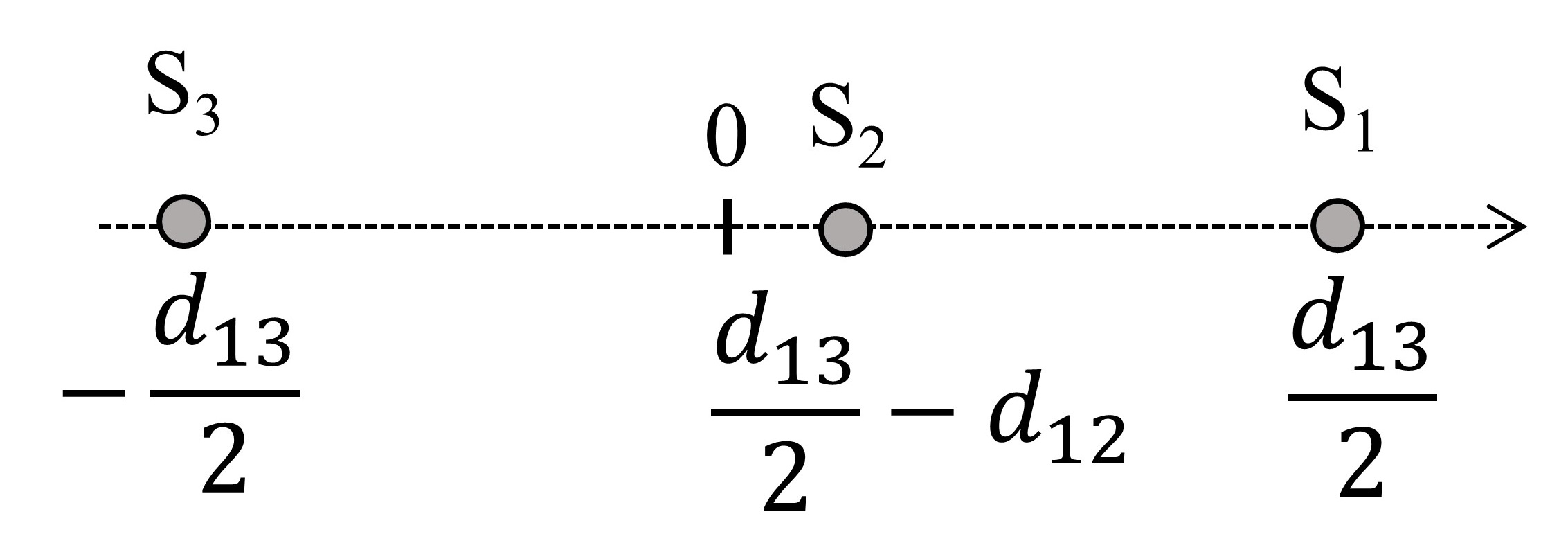}
\caption{Relative positions of three sources S$_1$, S$_2$ and S$_3$.}\label{8-three-slits}
\end{figure}

Equation (\ref{three-44}) can be simplified as
\begin{eqnarray}\label{three-2}
P_j(x)&=& \frac{3}{2}+\cos[\frac{2\pi d_{12}}{\lambda L}(x-\frac{d_{23}}{2})+(\varphi_{1j}-\varphi_{2j})]\nonumber\\
&&+\cos[\frac{2\pi d_{23}}{\lambda L}(x+\frac{d_{12}}{2})+(\varphi_{2j}-\varphi_{3j})]\nonumber\\
&&+\cos[\frac{2\pi d_{13}}{\lambda L} x+(\varphi_{1j}-\varphi_{3j})],
\end{eqnarray}
where paraxial approximation and one-dimension case are assumed to simplify the calculation, $d_{12}$ is the distance between S$_1$ and S$_2$, similar definitions hold for $d_{13}$ and $d_{23}$. The relative positions for the three sources are shown in Fig. \ref{8-three-slits}.

\begin{figure}[htb]
\centering
\includegraphics[width=85mm]{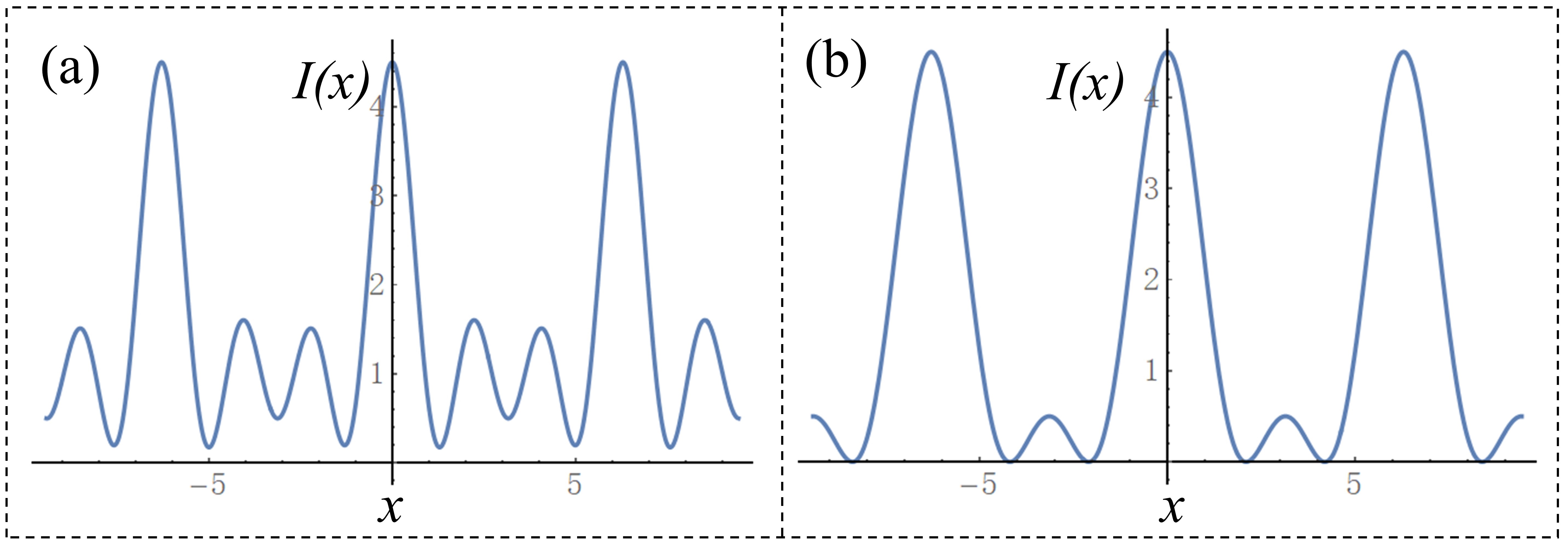}
\caption{Transient first-order interference patterns (a) $d_{12}\neq d_{23}$ and (b) $d_{12}= d_{23}$.}\label{9-three-pattern}
\end{figure}

Since the three lasers are independent, $\varphi_{j1}$, $\varphi_{j2}$ and $\varphi_{j3}$ are random. The relative phases are fixed if all the detected photons are within one coherence volume. The probability  function for $N$ detected photons is
\begin{equation}
P_N(x) \propto P_j(x),
\end{equation}
where transient first-order interference pattern of three independent laser light beams can be observed. Figure \ref{9-three-pattern} shows the transient first-order interference pattern of three laser light beams with equal initial phases. When $d_{12}\neq d_{23}$, the visibility of the transient first-order interference pattern can not reach 100\% as shown in Fig. \ref{9-three-pattern}(a). The visibility can reach 100\% when $d_{12}=d_{23}$ as shown in Fig. \ref{9-three-pattern}(b).

When the observation time is much longer than the coherence time, the probability function for $N$ detected photons is approximated to be a constant due to the sum of large number of random initial phases. 

The same method can be employed to calculate the first-order interference of three and more different types of independent light beams. It can be predicted that only when two or more laser light beams are employed, transient first-order interference pattern can be observed.

\section{Second-order interference of light}\label{sec-second}

As mentioned above, nearly all the studies of optical coherence are about the first-order interference of light before 1956 \cite{born2013principles}. In 1956, Hanbury Brown and Twiss discovered two-photon bunching of thermal light \cite{hanbury1956test,brown1956correlation}, which is regarded as one of the cornerstones of modern quantum optics. The history of second-order coherence of light can trace back to 1940s when several groups measured the correlation of photon pair generated in the annihilation of positron-electron pairs \cite{hanna1948polarization,bleuler1948correlation,wu1950angular}. The key element to measure the second-order interference of light is coincidence count detection system, which is proposed by Bothe in 1925 \cite{bothe1925wesen}.

The second-order interference of light plays an unique role in the history of the development of optical coherence. The first-order interference of light exists in nature. For instance, soap bubbles, transparent wings of insects, some butterfly wings, \textit{etc.} appearing colored streaks when exposed to sunlight is due to the first-order interference of light. The second-order interference of light does not exist in nature. It is a human-invented phenomenon. Special detection scheme is required to observe the second-order interference of light. The second-order interference of light is qualitatively different from the first-order interference of  light. It may be the reason why people took more than 200 years to observe the first phenomenon of the second-order interference of light since the first observation of the first-order interference of light. On the other hand, generalizing the second-oder interference of light to the third- and higher-order interference of light is straightforward. The third- and higher-order interference of light is quantitatively different from the second-order interference of light. 

\begin{figure}[htb]
\centering
\includegraphics[width=65mm]{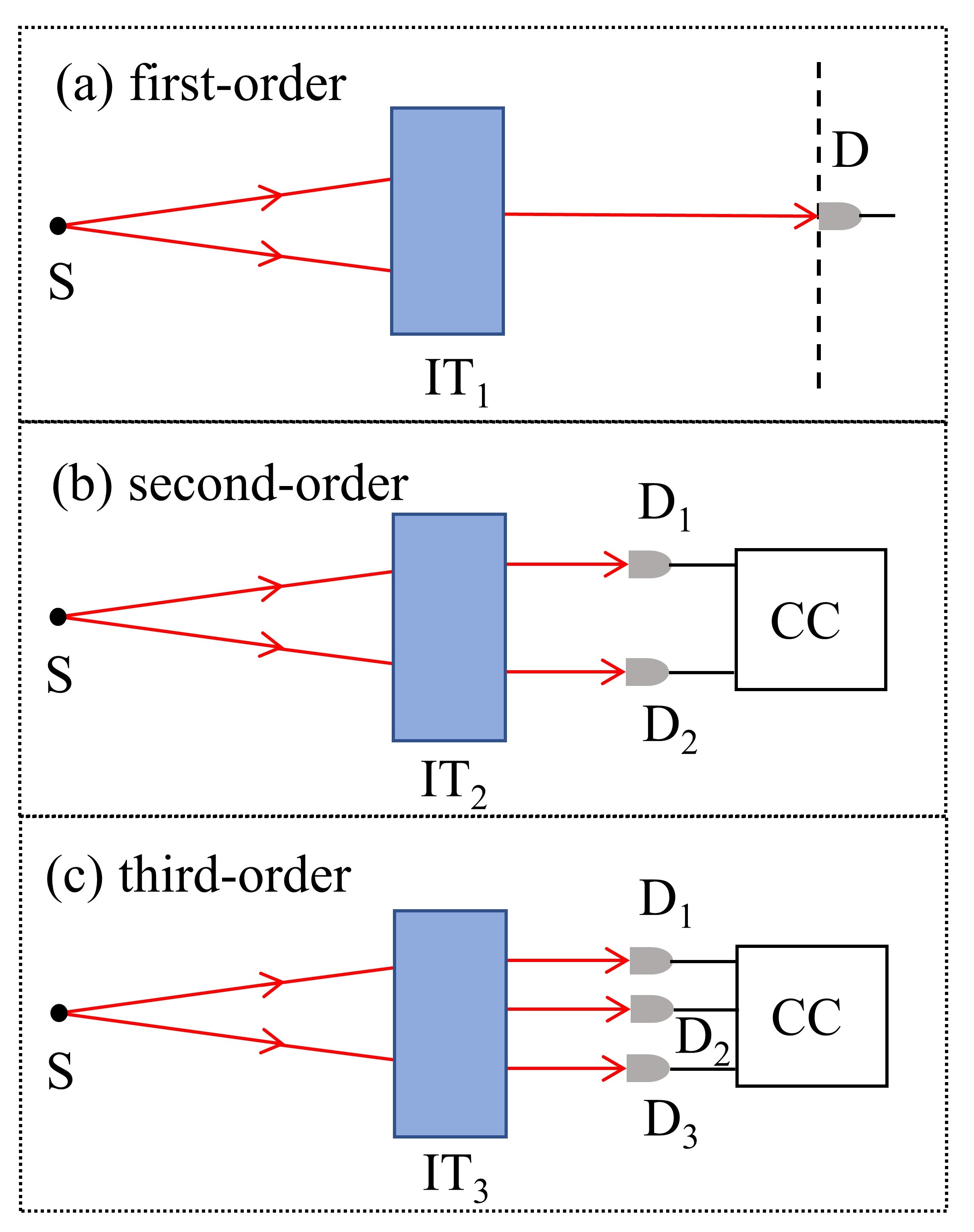}
\caption{(a) First-, (b) second- and (c) third-order interference of light. S is a light source. IT$_1$, IT$_2$, and IT$_3$ are first-, second- and third-order interferometers, respectively. D is a single-photon detector. CC is a coincidence count detection scheme.}\label{10-interferometer}
\end{figure}

Figure \ref{10-interferometer} shows the schemes for the first-, second- and third-order interference of light. For comparison, the first-order interference of light shown in Fig. \ref{10-interferometer}(a) is illustrated first. In a typical experiment of first-order interference of light, light generated by the light source, S, is split into more than one components in the first-order interferometer, IT$_1$. Then the split light beams are recombined to observe the intensity distribution over space or time by one detector. Young\rq{}s double-slit interferometer, Mach-Zehnder interferometer, Michelson interferometer are some typical first-order interferometers \cite{born2013principles}. The first-order interference of two and multiple independent light beams introduced in Sects. \ref{B-two} and \ref{B-three} belong to the first-order interference of light.

A typical scheme for the second-order interference of light is shown in Fig. \ref{10-interferometer}(b). Light emitted by S is split into two or more different components in the second-order interferometer, IT$_2$. Then the output light beams of IT$_2$ are detected by two detectors, D$_1$ and D$_2$. Hanbury Brown-Twiss (HBT) interferometer \cite{hanbury1956test,brown1956correlation}, Hong-Ou-Mandel (HOM) interferometer \cite{hong1987measurement,shih1988new,shih1986conference}, Franson interferometer \cite{franson1989bell} are some typical second-order interferometers. The output signals of these two detectors are sent into a coincidence count detection system, CC, to measure the correlation of these two signals. Based on what types of detector is employed, there are two work principles for CC. The first one is for normal detectors. The output of the detector is electric current or voltage proportional to the intensity of the incidental light. The measured correlation is given by the product of these two output signals,
\begin{equation}\label{intensity-f}
\Gamma^{(2)}(\vec{r}_1,t_1;\vec{r}_2,t_2)=\langle I_1(\vec{r}_1,t_1) I_2(\vec{r}_2,t_2) \rangle,
\end{equation}
where $I_1(\vec{r}_1,t_1)$ is the intensity of incidental light at space-time coordinate $(\vec{r}_1,t_1)$ detected by D$_1$, $I_2(\vec{r}_2,t_2)$ is the intensity of incidental light at space-time coordinate $(\vec{r}_2,t_2)$ detected by D$_2$, $\langle ... \rangle$ is ensemble average. For ergodic and stationary system, ensemble average can be approximated by time average \cite{mandel1995optical}.  Equation (\ref{intensity-f}) measures the correlation of intensity fluctuations detected by these two detectors. The CC system based on normal photo-electric detectors was employed by Hanbury Brown and Twiss in their experiments \cite{hanbury1956test,brown1956correlation}.

The second principle is for single-photon detectors. The output of a single-photon detector is a voltage pulse. Equation (\ref{intensity-f}) can not be employed to calculate the correlation of the outputs from two single-photon detectors. A different coincidence count principle should be employed. When the time difference between these two photon detection events is less than a time window $T$, these two photon detection events are a two-photon coincidence count. Otherwise, these two photon detection events are not a two-photon coincidence count. The time window $T$ should be less than the coherence time due to these two photons are indistinguishable within the same coherence volume. There are also other coincidence count detection system, such as two-photon absorption \cite{boitier2009measuring}, sum frequency generation \cite{pe2007broadband}, which are also based on the same principle as the one for single-photon detectors.

The second-order interference of light shown in Fig. \ref{10-interferometer}(b) can be extended to the third- and higher-order interference of light. Figure \ref{10-interferometer}(c) shows the scheme for the third-order interference of light. The light beam emitted by S is split into multiple components and detected by three detectors, D$_1$, D$_2$ and D$_3$. The output of these three detectors are sent into a three-photon coincidence count system, CC, to measure the third-order correlation. Based on different types of the employed detectors, there are also two different principles for three-photon coincidence count system. When normal detectors are employed, the third-order correlation is given by 
\begin{equation}\label{intensity-f3}
\Gamma^{(3)}(\vec{r}_1,t_1;\vec{r}_2,t_2;\vec{r}_3,t_3)=\langle I_1(\vec{r}_1,t_1) I_2(\vec{r}_2,t_2)I_3(\vec{r}_3,t_3) \rangle,
\end{equation}
where the meanings of the symbols in Eq. (\ref{intensity-f3}) are similar as the ones in Eq. (\ref{intensity-f}). When single-photon detectors are employed, the time differences between these three photon detection events are less than a time window $T$, the three photon detection events are a three-photon coincidence count. Otherwise, the three photon detection events are not a three-photon coincidence count. Similar method can be employed to define the fourth- and higher-order interference of light.

\subsection{Second-order interference of one light beam}

\begin{figure}[htb]
\centering
\includegraphics[width=55mm]{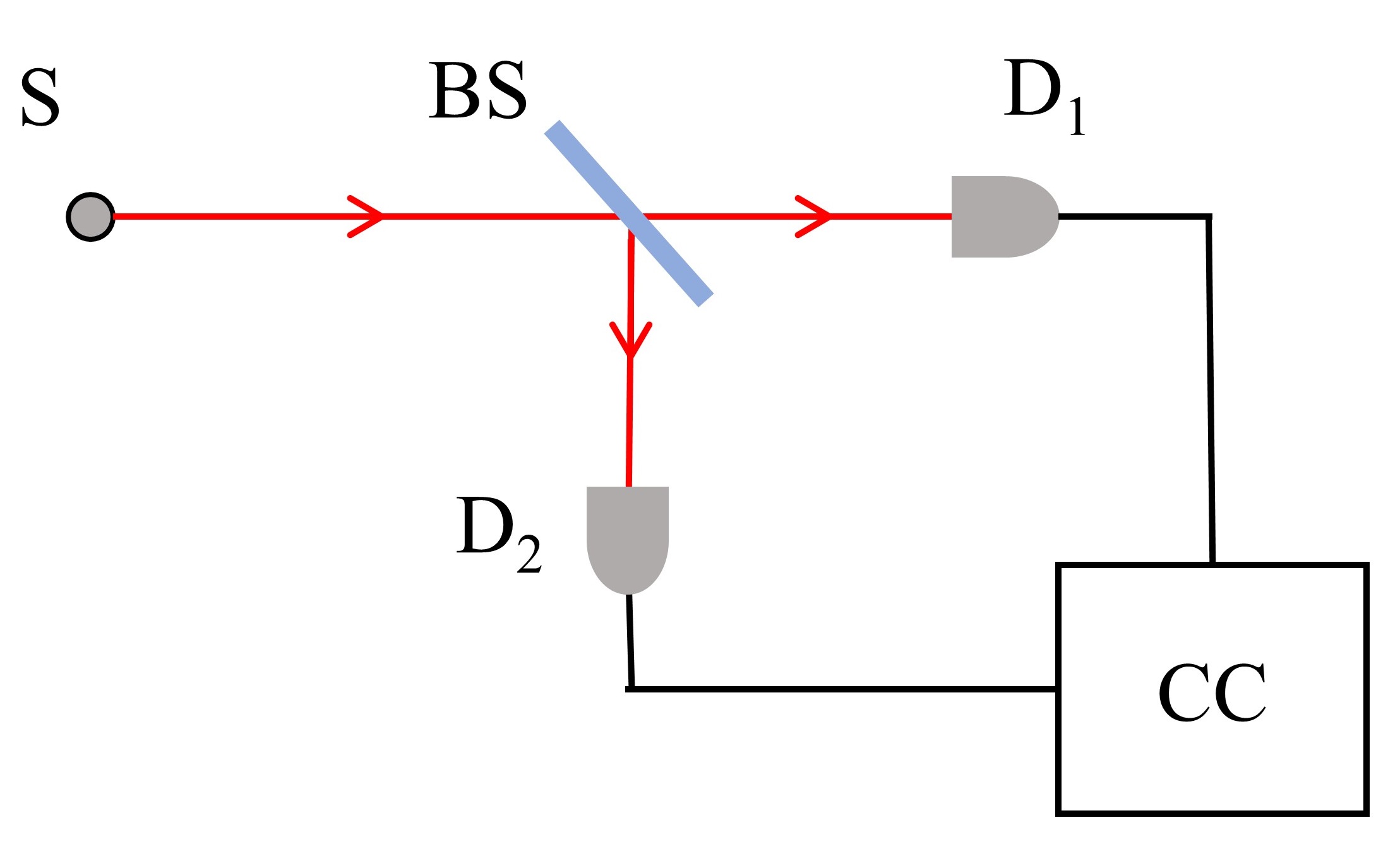}
\caption{Second-order interference of one light beam in a HBT interferometer. S is a light source. BS is a 1:1 non-polarizing beam splitter. D$_1$ and D$_2$ are two single-photon detectors. CC is a two-photon coincidence count detection system.}\label{11-HBT-interferometer}
\end{figure}

Figure \ref{11-HBT-interferometer} shows the scheme for the second-order interference one light beam in a HBT interferometer. The light beam emitted by source S is split into two by a 1:1 non-polarizing beam splitter, BS. The output light beams of BS are detected by two single-photon detectors, D$_1$ and D$_2$. The outputs of these two detectors are sent into a two-photon coincidence count detection system, CC, to measure the second-order correlation of light.

\subsubsection{Second-order interference of one thermal light beam}\label{second-thermal-one}

Assuming S is a thermal light source in Fig. \ref{11-HBT-interferometer}, the scheme shown in Fig. \ref{11-HBT-interferometer} is equivalent to the scheme employed by Hanbury Brown and Twiss in 1956 \cite{hanbury1956test,brown1956correlation}. There are two different paths for D$_1$ and D$_2$ detecting a two-photon coincidence count. The first path is that photon a is detected by D$_1$ and photon b is detected by D$_2$. The probability for the path is $1/2$ and the corresponding two-photon probability amplitude is $A(a,D_1;b,D_2)$. The second path is that photon a is detected by D$_2$ and photon b is detected by D$_1$. The probability for the path is $1/2$ and the corresponding two-photon probability amplitude is $A(a,D_2;b,D_1)$. When photons a and b are in the same coherence volume of thermal light, these two different paths to trigger a two-photon coincidence count are indistinguishable. The two-photon probability function for the $j$th detected photon pair is \cite{feynman2010quantum,feynman2011feynman}
\begin{eqnarray}\label{thermal-1}
&&P^{(2)}_j(\vec{r}_1,t_1;\vec{r}_2,t_2)\nonumber\\
&=&|\frac{1}{\sqrt{2}}[A_j(a,D_1;b,D_2)+A_j(a,D_2;b,D_1)]|^2\nonumber\\
&\propto&|e^{i(\varphi_{aj}+\varphi_{bj})}K_j(a,D_1;b,D_2)\nonumber\\
&&+e^{i(\varphi_{aj}+\varphi_{bj})}K_j(a,D_2;b,D_1)|^2\nonumber\\
&=& |K_j(a,D_1;b,D_2)+K_j(a,D_2;b,D_1)|^2,
\end{eqnarray}
where $\varphi_{aj}$ and $\varphi_{bj}$ are the initial phases of photons a and b in the $j$th detected photon pair, respectively, $K_j(a,D_1;b,D_2)$ is photon\rq{}s Feynman propagator for photon a emitted by S going to D$_1$ and photon b emitted by S going to D$_2$. The meaning of $K_j(a,D_2;b,D_1)$ is defined similarly. Based on the principle for calculating the probability amplitude of independent events introduced by Feynman \cite{feynman2011feynman}, $K_j(a,D_1;b,D_2)$ equals $K_j(a,D_1)K_j(b,D_2)$. Equation (\ref{thermal-1}) can be simplified as
\begin{eqnarray}\label{thermal-2}
&&P^{(2)}_j(\vec{r}_1,t_1;\vec{r}_2,t_2)\nonumber\\
&=& |K_j(a,D_1)K_j(b,D_2)+K_j(a,D_2)K_j(b,D_1)|^2.
\end{eqnarray}

The two-photon probability function of $N$ detected photon pairs is the sum of the probability functions of $N$ detected photon pairs,
\begin{eqnarray}\label{thermal-3}
&&P^{(2)}(\vec{r}_1,t_1;\vec{r}_2,t_2)\nonumber\\
&=&\sum_{j=1}^{N}P^{(2)}_j(\vec{r}_1,t_1;\vec{r}_2,t_2)\nonumber\\
&\propto&\langle P^{(2)}_j(\vec{r}_1,t_1;\vec{r}_2,t_2)\rangle \\
&=&\langle|K_j(a,D_1)K_j(b,D_2)+K_j(a,D_2)K_j(b,D_1)|^2\rangle, \nonumber
\end{eqnarray}
where $\langle...\rangle$ is ensemble average. The normalized two-photon probability function,
\begin{equation}\label{thermal-4}
p^{(2)}(\vec{r}_1,t_1;\vec{r}_2,t_2)=\frac{P^{(2)}(\vec{r}_1,t_1;\vec{r}_2,t_2)}{P^{(1)}(\vec{r}_1,t_1)P^{(1)}(\vec{r}_2,t_2)},
\end{equation}
is employed to eliminate the influence of the first-order interference pattern on the second-order interference pattern, where $P^{(1)}(\vec{r}_1,t_1)$ and $P^{(1)}(\vec{r}_2,t_2)$ are single-photon probability functions at space-time coordinates $(\vec{r}_1,t_1)$ and $(\vec{r}_2,t_2)$, respectively. $P^{(1)}(\vec{r}_1,t_1)$ is usually written as $P(\vec{r}_1,t_1)$ for short if no possible misunderstanding exists. The two-photon probability function, $P^{(2)}(\vec{r}_1,t_1;\vec{r}_2,t_2)$, and the normalized two-photon probability function, $p^{(2)}(\vec{r}_1,t_1;\vec{r}_2,t_2)$, correspond to the second-order coherence function, $G^{(2)}(\vec{r}_1,t_1;\vec{r}_2,t_2)$, and the normalized second-order coherence function, $g^{(2)}(\vec{r}_1,t_1;\vec{r}_2,t_2)$, in Glauber\rq{}s quantum optical coherence theory, respectively \cite{glauber1963coherent,glauber1963quantum}.

Let us first calculate temporal two-photon bunching of thermal light. Assume S is a point thermal light source. D$_1$ and D$_2$ are at symmetrical positions of BS to simplify the calculation. In this case, photon\rq{}s Feynman propagator is
\begin{equation}\label{thermal-5}
K(a,D_1)=\int_{\omega_0-\Delta \omega/2}^{\omega_0+\Delta\omega/2}f(\omega)e^{-j\omega(t_1-t_a)}d\omega,
\end{equation}
where the spatial part is ignored, $\omega_0$ is the central angular frequency, $\Delta \omega$ is the angular frequency distribution function of thermal light, $t_1$ is the time of photon detection event at D$_1$, $t_a$ is the emitting time of photon a. Substituting Eq. (\ref{thermal-5}) into Eq. (\ref{thermal-3}) and assuming $f(\omega)$ equals 1 within $[\omega_0-\Delta\omega/2,\omega_0+\Delta\omega/2]$ and equals 0 in other domains, the temporal two-photon probability function is \cite{shih2020introduction}
\begin{equation}\label{thermal-6}
P^{(2)}(t_1-t_2)\propto 1+\text{sinc}^2\frac{\Delta \omega (t_1-t_2)}{2}.
\end{equation}
It is easy to prove that $P^{(1)}(t_1)$ and $P^{(1)}(t_2)$ are constant for thermal light. The normalized temporal two-photon probability function, $p^{(2)}(t_1-t_2)$ is proportional to $P^{(2)}(t_1-t_2)$. In the following parts, if the normalized probability function is proportional to the corresponding probability function, only probability function is calculated.

\begin{figure}[htb]
\centering
\includegraphics[width=65mm]{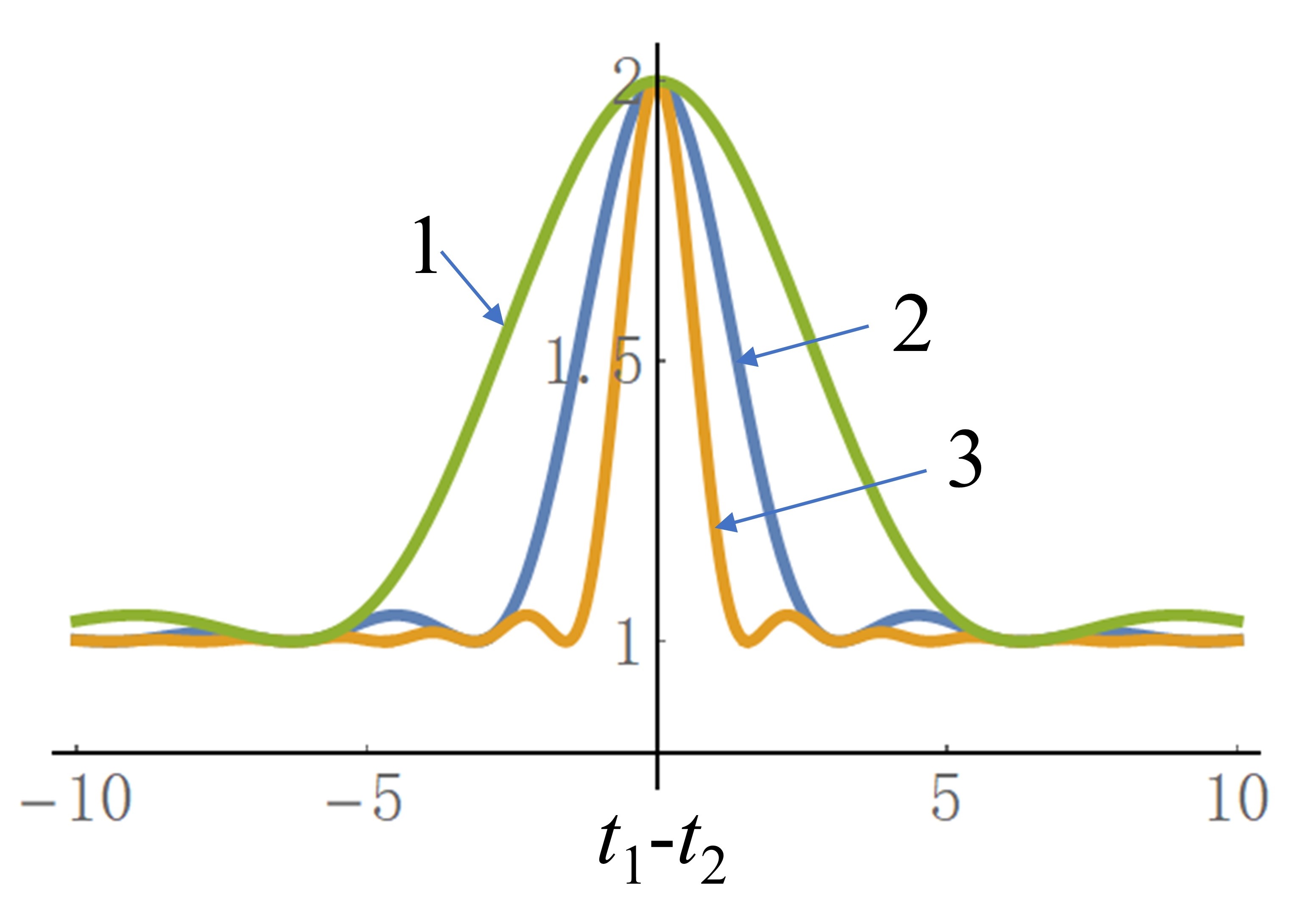}
\caption{Temporal two-photon bunching of thermal light. $t_1-t_2$ is the time difference between photon detection events at D$_1$ and D$_2$. Lines 1, 2, 3 correspond to the frequency bandwidths equaling $\Delta \omega$, $2\Delta \omega$, and $4\Delta \omega$, respectively. }\label{12-HBT-temporal}
\end{figure}

Figure \ref{12-HBT-temporal} shows the temporal two-photon probability functions of thermal light with different frequency bandwidths. When $t_1$ equals $t_2$, the probability of detecting two photons equals 2. As the time difference, $|t_1-t_2|$, increases, the probability of detecting two photons decreases. When the time difference between two photon detection events is much longer than the coherence time, the probability of detecting two photons equals 1, which indicates that these two photon detection events are independent. 

The spatial two-photon bunching of thermal light can be calculated in the same way. Assume one dimension thermal light source is employed. The light emitted by S is single-frequency light. The distance between the source plane and D$_1$ equals the one between source plane and D$_2$. The photon\rq{}s Feynman propagator in this case can be approximated by Green function in classical optics \cite{goodman2005introduction}
\begin{equation}\label{thermal-7}
K(a,D_1)=\frac{L}{i\lambda}\int_{-d/2}^{d/2}s(x_s)\frac{e^{j\vec{k}_{s1}\cdot(\vec{r}_1-\vec{r}_s)}}{|\vec{r}_1-\vec{r}_s|^2} dx_s
\end{equation}
where $d$ is the length of thermal light source, $s(x_s)$ is the intensity distribution function of S, $\vec{r}_1$ is the position vector of D$_1$, $\vec{r}_s$ is the position vector of S emitting photon a, $\vec{k}_{s1}$ is the wave vector of light. Substituting Eq. (\ref{thermal-7}) into Eq. (\ref{thermal-3}) and assuming $s(x_s)$ equals 1 within $[-d/2,d/2]$ and equals 0 in other domains, spatial two-photon probability function of thermal light in a HBT interferometer is \cite{shih2020introduction}
\begin{equation}\label{thermal-8}
P^{(2)}(x_1-x_2)\propto 1+\text{sinc}^2\frac{ d}{L \lambda} (x_1-x_2).
\end{equation}

Based on the results above, two-photon bunching of thermal light can be interpreted as the result of two-photon interference. It is the result of superposing two-photon probability amplitudes for two different and indistinguishable paths to trigger a two-photon coincidence count \cite{fano1961quantum,feynman2006qed,glauber1995dirac}. Two-photon interference in the second-order interference of light can not be used to disprove Dirac\rq{}s statement that different photons never interferes with each other, which was meant for the first-order interference of light \cite{paul1986interference}. By analogy, Dirac\rq{}s statement about the first-order interference of light can be generalized to the second-order interference of light: Each photon pair only interferes with itself and interferes between two different photon pairs never occurs \cite{shih2020introduction}.

\subsubsection{Second-order interference of one superbunching light beam}

$g^{(2)}(0)$ measured in a HBT interferometer is the degree of second-order coherence of light, which is usually employed to categorize different types of light \cite{loudon2000quantum}. $g^{(2)}(0)$ equals 1 for single-mode continuous-wave laser light. When $g^{(2)}(0)$ is less than 1, the light belongs to non-classical light.  $g^{(2)}(0)$ equals 2 for thermal light.  When $g^{(2)}(0)$ is larger than 2, it belongs to superbunching light \cite{ficek2005quantum}. Superbunching light can be either classical or non-classical. For instance, entangled photon pairs is non-classical superbunching light \cite{klyshko1996nonclassical} and superbunching pseudothermal light is classical light \cite{zhou2017superbunching}. This section will concentrate on the second-order interference of a classical superbunching light beam in a HBT interferometer. 

There are two different methods to increase the degree of second-order coherence of classical light. The first one is adding more two-photon paths. The second one is the two-photon probability amplitudes changing with time. 

\begin{figure}[htb]
\centering
\includegraphics[width=75mm]{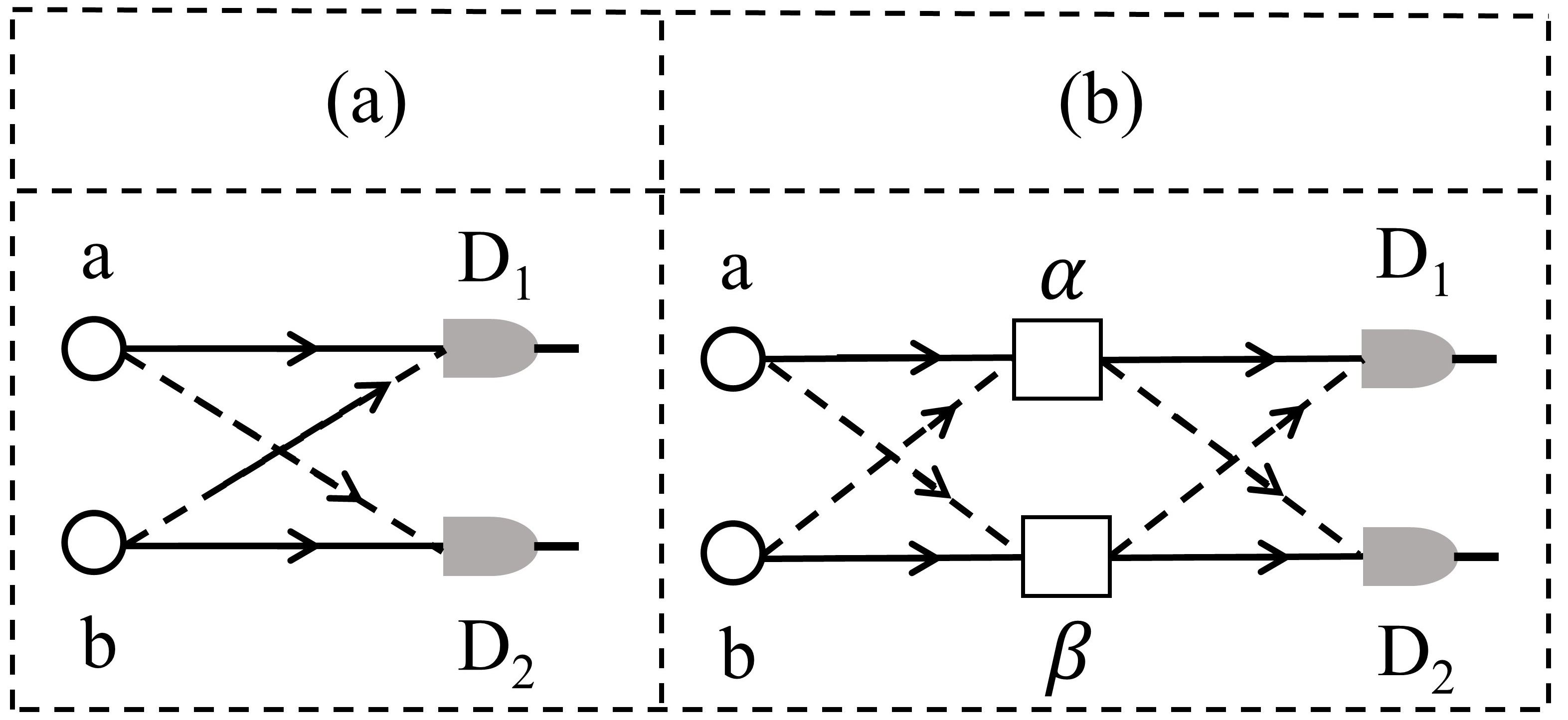}
\caption{Increase the degree of second-order coherence by adding more two-photon paths. a and b are two photons. D$_1$ and D$_2$ are two single-photon detectors. $\alpha$ and $\beta$ are two nodes for photons.}\label{13-multiple-paths}
\end{figure}

Figure \ref{13-multiple-paths} shows how to increase the degree of second-order coherence by adding more two-photon paths. Figure \ref{13-multiple-paths}(a) shows there are two different paths for photons a and b to trigger a two-photon coincidence count in D$_1$ and D$_2$. The first path is that photon a is detected by D$_1$ and photon b is detected by D$_2$. The corresponding two-photon probability amplitude is $A(a,D_1;b,D_2)$. The second path is that photon a is detected by D$_2$ and photon b is detected by D$_1$. The corresponding two-photon probability amplitude is $A(a,D_2;b,D_1)$. If these two different paths to trigger a two-photon coincidence count are indistinguishable, the two-photon probability function is
\begin{eqnarray}\label{path-1}
&&P^{(2)}(\vec{r}_1,t_1;\vec{r}_2,t_2)\nonumber\\
&=&\langle|\frac{1}{\sqrt{2}}[A(a,D_1;b,D_2)+A(a,D_2;b,D_1)]|^2 \rangle.
\end{eqnarray}
When these two different paths are distinguishable, the two-photon probability function is
\begin{eqnarray}\label{path-2}
&&P^{(2)}(\vec{r}_1,t_1;\vec{r}_2,t_2)\nonumber\\
&=&\frac{1}{2}[\langle|A(a,D_1;b,D_2)|^2 \rangle+\langle|A(a,D_2;b,D_1)|^2 \rangle],
\end{eqnarray}
which consists of the background of two-photon probability function. Figure \ref{13-multiple-paths}(a) and Eqs. (\ref{path-1}) are valid for thermal light in a HBT interferometer. $g^{(2)}(0)$ equals 2 for thermal light. There is another way to estimate the value of $g^{(2)}(0)$. Assuming all the photons have equal probability reaching D$_1$ and D$_2$, $|A(a,D_1;b,D_2)|$ equals $|A(a,D_2;b,D_1)|$. The are four terms in Eq. (\ref{path-1}) and $P^{(2)}(0)$ equals $2|A|^2$, where $A$ is short for $|A(a,D_1;b,D_2)|$. There are 2 terms in Eq. (\ref{path-2}) and $P^{(2)}(0)$ equals $|A|^2$. The degree of second-order coherence equals the normalized two-photon probability function when these two detectors are at the same space-time coordinate,
\begin{eqnarray}\label{path-3}
g^{(2)}(0)=p^{(2)}(0)=\frac{2|A|^2}{|A|^2}=2.
\end{eqnarray}
The reason why Eq. (\ref{path-2}) can be treated as the background is that these two paths are distinguishable. The two photon detection events are independent and $P^{(2)}(\vec{r}_1,t_1;\vec{r}_2,t_2)$ equals $P^{(1)}(\vec{r}_1,t_1)P^{(1)}(\vec{r}_2,t_2)$ in this case.

Similar method can be employed to calculate the degree of second-order coherence of light in Fig. \ref{13-multiple-paths}(b). There are four different paths for photons a and b to trigger a two-photon coincidence count. The first path is that photon a transmitting through $\alpha$ is detected by D$_1$ and photon b transmitting through $\beta$ is detected by D$_2$. The corresponding two-photon probability amplitude is $A(a,\alpha,D_1;b,\beta,D_2)$. The other three paths can be defined similarly and the corresponding two-photon probability amplitudes are $A(a,\alpha,D_2;b,\beta,D_1)$, $A(a,\beta,D_1;b,\alpha,D_2)$ and $A(a,\beta,D_2;b,\alpha,D_1)$. When these four different paths to trigger a two-photon coincidence count are indistinguishable, the two-photon probability function is
\begin{eqnarray}\label{path-4}
&&P^{(2)}(\vec{r}_1,t_1;\vec{r}_2,t_2)\\
&=&\langle| \frac{1}{\sqrt{4}}[A(a,\alpha,D_1;b,\beta,D_2)+A(a,\alpha,D_2;b,\beta,D_1)\nonumber\\
&&+A(a,\beta,D_1;b,\alpha,D_2)+A(a,\beta,D_2;b,\alpha,D_1)]|^2 \rangle.\nonumber
\end{eqnarray}
When these four different paths are distinguishable, the two-photon probability function equals
\begin{eqnarray}\label{path-5}
&&P^{(2)}(\vec{r}_1,t_1;\vec{r}_2,t_2)\\
&=&\frac{1}{4}[\langle|A(a,\alpha,D_1;b,\beta,D_2)|^2\rangle+\langle |A(a,\alpha,D_2;b,\beta,D_1)|^2\rangle \nonumber\\
&&+\langle|A(a,\beta,D_1;b,\alpha,D_2)|^2\rangle+\langle |A(a,\beta,D_2;b,\alpha,D_1)|^2 \rangle] \nonumber
\end{eqnarray}
Again, assume photons have equal probability being detected by D$_1$ and D$_2$. The ratio between Eqs. (\ref{path-4}) and (\ref{path-5}) at the same space-time coordinate equals the degree of second-order coherence,
\begin{equation}\label{4-four-3}
g^{(2)}(0)=p^{(2)}(0)=\frac{4|A\rq{}|^2}{|A\rq{}|^2}=4,
\end{equation}
where $|A\rq{}|$ is short for $|A(a,\alpha,D_1;b,\beta,D_2)|$. 

Similar method can be employed to calculate the degree of second-order coherence when there are $2N$ different paths to trigger a two-photon coincidence count. The corresponding degree of second-order coherence equals $2^N$, where $N$ is a positive integer \cite{zhou2017superbunching}.

\begin{figure}[htb]
\centering
\includegraphics[width=75mm]{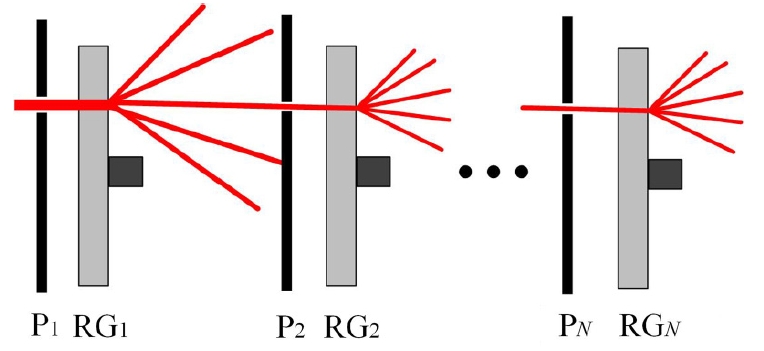}
\caption{Scheme to observe two-photon superbunching with $2N$ different paths \cite{zhou2017superbunching}. P$_1$, P$_2$, ..., P$_N$ are $N$ pinholes. RG$_1$, RG$_2$, ..., RG$_N$ are $N$ rotating groundglasses.}\label{14-superbunching-scheme}
\end{figure}

Figure \ref{14-superbunching-scheme} shows the experimental scheme to observe two-photon superbunching with $2N$ different paths. A single-mode continuous-wave laser is employed as the light source. After passing through pinhole P$_1$, the light is scattered by a rotating groundglass, RG$_1$. Then the scattered light is filtered by another pinhole, P$_2$. The size of P$_2$ should be less than the coherence area of the scattered light at P$_2$ plane. Then the filtered light is scattered by RG$_2$. Repeating the process $N$ times and superbunching pseudothermal light with $g^{(2)}(0)$ equaling $2^N$ can be generated.

\begin{figure}[htb]
\centering
\includegraphics[width=75mm]{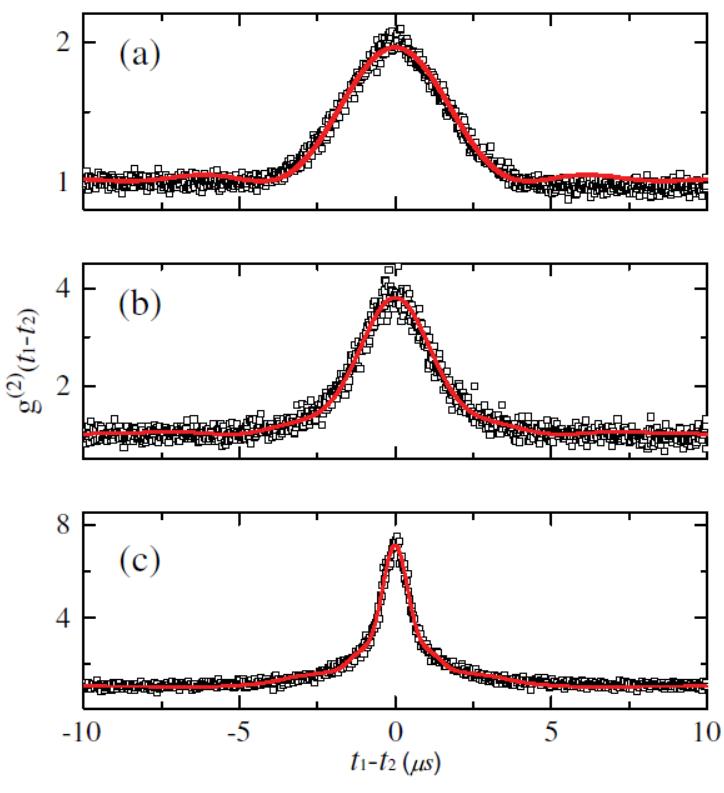}
\caption{Normalized second-order coherence functions with one (a), two (b), and three (c) rotating groundglasses \cite{bai2017photon}. $t_1-t_2$ is the time difference between two photon detection events at D$_1$ and D$_2$. $g^{(2)}(t_1-t_2)$ is the normalized second-order coherence function in Glauber\rq{}s quantum optical coherence theory, which is equivalent to the normalized two-photon probability function in quantum optical coherence theory based on Feynman\rq{}s path integral.}\label{15-superbunching-results}
\end{figure}

Figure \ref{15-superbunching-results}(a)-(c) shows the measured temporal second-order coherence functions of light with one, two, and three rotating groundglasses, respectively \cite{bai2017photon}. $g^{(2)}(0)$ equals $1.96\pm0.01$ in Fig. \ref{15-superbunching-results}(a), in which there is one rotating groundglass and two-photon bunching is observed. The reason why it is less than 2 is due to that part of the un-scattered laser light contributes to the two-photon coincidence count. The source with one rotating groundglass is pseudothermal light source invented by Martienssen and Spiller \cite{martienssen1964coherence}. $g^{(2)}(0)$ equals $3.80\pm0.04$ in Fig. \ref{15-superbunching-results}(b) when there are two rotating groundglasses. When there are three rotating groundglasses, $g^{(2)}(0)$ equals $7.10\pm0.07$ in Fig. \ref{15-superbunching-results}(c). Two-photon superbunching is observed in  Figs. \ref{15-superbunching-results}(b) and  \ref{15-superbunching-results}(c). 

\begin{figure}[htb]
\centering
\includegraphics[width=45mm]{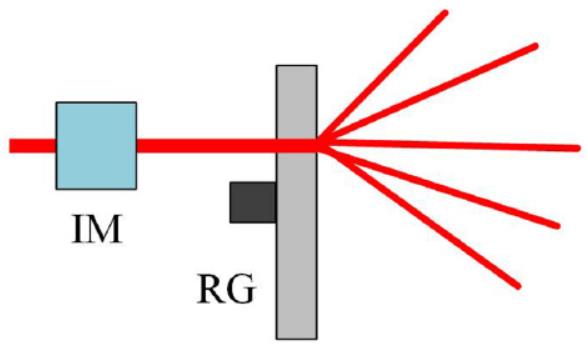}
\caption{Simplified superbunching pseudothermal light source \cite{zhou2019superbunching}. IM is intensity modulator and RG is rotating groundglass.}\label{16-superbunching-2}
\end{figure}

Another method to increase the degree of second-order coherence of light is that the two-photon probability amplitudes changes with time. Figure \ref{16-superbunching-2} shows the scheme for observing two-photon superbunching by changing the two-photon probability amplitudes. The light source shown in Fig. \ref{16-superbunching-2} is the light source in a HBT interferometer shown in Fig. \ref{11-HBT-interferometer}. There are two different paths for photons a and b to trigger a two-photon coincidence count in a HBT interferometer. The two-photon probability amplitudes are $A(a,D_1;b,D_2)$ and $A(a,D_2;b,D_1)$. 

The difference between usual pseudothermal light source and the light source shown in Fig. \ref{16-superbunching-2} in a HBT interferometer is that $A(a,D_1;b,D_2)$ and $A(a,D_2;b,D_1)$ will change with time in the latter case. An intensity modulator (IM) is employed to modulate the intensity of incidental laser light. When these two different paths to trigger a two-photon coincidence count are indistinguishable, the temporal two-photon probability function is
\begin{eqnarray}\label{intensity-1}
P^{(2)}(t_1;t_2)&=&\langle|\sqrt{P_{a1}(t_{a1})P_{b2}(t_{b2})}K_{a1}K_{b2}\nonumber\\
&&+\sqrt{P_{a2}(t_{a2})P_{b1}(t_{b1})}K_{a2}K_{b1}|^2\rangle,
\end{eqnarray}
where $P_{a1}(t_{a1})$ is the probability that photon a is the detected by D$_1$ at time $t_{a1}$ and the meanings of other symbols are defined similarly. 

Assuming point light source is employed and with the same method above, Eq. (\ref{intensity-1}) can be simplified as \cite{zhou2019superbunching}
\begin{eqnarray}\label{intensity-2}
P^{(2)}(t_1-t_2)  &\propto&  p^{(2)}(t_1-t_2) \\
&\propto & \gamma(t_1-t_2)[1+\text{sinc}^2\frac{\Delta \omega (t_2-t_1)}{2}], \nonumber
\end{eqnarray}
where $\gamma(t_1-t_2)=\langle I(t_1)I(t_2)\rangle /(\langle I(t_1)\rangle \langle I(t_2)\rangle)$ is the correlation function of laser light introduced by IM, $I(t_1)$ and $I(t_2)$ are the intensities of light at time $t_1$ and $t_2$ before RG, respectively. When $\gamma(t_1-t_2)$ equals 1, Eq. (\ref{intensity-2}) describes normal two-photon bunching of thermal light. When $\gamma(t_1-t_2)$ is larger than 1, two-photon superbunching can be observed. 

\begin{figure}[htb]
\centering
\includegraphics[width=70mm]{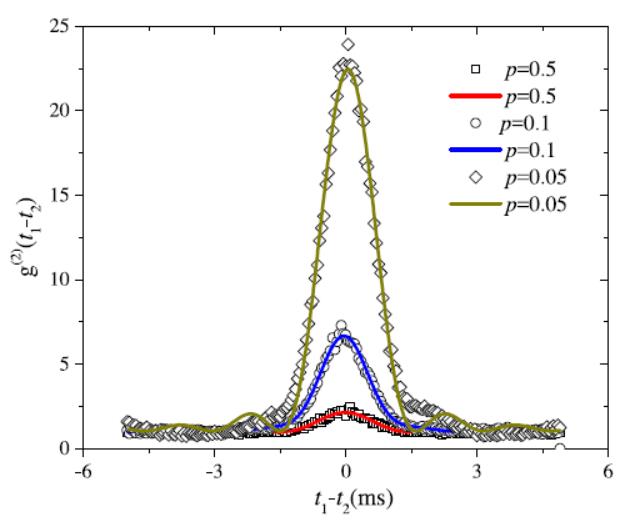}
\caption{Two-photon superbunching with changing probability amplitudes \cite{zhou2019superbunching}. }\label{17-superbunching-results}
\end{figure}

Figure \ref{17-superbunching-results} shows the experimental measured temporal second-order coherence function of superbunching pseudothermal light with the scheme shown in Fig. \ref{16-superbunching-2}. Two-photon superbunching with $g^{(2)}(0)$ equaling 20.45 is observed \cite{zhou2019superbunching}.

There are also other different schemes to increase the degree of second-order coherence of light \cite{hong2012two,straka2018generator,zhang2019superbunching,zhang2020super}, which can also be categorized into the above two methods.

\subsubsection{Second-order interference of one laser light beam}

Assume S is a single-mode continuous-wave laser in Fig. \ref{11-HBT-interferometer}. Same method can be employed to calculate two-photon probability function of laser light in a HBT interferometer. There are two different paths for photons a and b to trigger a two-photon coincidence count. The probability amplitudes for these two paths are $A(a,D_1;b,D_2)$ and $A(a,D_2;b,D_1)$. If these two different paths are indistinguishable, the two-photon probability function is
\begin{eqnarray}\label{laser-1}
&&P^{(2)}(\vec{r}_1,t_1;\vec{r}_2,t_2)\nonumber\\
&=&\langle |\frac{1}{\sqrt{2}}[A(a,D_1;b,D_2)+A(a,D_2;b,D_1)]|^2\rangle.
\end{eqnarray} 
It seems that two-photon bunching should be observed for single-mode continuous-wave laser light in a HBT interferometer. However, it was proved that photon detection events are independent for a single-mode continuous-wave laser light beam in a HBT interferometer \cite{arecchi1965measurement,freed1965photoelectron}. It is predicted that the degree of second-order coherence of coherent state equals 1 in Glauber\rq{}s quantum optical coherence theory \cite{glauber1963coherent}, which indicates that the photon detection events are independent in single-mode continuous-wave laser light.

The reason why there is no two-photon bunching of single-mode continuous-wave laser light lies at the difference between the superposition principles in classical and quantum theories. In classical theory, if there are two fields at space-time coordinate $(\vec{r},t)$, the final field is the superposition of these two fields,
\begin{equation}
\vec{E}(\vec{r},t)=\vec{E}_1(\vec{r},t)+\vec{E}_2(\vec{r},t).
\end{equation}
If $\vec{E}_1(\vec{r},t)$ is identical to $\vec{E}_2(\vec{r},t)$, the superposition will result in $2\vec{E}_1(\vec{r},t)$. The intensity at the point will be four times of the one of $\vec{E}_1(\vec{r},t)$.

This is not the case for the superposition principle in quantum theory. Assuming there are two wave functions at space-time coordinate $(\vec{r},t)$, the final wave function is the superposition of these two wave functions,
\begin{equation}
\Psi(\vec{r},t)=\Psi_1(\vec{r},t)+\Psi_2(\vec{r},t).
\end{equation}
If $\Psi_1(\vec{r},t)$ is identical to $\Psi_2(\vec{r},t)$, the superposition will not give $2\Psi_1(\vec{r},t)$. The result of superposition of two identical states equals the individual state, not two times of the individual state \cite{bohm2012quantum}. The wave function can be related to the probability amplitude \cite{feynman2010quantum},
\begin{equation}
\Psi(\vec{r},t)=\int \Psi(\vec{r}_1,t_1) K(\vec{r}_1,t_1;\vec{r},t) d\vec{r}_1 d t_1.
\end{equation}
The superposition of two probability amplitudes for two identical paths will not get two times of each probability amplitude, but only one probability amplitude. The two different paths in Eq. (\ref{laser-1}) are identical for single-mode continuous-wave laser light. The two-photon probability function is
\begin{eqnarray}\label{laser-2}
P^{(2)}(\vec{r}_1,t_1;\vec{r}_2,t_2)=\langle |A(S,D_1;S,D_2)|^2\rangle,
\end{eqnarray} 
no second-order interference pattern exists and $A(S,D_1;S,D_2)$ is the probability amplitude for two photons emitted by S going to D$_1$ and D$_2$, respectively.

Comparing the two-photon probability functions of thermal and laser light in a HBT interferometer, the reason why these two cases are different is due to the initial phases of photons are random in thermal light and the initial phases of photons are identical in laser light within one coherence volume. This difference makes two different paths, $A(a,D_1;b,D_2)$ and $A(a,D_2;b,D_1)$, different for thermal light and identical for laser light. There is two-photon bunching for thermal light and no two-photon bunching for single-mode continuous-wave laser light. 

\subsection{Second-order interference of two light beams}

The first second-order interference experiment with two light beams is correlation measurement of photon pairs generated in the annihilation of positron-electron pairs \cite{hanna1948polarization,bleuler1948correlation,wu1950angular}. Pfleegor and Mandel employed the second-order correlation of light to prove that there exists transient first-order interference pattern for two independent lasers at low light level \cite{pfleegor1967interference}. The most famous second-order interference of two light beams is the second-order interference of entangled photon pairs in a HOM interferometer \cite{hong1987measurement,shih1988new,shih1986conference}. Many other second-order interference experiments with light were performed, please see Refs. \cite{belinskiui1993interference,paul1986interference,ou2007multi,shih2020introduction} for details. This section will concentrate on applying Feynman\rq{}s path integral to discuss the second-order interference of two light beams in a HOM interferometer.

\begin{figure}[htb]
\centering
\includegraphics[width=50mm]{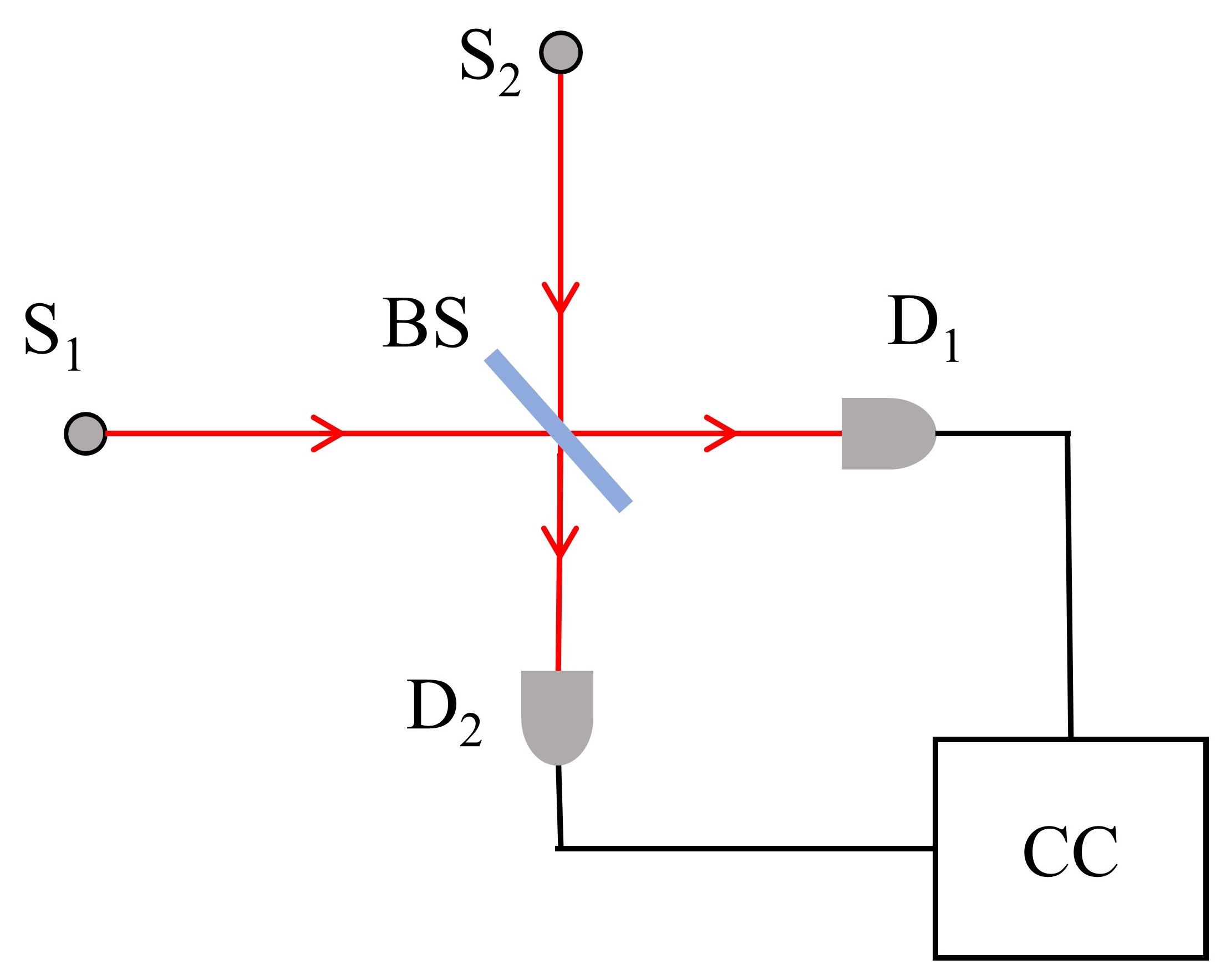}
\caption{Second-order interference of two light beams in a HOM interferometer. S$_1$ and S$_2$ are two light sources. BS is a 1:1 non-polarizing beam splitter. D$_1$ and D$_2$ are two single-photon detectors. CC is a two-photon coincidence count detection system.}\label{18-HOM-interferomter}
\end{figure}

The scheme for the second-order interference of two light beams in a HOM interferometer is shown in Fig. \ref{18-HOM-interferomter}. Two light beams emitted by two sources, S$_1$ and S$_2$, are incident into two input ports of a 1:1 non-polarizing beam splitter, BS. The output light beams from BS are detected by two single-photon detectors, D$_1$ and D$_2$. The output of these two detectors are sent into a two-photon coincidence count detection system to measure the second-order correlation.

\subsubsection{Second-order interference of entangled photon pairs}

Assume two photons in one entangled photon pair are in the scheme shown in Fig. \ref{18-HOM-interferomter}, which is similar as the original second-order interference of entangled photon pairs \cite{hong1987measurement,shih1988new,shih1986conference}. There are two different paths for photons a and b to trigger a two-photon coincidence count in D$_1$ and D$_2$. For simplicity, assume photon a is emitted by S$_1$ and photon b is emitted by S$_2$. The first path is that photon a is detected by D$_1$ and photon b is detected by D$_2$. The probability for the path is 1/2 and the corresponding two-photon probability amplitude is $A_{a,D_1;b,D_2}$. The second path is that photon a is detected by D$_2$ and photon b is detected by D$_1$. The probability for the path is 1/2 and the corresponding two-photon probability amplitude is $A_{a,D_2;b,D_1}$. If these two different paths are indistinguishable, the two-photon probability function is
\begin{eqnarray}\label{entangle-1}
&&P^{(2)}(\vec{r}_1,t_1;\vec{r}_2,t_2)\nonumber\\
&=& \langle |\frac{1}{\sqrt{2}}[A_{a,D_1;b,D_2}+A_{a,D_2;b,D_1}]|^2\rangle.
\end{eqnarray}
There is an extra $\pi/2$ phase for the reflected photon compared to the transmitted photon introduced by BS \cite{loudon2000quantum}, which can not be ignored in HOM interferometer. Equation (\ref{entangle-1}) can be simplified as
\begin{eqnarray}\label{entangle-2}
&&P^{(2)}(\vec{r}_1,t_1;\vec{r}_2,t_2)\nonumber\\
&\propto&\langle|e^{i\varphi_{a}}K_{aD_1}e^{i\varphi_{b}}K_{bD_2}+e^{i\varphi_{a}}e^{i\pi/2}K_{aD_2}e^{i\varphi_{b}}e^{i\pi/2}K_{bD_1}|^2\rangle \nonumber\\
&=& \langle |K_{aD_1}K_{bD_2}-K_{aD_2}K_{bD_1}|^2 \rangle,
\end{eqnarray} 
where $\varphi_a$ and $\varphi_b$ are the initial phases of photons a and b, respectively, $K_{aD_1}$ is the photon\rq{}s Feynman propagator for photon a emitted by S$_1$ goes to D$_1$ and the meanings of other symbols are defined similarly.  $\varphi_a$ and $\varphi_b$ are random for entangled photon pairs, the probability to detect one photon at D$_1$ or D$_2$ is a constant. The normalized two-photon probability function is proportional to the two-photon probability function, $P^{(2)}(\vec{r}_1,t_1;\vec{r}_2,t_2)$.  

Both temporal and spatial second-order interference patterns can be calculated. Let us first calculate the temporal case. Assume D$_1$ and D$_2$ are in the symmetrical positions of the BS. All the properties such as polarization, frequency, are the same for photons a and b. With the same method as the one in Sect. \ref{sec-second}, Eq. (\ref{entangle-2}) can be simplified as
\begin{equation}\label{entangle-3}
P^{(2)}(t_1-t_2)\propto 1-\text{sinc}^2\frac{\Delta \omega (t_2-t_1)}{2},
\end{equation}
where $t_1$ and $t_2$ are the photon detection time at D$_1$ and D$_2$, respectively, $\Delta \omega$ is frequency bandwidth of entangle photon pairs. Point light sources are assumed to simplify the calculation. 

\begin{figure}[htb]
\centering
\includegraphics[width=55mm]{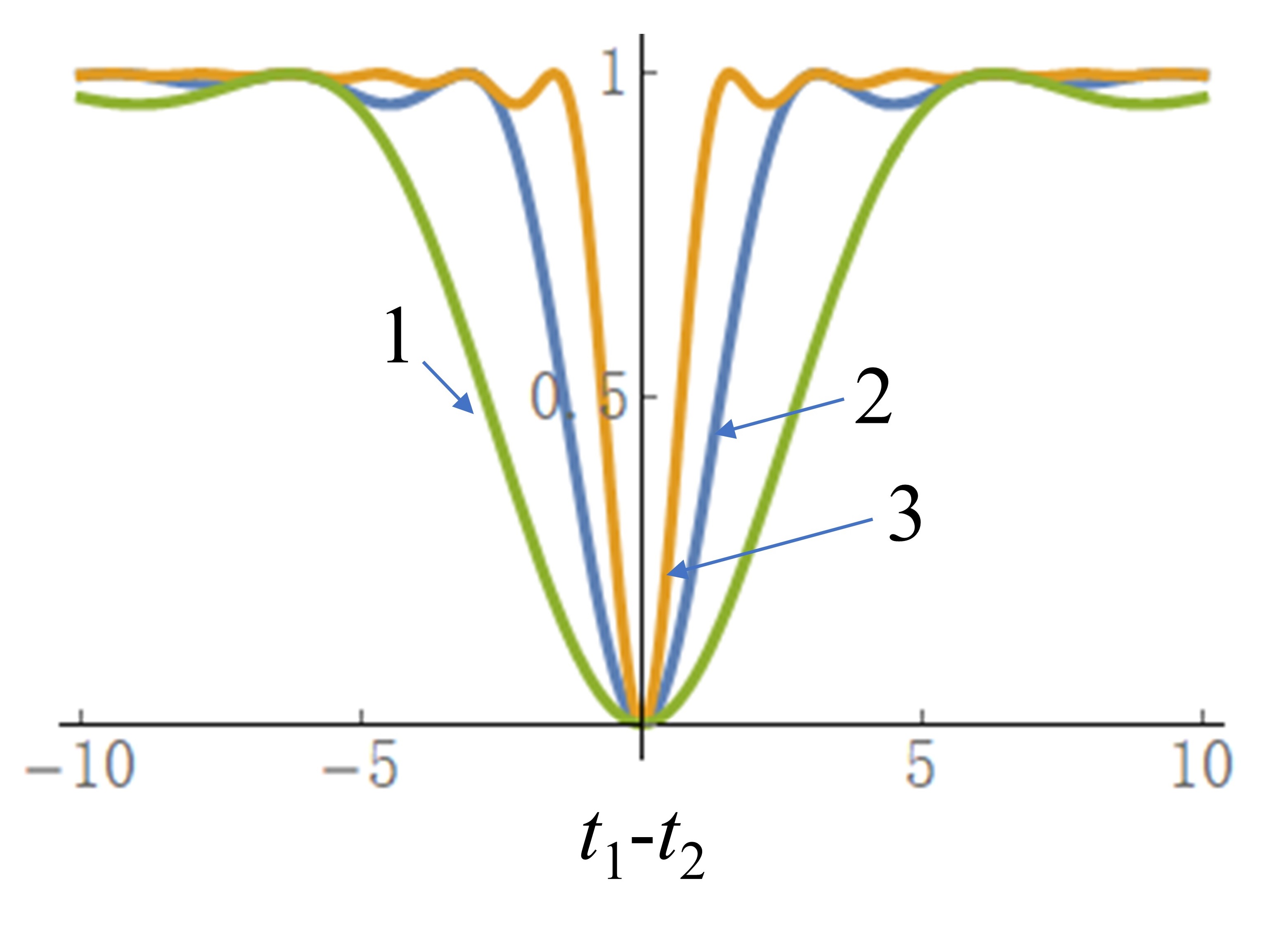}
\caption{Simulated HOM dip for entangled photon pairs with different frequency bandwidths. 1, 2, and 3 correspond to $\Delta \omega$, $2\Delta \omega$, and $4\Delta \omega$, respectively. }\label{19-HOM-simulation}
\end{figure}

When the time difference between these two photon detection events is zero, $P^{(2)}(t_1-t_2)$ equals zero, which means that it is impossible to detect two photons at D$_1$ and D$_2$ simultaneously. These two photons are in the same output port of BS. When $|t_1-t_2|$ increases, $P^{(2)}(t_1-t_2)$ increases from zero to one.  $P^{(2)}(t_1-t_2)$ equals 1 when $|t_1-t_2|$ is much longer than the coherence time, which indicates that the detections of these two photons are independent.

Similar method can be employed to calculate the spatial two-photon probability function,
\begin{equation}\label{entangle-4}
P^{(2)}(x_1-x_2) \propto 1- \text{sinc}^2\frac{d}{\lambda L}(x_1-x_2),
\end{equation}
where $x_1$ and $x_2$ are the transverse coordinates of D$_1$ and D$_2$, respectively, $d$ is the size of one dimension source. Paraxial approximation and one-dimension case are assumed to simplify the results. Experiments with entangled photon pairs confirmed the result in Eq. (\ref{entangle-4}) \cite{kim2006spatial,lee2006spatial,devaux2019stochastic}.

It is interesting to compare two-photon bunching of thermal light in a HBT interferometer and HOM dip of entangled photon pair in a HOM interferometer. Two-photon bunching of thermal light shown in Fig. \ref{12-HBT-temporal} looks like a contrary effect of HOM dip shown in Fig. \ref{19-HOM-simulation}. Photons have the tendency to bunch together when these two detectors are at symmetrical positions in Fig. \ref{12-HBT-temporal}.  Photons have the tendency to leave the BS from the same output port when these two detectors are at symmetrical positions in Fig. \ref{19-HOM-simulation}. Even through different phenomena are observed with different types of light in different interferometers, the physics of these two phenomena are the same. As shown in Eqs. (\ref{thermal-3}) and (\ref{entangle-2}), they are results of two-photon interference. The only difference is that two-photon bunching of thermal light is caused by constructive two-photon interference and HOM dip with entangled photon pairs is caused by destructive two-photon interference.

Entangled photon pairs are not necessary to observe HOM dip shown in Fig. \ref{19-HOM-simulation}. The same HOM dip with 100\% visibility can be observed with two independent and identical single-photon states. Theoretical calculations are exactly the same as the one for entangled photon pairs and the results are experimentally verified by several groups \cite{legero2004quantum,beugnon2006quantum,flagg2010interference,qian2016temporal}. 

\subsubsection{Second-order interference of two classical light beams}

The same method can be employed to calculate the second-order interference of two classical light beams in a HOM interferometer. Assume S$_1$ and S$_2$ are two independent and identical single-mode continuous-wave lasers. The polarizations and intensities of these two laser light beams are the same. There are three different ways to trigger a two-photon coincidence count in this case. The first way is these two photons are both emitted by S$_1$. The probability is 1/4. There is one path and the corresponding two-photon probability amplitude is $A_{11,12}$. The second way is these two photons are both emitted by S$_2$. The probability is 1/4. There is one path and the corresponding two-photon probability amplitude is $A_{21,22}$. The third way is the detected two photons are emitted by S$_1$ and S$_2$ one each. The probability is 1/2 and there are two different paths. The corresponding two-photon probability amplitudes are $A_{11,22}$ and $A_{12,21}$. If these four different paths to trigger a two-photon coincidence count are indistinguishable, the two-photon probability function is
\begin{eqnarray}\label{laser-laser-1}
&&P^{(2)}(\vec{r}_1,t_1;\vec{r}_2,t_2)\nonumber\\
&=&\langle |\frac{1}{\sqrt{4}}A_{11,12}+\frac{1}{\sqrt{4}}A_{21,22}+\frac{1}{\sqrt{4}}(A_{11,22}+A_{12,21})|^2\rangle\nonumber\\
&=& \langle |\frac{1}{\sqrt{4}}e^{i\varphi_{11,12}}K_{11,12}+\frac{1}{\sqrt{4}}e^{i\varphi_{21,22}}K_{21,22}\nonumber\\
&&+\frac{1}{\sqrt{4}}(e^{i\varphi_{11,22}}K_{11,22}+e^{i\varphi_{12,21}}K_{12,21})|^2\rangle,
\end{eqnarray}
where $\varphi_{11,12}$ is the phase related to the initial phase of these two photons and extra $\pi/2$ phase due to the reflection on BS. The last $1/\sqrt{4}$ in Eq. (\ref{laser-laser-1}) is the product of $1/\sqrt{2}$ times $1/\sqrt{2}$, where the first $1/\sqrt{2}$ is due to the probability of these two paths equals 1/2 and the second $1/\sqrt{2}$ is the normalization parameter. Table \ref{extra-phase} summarizes the phases for these four different paths, where $\varphi_1$ and $\varphi_2$ are the initial phases of photons emitted by S$_1$ and S$_2$, respectively.

\begin{table}[htb]
\caption{Phases for different two-photon paths.}\label{extra-phase}
\centering
\begin{tabular}[c]{|c|c|}
\hline
$\varphi_{11,12}$&$\varphi_{1}+\varphi_{1}+\pi/2$\\
\hline
$\varphi_{21,22}$&$\varphi_{2}+\varphi_{2}+\pi/2$\\
\hline
$\varphi_{11,22}$&$\varphi_{1}+\varphi_{2}$\\
\hline
$\varphi_{12,21}$&$\varphi_{1}+\varphi_{2}+\pi$\\
\hline
\end{tabular}
\end{table}

With the phases summarized in Table \ref{extra-phase}, Eq. (\ref{laser-laser-1}) can be simplified as
\begin{eqnarray}\label{laser-laser-2}
&&P^{(2)}(\vec{r}_1,t_1;\vec{r}_2,t_2)\nonumber\\
&\propto&\langle | e^{i(2\varphi_{1}+\pi/2)}K_{11,12}+e^{i(2\varphi_{2}+\pi/2)}K_{21,22}\nonumber\\
&&+e^{i(\varphi_{1}+\varphi_{2})}(K_{11,22}-K_{12,21}|^2 \rangle.
\end{eqnarray}
Since these two lasers are independent, Eq. (\ref{laser-laser-2}) can be simplified as
\begin{eqnarray}\label{laser-laser-3}
&&P^{(2)}(\vec{r}_1,t_1;\vec{r}_2,t_2)\\
&=&\langle |K_{11,12}|^2 \rangle + \langle |K_{21,22}|^2 \rangle + \langle |K_{11,22}-K_{12,21}|^2 \rangle. \nonumber
\end{eqnarray}
Assuming S$_1$ and S$_2$ are point laser light sources, temporal two-photon probability function can be obtained by simplifying Eq. (\ref{laser-laser-3}),
\begin{eqnarray}\label{laser-laser-4}
P^{(2)}(t_1-t_2) & \propto & 1+1+2[1-\cos \Delta \omega(t_1-t_2)] \nonumber\\
&\propto & 1-\frac{1}{2}\cos \Delta \omega(t_1-t_2),
\end{eqnarray}
where $\Delta \omega=\omega_1-\omega_2$ is frequency difference between these two laser light beams. There is second-order interference pattern by superposing two independent laser light beams. Unlike the  transient first-order interference pattern of two independent laser light beams is observed within the coherence time, the second-order interference pattern in Eq. (\ref{laser-laser-4}) is observed over long time average. 

\begin{figure}[htb]
\centering
\includegraphics[width=80mm]{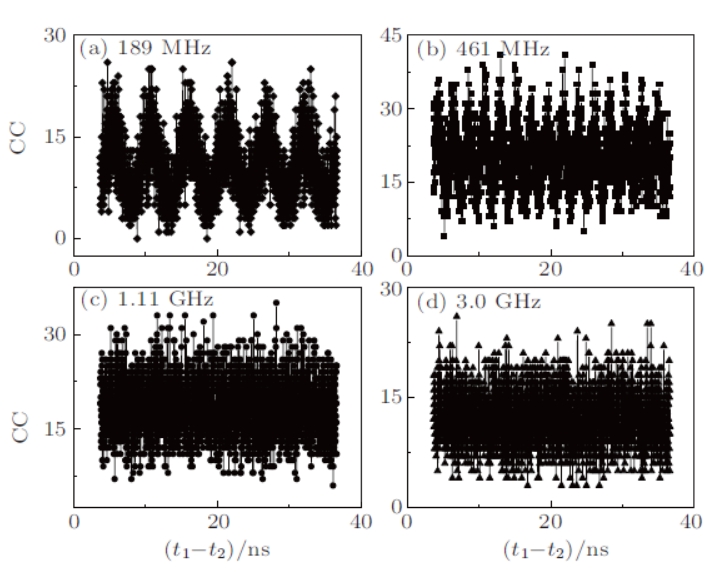}
\caption{Temporal second-order interference patterns of two independent laser light beams in a HOM interferometer \cite{liu2016second}}\label{20-laser-laser}
\end{figure}

Figure \ref{20-laser-laser} shows the temporal second-order interference patterns with two tunable single-mode continuous-wave lasers. The frequency bandwidth of laser light is 100 kHz. The frequency difference between these two lasers are 189 MHz, 461 MHz, 1.11 GHz, and 3.0 GHz in Figs. \ref{20-laser-laser}(a)-(d), respectively. Periodic interference patterns are observed in Figs. \ref{20-laser-laser}(a)-(b). There are also periodic structure in Figs. \ref{20-laser-laser}(c)-(d). The reason why it can not be observed is that the period of the interference pattern is too small to be observed in the time scale \cite{liu2016second}.

The same method can be employed to calculate the second-order interference of two independent thermal and laser light beams. Assume S$_1$ is a single-mode continuous-wave laser and S$_2$ is a thermal light source in Fig. \ref{18-HOM-interferomter}. The intensities, polarizations, and frequencies of the light beams emitted by these two sources are the same. There are three different ways for D$_1$ and D$_2$ detecting a two-photon coincidence count. The first way is these two photons are emitted by laser, S$_1$. The probability is 1/4 and there is one path, $A_{11,12}$. The second way is these two photons are emitted by thermal light source, S$_2$. The probability is 1/4 and there are two paths, $A_{2a1,2b2}$ and $A_{2a2,2b1}$. The third way is these two photons are emitted by S$_1$ and S$_2$ each. The probability is 1/2 and there are two different paths, $A_{11,22}$ and $A_{12,21}$. If these five different paths are indistinguishable, the two-photon probability function equals
\begin{eqnarray}\label{laser-thermal-1}
&&P^{(2)}(\vec{r}_1,t_1;\vec{r}_2,t_2)\nonumber\\
&=&\langle |\frac{1}{\sqrt{4}}A_{11,12}+\frac{1}{\sqrt{8}}(A_{2a1,2b2}+A_{2a2,2b1})\nonumber\\
&&+\frac{1}{\sqrt{4}}(A_{11,22}+A_{12,21})|^2\rangle,
\end{eqnarray}
With the same method as the one for Eq. (\ref{laser-laser-1}), Eq. (\ref{laser-thermal-1}) can be simplified as \cite{liu2014second}
\begin{eqnarray}\label{laser-thermal-2}
&&P^{(2)}(\vec{r}_1,t_1;\vec{r}_2,t_2)\nonumber\\
&=&2\langle |K_{11,12}|^2\rangle+\langle|K_{2a1,2b2}+K_{2a2,2b1}|^2\rangle\nonumber\\
&&+2\langle |K_{11,22}-K_{12,21}|^2\rangle.
\end{eqnarray}
With paraxial approximation, one dimension spatial two-photon probability function can be obtained via Eq. (\ref{laser-thermal-2}),
\begin{eqnarray}\label{laser-thermal-3}
&&p^{(2)}(x_1-x_2)\\
&\propto& 1+\frac{1}{4} \text{sinc}^2 \frac{\pi l}{L\lambda}(x_1-x_2)[1-2\cos\frac{2\pi d}{L\lambda}(x_1-x_2)] \nonumber,
\end{eqnarray}
where $l$ is the length of thermal light source, $d$ is the distance between the central points of  thermal light source and the virtual source of laser relative to BS \cite{liu2014second}. 

\begin{figure}[htb]
\centering
\includegraphics[width=70mm]{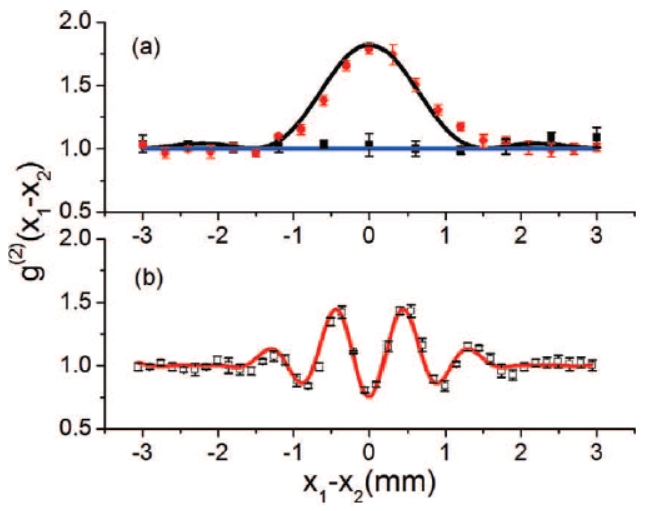}
\caption{Second-order interference pattern of thermal and laser light (b) and the second-order coherence functions of laser and thermal light beams in the interferometer alone (a) \cite{liu2014second}. }\label{21-laser-thermal}
\end{figure}

Figure \ref{21-laser-thermal}(b) shows the spatial second-order interference pattern of thermal and laser light beams in a HOM interferometer, which is consistent with Eq. (\ref{laser-thermal-3}). For comparison, Fig. \ref{21-laser-thermal}(a) shows the measured second-order coherence functions when one of the incidental light beams is blocked, which is equivalent to laser and thermal light in a HBT interferometer \cite{liu2014second}. 

The second-order interference of two independent thermal light beams \cite{liu2013spatial}, classical and non-classical light beams \cite{hong1988interference} and other two independent light beams can also be calculated in the same method above. The main difference between the second-order of two independent classical and non-classical light beams is that the visibility of the second-order interference pattern of two independent non-classical light beams can reach 100\%, while the maximal visibility of the second-order interference pattern of two independent classical light beam is 50\% \cite{mandel1983photon}. The reason is clearly illustrated by Eqs. (\ref{entangle-1}), (\ref{laser-laser-1}), and (\ref{laser-thermal-1}). In the second-order interference of two independent classical light beams, it is impossible to distinguish  two sources emitting a photon each from one source emitting two photons, which will contribute to the background coincidence counts. If special two-photon coincidence count detection system can distinguish these two types of coincidence count, the visibility of the second-order interference pattern of two independent classical light beams can exceed 50\%. For instance, Kaltenbaek \textit{et al.} observed the visibility of second-order interference pattern with two independent laser light beams exceeding 85\% by employing sum frequency generation to distinguish these two different types of coincidence count \cite{kaltenbaek2008quantum}. With two non-independent classical light beams and normal two-photon coincidence count system, Sadana \textit{et al.} observed the second-order interference pattern with near 100\% visibility \cite{sadana2019near}. Their results \cite{sadana2019near,kaltenbaek2008quantum} do not conflict with Mandel\rq{}s conclusion due to Mandel\rq{}s conclusion is valid for two independent light beams and normal two-photon coincidence count detection system \cite{mandel1983photon}.

\subsection{Second-order interference of multiple light beams}

\begin{figure}[htb]
\centering
\includegraphics[width=50mm]{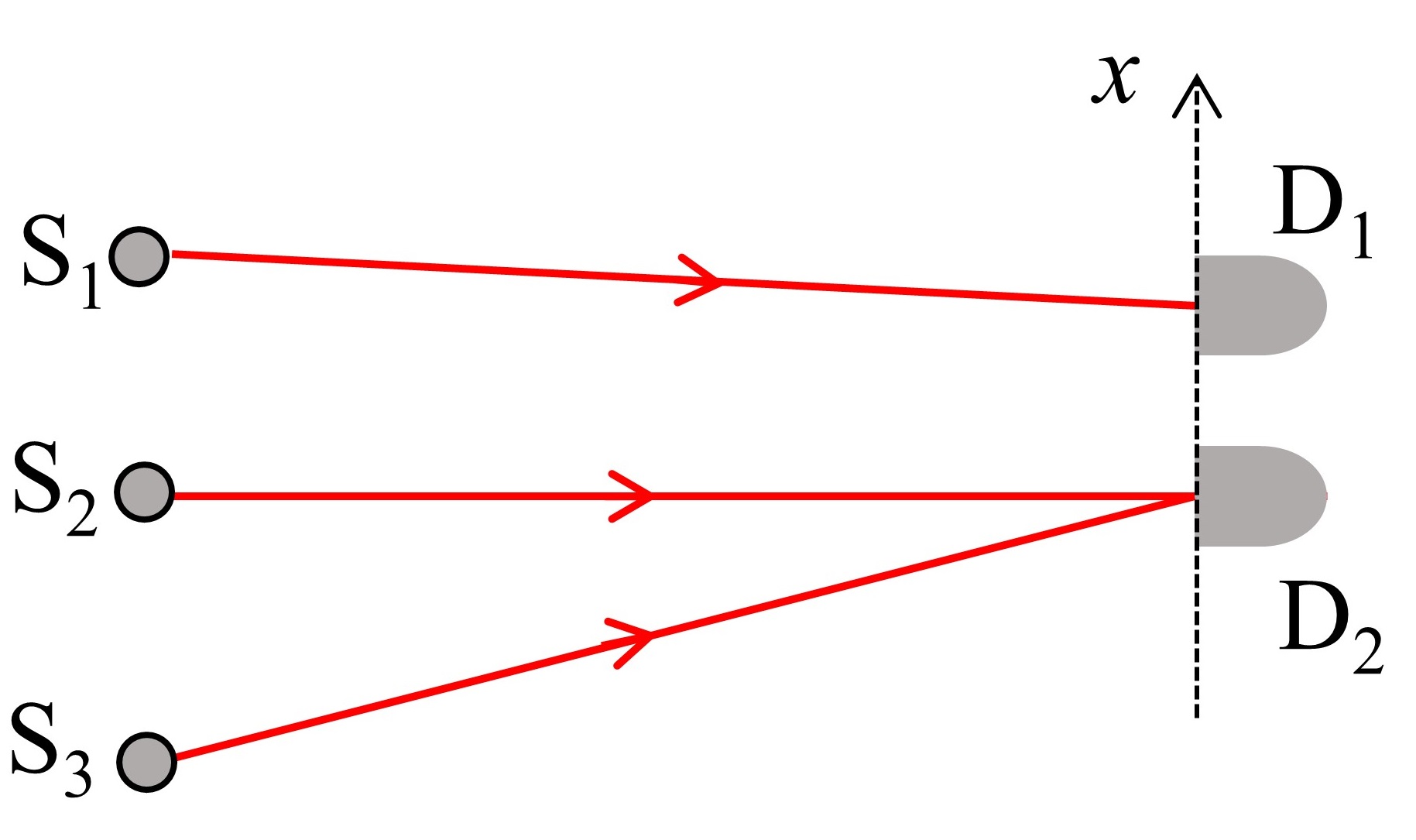}
\caption{Second-order interference of three independent light beams. S$_1$, S$_2$ and S$_3$  are three light sources. D$_1$ and D$_2$ are two single-photon detectors. The output of two detectors are sent into a two-photon coincidence count system, which is not shown.}\label{22-three-sources}
\end{figure}

Feyman\rq{}s path integral can be employed to calculate the second-order interference of more than two independent light beams. Figure \ref{22-three-sources} shows the scheme for the second-order interference of three independent light beams. S$_1$, S$_2$ and S$_3$ are three independent light sources. D$_1$ and D$_2$ are two single-photon detectors. The output of these two detectors are sent into a two-photon coincidence count detection system to measure the second-order correlation, which is not shown in the figure.

Assume S$_1$, S$_2$ and S$_3$  are three independent and identical single-photon sources. There are three different ways to trigger a two-photon coincidence count. The first way is that the two detected photons are emitted by S$_1$ and S$_2$. The probability is 1/3. There are two different paths and the corresponding two-photon probability amplitudes are $A_{11,22}$ and $A_{12,21}$. The second way is that these two photons are emitted by S$_1$ and S$_3$. The probability is 1/3. There are two paths and the corresponding two-photon probability amplitudes are $A_{11,32}$ and $A_{12,31}$. The third way is that these two photons are emitted by S$_2$ and S$_3$. The probability is 1/3. There are two paths and the corresponding two-photon probability amplitudes are $A_{21,32}$ and $A_{22,31}$. If only one single-photon source emits a photon, it is impossible to trigger a two-photon coincidence count. If two single-photon sources emit a photon simultaneously, these three different ways are distinguishable. One can measure the status of the single-photon sources to ensure which two sources emitted photons. The probability for three independent single-photon sources simultaneously emitting a photon is much smaller than the probability of two single-photon sources simultaneously emitting a photon. Hence three single-photon sources simultaneously emitting a photon can be ignored in the calculation.

If every two different paths to trigger a two-photon coincidence count are indistinguishable, the two-photon probability function is
\begin{eqnarray}\label{three-1}
&&P^{(2)}(\vec{r}_1,t_1;\vec{r}_2,t_2)\nonumber\\
&=&\langle | \frac{1}{\sqrt{6}}(A_{11,22}+A_{12,21})|^2+|\frac{1}{\sqrt{6}}(A_{11,32}+A_{12,31})|^2 \nonumber\\
&&+|\frac{1}{\sqrt{6}}(A_{21,32}+A_{22,31})|^2\rangle.
\end{eqnarray}
With the same method above, Eq. (\ref{three-1}) can be simplified as,
\begin{eqnarray}\label{three-2}
&&P^{(2)}(\vec{r}_1,t_1;\vec{r}_2,t_2)\nonumber\\
&\propto&\langle |K_{11}K_{22}+K_{12}K_{21}|^2 \rangle + \langle |K_{11}K_{32}+K_{12}K_{31}|^2 \rangle \nonumber\\
&& +\langle |K_{21}K_{32}+K_{22}K_{31}|^2 \rangle,
\end{eqnarray}
where $K_{11}$ is the photon\rq{}s Feynman propagator from S$_1$ to D$_1$ and the meanings of other symbols are defined similarly. Assuming S$_1$, S$_2$ and S$_3$  are point single-photon sources in the same plane, the distance between the source and the observation planes is much larger than one between sources, the spatial two-photon probability function can be obtained via Eq. (\ref{three-2}),
\begin{eqnarray}\label{three-3}
&&P^{(2)}(x_1-x_2)\nonumber\\
&=&3+\cos\frac{2\pi d_{12}}{L \lambda}(x_1-x_2-\frac{d_{23}}{2})\\
&&+ \cos\frac{2\pi d_{13}}{L \lambda}(x_1-x_2) + \cos\frac{2\pi d_{23}}{L \lambda}(x_1-x_2+\frac{d_{12}}{2}),\nonumber
\end{eqnarray}
where paraxial approximation and one-dimension case have been assumed, $\lambda$ is the wavelength of light and single-frequency light is employed to simplify the calculation, $d_{12}$ is the distance between S$_1$ and S$_2$ and the relative positions of these three sources are shown in Fig. \ref{8-three-slits}. The meanings of other symbols are defined similarly. 

\begin{figure}[htb]
\centering
\includegraphics[width=90mm]{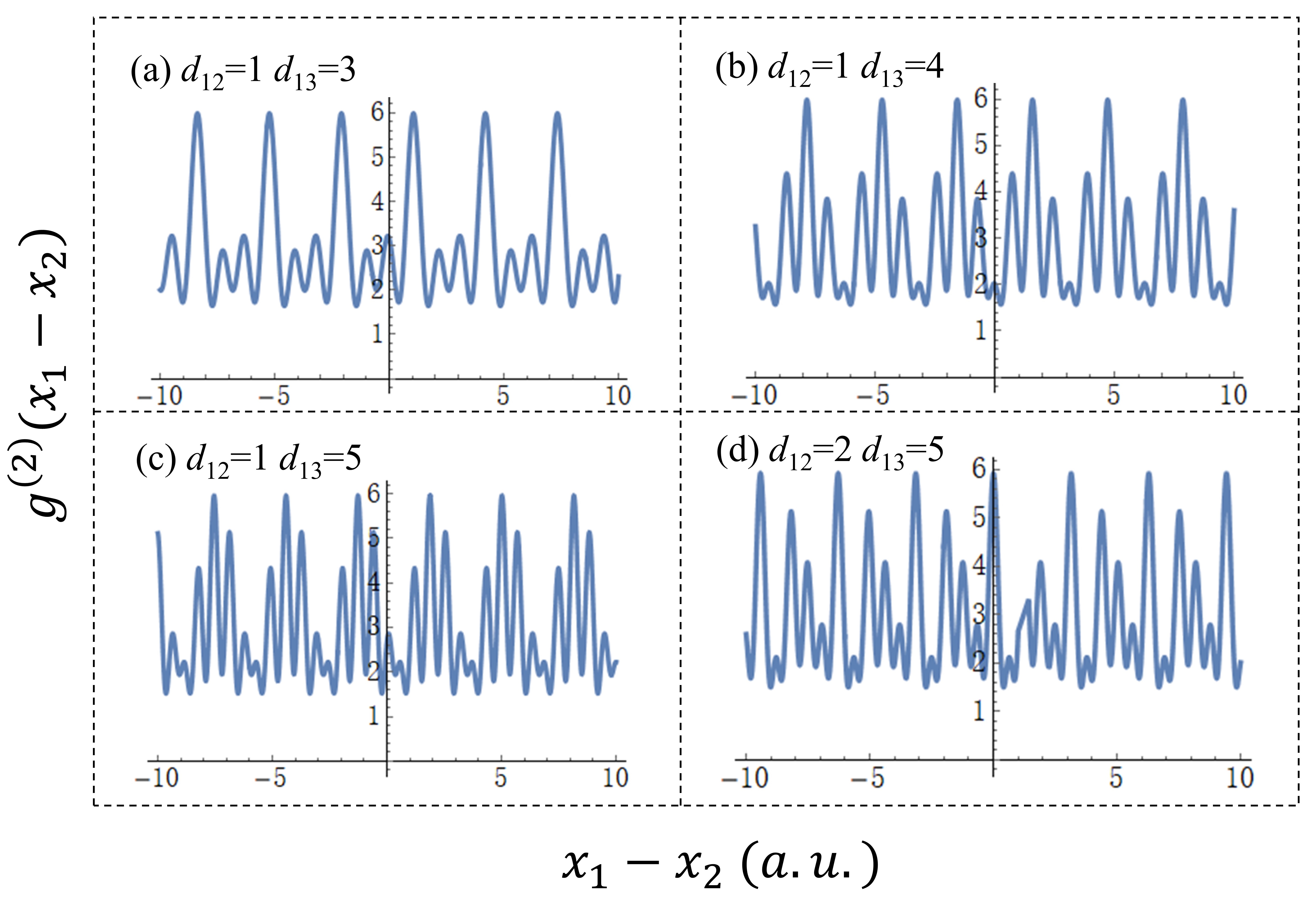}
\caption{Second-order interference patterns of three independent light beams. $x_1-x_2$ is the transverse position difference of D$_1$ and D$_2$.}\label{23-three-pattern}
\end{figure}

Figure \ref{23-three-pattern} shows the second-order interference patterns of three single-photon sources. The patterns will change as the distances between these three sources change. Compared to the first-order interference pattern of three light beams shown in Fig. \ref{9-three-pattern}, the visibility of the second-order interference pattern with three independent light beams can not reach 100\% even when non-classical light are employed.

The second-order interference of three independent laser light beams can be calculated in the same method. Assume  S$_1$, S$_2$ and S$_3$ are three independent and single-mode continuous-wave lasers and the polarizations of these three laser light beams are identical. There are six different ways to trigger a two-photon coincidence count in Fig. \ref{22-three-sources}. The first way is that these two photons are both emitted by S$_1$. The probability is $I_1^2/(I_1+I_2+I_3)^2$, where $I_1$, $I_2$, and $I_3$ are the intensities of laser light beams emitted by S$_1$, S$_2$ and S$_3$, respectively. There is one path and the corresponding two-photon probability amplitude is $A_{11,12}$. The second ways is that these two photons are both emitted by S$_2$. The probability is $I_2^2/(I_1+I_2+I_3)^2$. There is one path and the two-photon probability amplitude is $A_{21,22}$. The third way is that these two photons are both emitted by S$_3$. The probability is $I_3^2/(I_1+I_2+I_3)^2$. There is one path and the probability amplitude is $A_{31,32}$. The fourth way is that these two photons are emitted by S$_1$ and S$_2$ one each. The probability is $2I_1I_2/(I_1+I_2+I_3)^2$. There are two paths and the probability amplitudes are $A_{11,22}$ and $A_{12,21}$. The fifth way is that these two photons are emitted by S$_1$ and S$_3$ one each. The probability is $2I_1I_3/(I_1+I_2+I_3)^2$ and the two paths are $A_{11,32}$ and $A_{12,31}$. The sixth way is that these two photons are emitted by S$_2$ and S$_3$ one each. The probability is $2I_2I_3/(I_1+I_2+I_3)^2$ and the two paths are $A_{21,32}$ and $A_{22,31}$. If all the nine different paths to trigger a two-photon coincidence count are indistinguishable, the two-photon probability function is
\begin{eqnarray}\label{three-4}
&&P^{(2)}(\vec{r}_1,t_1;\vec{r}_2,t_2)\nonumber\\
&\propto& \langle |I_1 A_{11,12}+I_2 A_{21,22}+ I_3 A_{31,32}\nonumber\\
&&+\sqrt{I_1I_2}(A_{11,22}+A_{12,21})\\
&&+\sqrt{I_1I_3}(A_{11,32}+A_{12,31})+\sqrt{I_2I_3}(A_{21,32}+A_{22,31})|^2 \rangle,\nonumber
\end{eqnarray}
where the common factor $1/(I_1+I_2+I_3)$ has been ignored, $\sqrt{I_1I_2}$ equals $\sqrt{2I_1I_2}\times 1/\sqrt{2}$.

Considering three lasers are independent, Eq. (\ref{three-4}) can be simplified as
\begin{eqnarray}\label{three-5}
&&P^{(2)}(\vec{r}_1,t_1;\vec{r}_2,t_2)\nonumber\\
&=& I_1^2 \langle | K_{11}K_{12}|^2 \rangle +I_2^2 \langle | K_{21}K_{22}|^2 \rangle+I_3^2 \langle | K_{31}K_{32}|^2 \rangle\nonumber\\
&& +I_1I_2\langle |K_{11}K_{22}+K_{12}K_{21}|^2 \rangle \nonumber \\
&&+I_1I_3\langle |K_{11}K_{32}+K_{12}K_{31}|^2 \rangle  \nonumber \\
&&+I_2I_3\langle |K_{21}K_{32}+K_{22}K_{31}|^2 \rangle.
\end{eqnarray}
Assuming the intensities of these three light beams are equal and single-frequency light is emitted, Eq. (\ref{three-5}) can be simplified as
\begin{eqnarray}\label{three-6}
&&P^{(2)}(x_1-x_2) \nonumber\\
&\propto& 4.5+\cos\frac{2\pi d_{12}}{L \lambda}(x_1-x_2-\frac{d_{23}}{2})  \nonumber\\
&&+ \cos\frac{2\pi d_{13}}{L \lambda}(x_1-x_2)\nonumber\\
&& + \cos\frac{2\pi d_{23}}{L \lambda}(x_1-x_2+\frac{d_{12}}{2}),
\end{eqnarray}
where paraxial approximation and one-dimension case are assumed to simplify the calculations. The second-order interference patterns of three independent single-photon states and three independent single-mode continuous-wave lasers are the same except the constant changes from 3 to 4.5. The reason for this difference is that laser light can not avoid two photons emitted by the same source with normal two-photon coincidence count detection system, which is the same as the one for the difference between the visibilities of the second-order interference patterns of two independent classical and non-classical light beams \cite{mandel1983photon}.

\subsection{Subwavelength interference}

In the above parts of Sect. \ref{sec-second}, we have introduced how to employ Feynman\rq{}s path integral to calculate the second-order interference of one, two, and multiple light beams. As an example of how quantum optical coherence theory based on Feynman\rq{}s path integral can be helpful to understand the physics of optical coherence, subwavelength interference of light is analyzed in Feynman\rq{}s path integral.

Take the first- and second-order interference of light emitted by two independent sources shown in Fig. \ref{5-two-lasers} as an example to illustrate subwavelength interference. The transient first-order interference pattern of two independent single-mode continuous-wave lasers is 
\begin{equation}\label{sub-1}
P^{(1)}(x)\propto 1+\cos (\frac{2\pi d }{\lambda L} x+\varphi).
\end{equation}
Putting two detectors in the detection plane in Fig. \ref{5-two-lasers}, the second-order interference pattern of two independent single-mode continuous-wave lasers is
\begin{equation}\label{sub-2}
p^{(2)}(x_1-x_2)\propto 1+ \frac{1}{2}\cos \frac{2\pi d}{\lambda L} (x_1-x_2).
\end{equation}
Comparing Eq. (\ref{sub-2}) with Eq. (\ref{sub-1}) , if these two detectors are moving in the opposite directions with the same speed, the period of the second-order interference pattern will be half of the period of the first-order interference pattern, which is called subwavelength interference. 

Subwavelength interference can be understood with the concept of de Broglie wavelength. The de Broglie wavelength of a photon equals $h/p$, where $h$ is Planck constant and $p$ is the momentum of the photon. If two photons can be treated as a whole, the corresponding de Broglie wavelength equals $h/(2p)$, which is only half of the wavelength of one photon. The de Broglie wavelength of $N$ photon equals $h/(Np)$, where $N$ is a positive integer larger than 1 \cite{jacobson1995photonic}. Two-photon subwavelength interference was first observed with entangled photon pairs \cite{fonseca1999measurement,boto2000quantum,d2001two}. Then it was found that two-photon subwavelength interference can also be observed with thermal light \cite{scarcelli2004two,xiong2005experimental,zhai2005two} and laser light \cite{liu2010unified}.

\begin{figure}[htb]
\centering
\includegraphics[width=60mm]{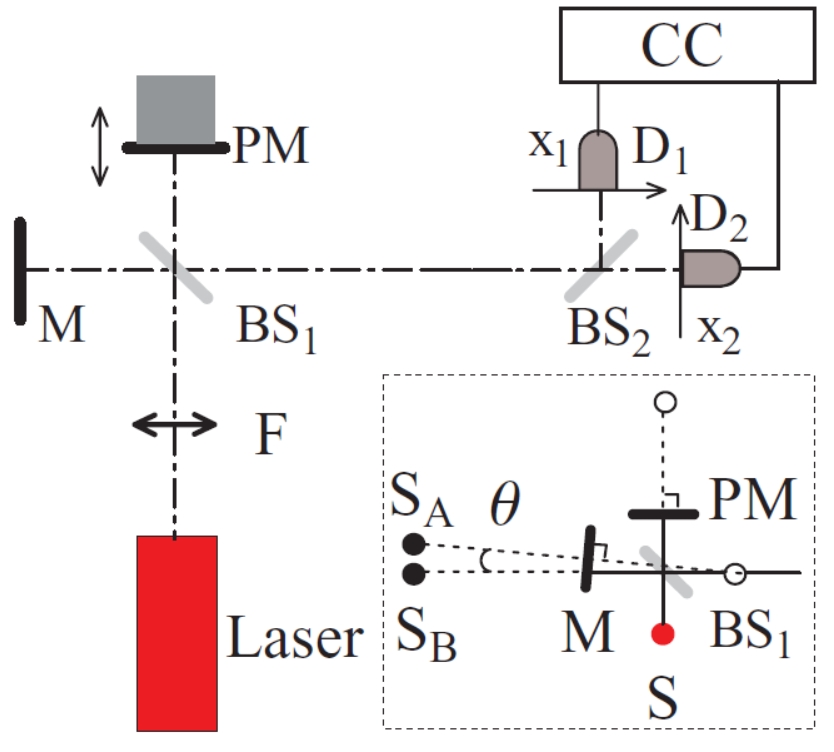}
\caption{Scheme for two-photon subwavelength interference with two laser light beams \cite{liu2010unified}. Laser is a single-mode continuous-wave laser. M is a mirror. PM is mirror mounted on a piezoelectric ceramic. S$_A$ and S$_B$ are two virtual sources of source S of M and PM, respectively. The meanings of other symbols are similar as the ones above.}\label{24-sub-scheme}
\end{figure}

Figure \ref{24-sub-scheme} shows the scheme for observing two-photon subwavelength interference with two independent laser light beams \cite{liu2010unified}, which is equivalent to a two point-source interferometer. Two independent laser light beams was obtained by splitting a single-mode continuous-wave laser light beam with a beam splitter (BS$_1$) and randomizing the phase of one beam by a mirror mounted on a piezoelectric ceramic (PM). One beam splitter (BS$_2$), two single-photon detectors (D$_1$ and D$_2$), and a two-photon coincidence count detection system (CC) are employed to measure the second-order interference pattern.

\begin{figure}[htb]
\centering
\includegraphics[width=70mm]{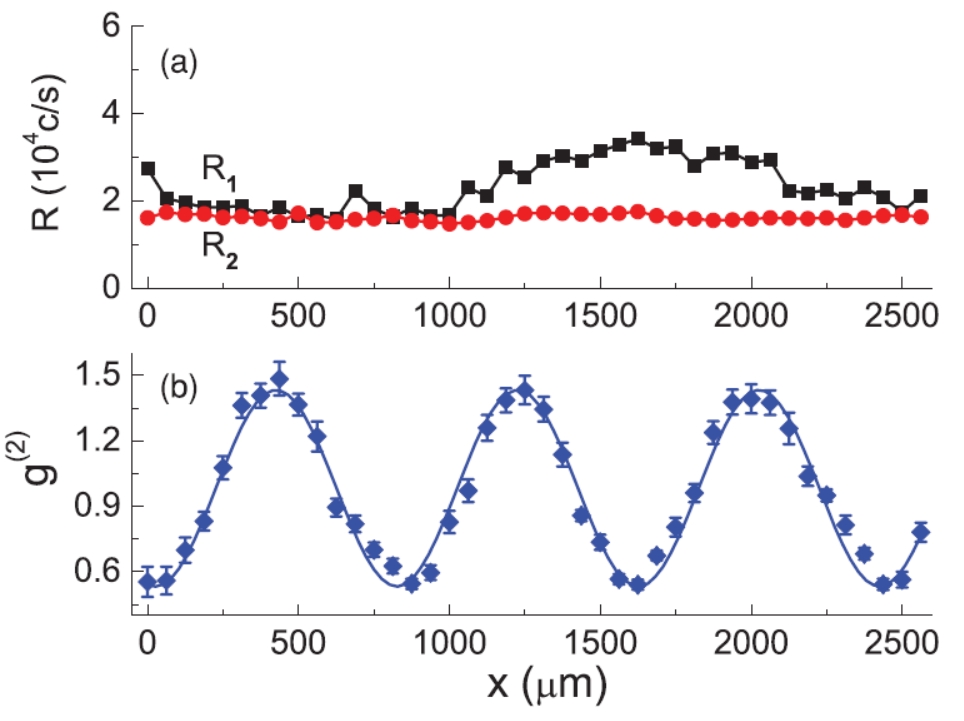}
\caption{The observed normal second-order interference pattern (b) when the position of D$_1$ is scanning why the position of D$_2$ is fixed \cite{liu2010unified}. R$_1$ and R$_2$ in (a) are single-photon counting rates of D$_1$ and D$_2$, respectively.}\label{25-normal}
\end{figure}

Figure \ref{25-normal}(b) shows the normal second-order interference pattern by fixing the position of D$_2$ and scanning the position of D$_1$. The period of the observed second-order interference pattern is 800 $\mu$m.  The visibility is $(46\pm2)\%$, which does not exceed 50\%. Figure \ref{25-normal}(a) shows the single-photon counting rates observed by two single-photon detectors, respectively. There is no first-order interference pattern.

\begin{figure}[htb]
\centering
\includegraphics[width=70mm]{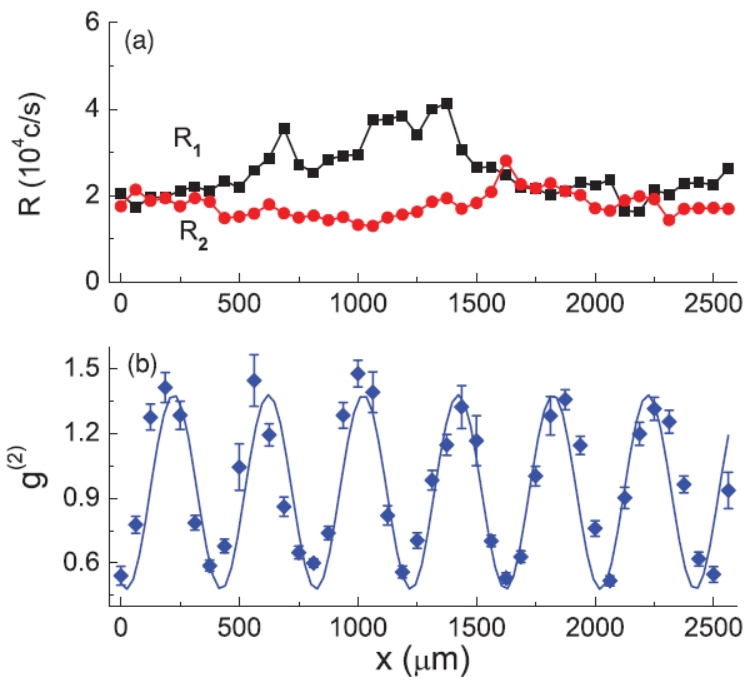}
\caption{Two-photon subwavelength interference pattern (b) when the positions of D$_1$ and D$_2$ are scanned in the opposite directions \cite{liu2010unified}. R$_1$ and R$_2$ in (a) are single-photon counting rates of D$_1$ and D$_2$, respectively. }\label{26-subwavelenth}
\end{figure}

Figure \ref{26-subwavelenth}(b) shows the observed two-photon subwavelength interference pattern by scanning D$_1$ and  D$_2$ in the opposite directions. The period of the second-order interference pattern is 400 $\mu$m, which is exactly half of the one shown in Fig. \ref{25-normal}(b) and indicates two-photon subwavelength interference is observed. The visibility of the second-order interference pattern equals $(48\pm4)\%$. One may argue that it is meaningless to observe subwavelength interference by scanning two detectors in the opposite directions \cite{liu2014super}. However, two-photon subwavelength interference with two detectors scanning in the same direction for classical light can be observed with suitable changes on the experimental scheme \cite{cao2010two}, which is similar as the one with entangled photon pairs \cite{fonseca1999measurement,d2001two}. As shown below, there is no intrinsic difference between the observed subwavelength interference patterns for detectors scanning in the same direction and opposite directions. 

In order to understand the physics of subwavelength interference, take the second-order interference of two single-mode continuous-wave laser light beams with the same phase as an example. Assume the intensities, frequency spectra and polarizations of these two laser light beams are the same. There are four different paths to trigger a two-photon coincidence count in Fig. \ref{24-sub-scheme}. If these four difference paths are indistinguishable , the two-photon probability function is
\begin{eqnarray}\label{sub-3}
&&P^{(2)}(x_1;x_2)\nonumber\\
&=& \langle |\frac{1}{\sqrt{4}}[A_{11,12}+A_{21,22}+A_{11,22}+A_{12,21}]|^2 \rangle\nonumber\\
&\propto&|e^{i2\varphi}(K_{11}K_{12}+K_{21}K_{22}+K_{11}K_{22}+K_{12}K_{21}|^2\rangle\nonumber\\
&=& |K_{11}K_{12}+K_{21}K_{22}+K_{11}K_{22}+K_{12}K_{21}|^2,
\end{eqnarray}
where $\varphi$ is the initial phases of photons in these two beams. Equation (\ref{sub-3}) can be re-arranged as
\begin{eqnarray}\label{sub-4}
&&P^{(2)}(x_1;x_2)\nonumber\\
&\propto&  4I_1I_2\nonumber\\
&&+I_1(K_{12}K_{22}^*+K_{22}K_{12}^*+K_{22} K_{12}^*+K_{12}K_{22}^*)\nonumber\\
&&+I_2(K_{11} K_{21}^*+K_{11} K_{21}^*+K_{11}K_{21}^*+K_{21}K_{11}^*)\nonumber\\
&&+K_{11} K_{12}K_{21}^*K_{22}^*+K_{21} K_{22}K_{11}^*K_{12}^*\nonumber\\
&&+K_{11}K_{22}K_{12}^*K_{21}^*+K_{12}K_{21}K_{11}^*K_{22}^*,
\end{eqnarray}
where $I_1=|K_{11}|^2=|K_{21}|^2$ and $I_2=|K_{12}|^2=|K_{22}|^2$, which equals half of the intensities detected by D$_1$ and D$_2$, respectively. Assume point laser light sources are employed, Eq. (\ref{sub-4}) can be simplified as \cite{liu2010unified}
\begin{eqnarray}\label{sub-5}
&&P^{(2)}(x_1;x_2)\nonumber\\
&\propto&  4I_1I_2\nonumber\\
&&+4I_1I_2\cos\frac{2\pi d}{\lambda L}x_2\nonumber\\
&&+4I_1I_2\cos\frac{2\pi d}{\lambda L}x_1\nonumber\\
&&+2I_1I_2\cos\frac{2\pi d}{\lambda L}(x_1+x_2)\nonumber\\
&&+2I_1I_2\cos\frac{2\pi d}{\lambda L}(x_1-x_2).
\end{eqnarray}
The last two lines on the right hand side of Eq. (\ref{sub-5}) correspond to two-photon subwavelength when two detectors scanning in the same direction and opposition directions, respectively. $2I_1I_2\cos\frac{2\pi d}{\lambda L}(x_1+x_2)$ corresponds to $K_{11} K_{12}K_{21}^*K_{22}^*+K_{21} K_{22}K_{11}^*K_{12}^*$ in Eq. (\ref{sub-4}), which is a result of two-photon interference for these two photons emitted by the same source. It is consistent with the condition for observing subwavelength interference with entangled photon pairs \cite{fonseca1999measurement,d2001two}.  $2I_1I_2\cos\frac{2\pi d}{\lambda L}(x_1-x_2)$ corresponds to $K_{11}K_{22}K_{12}^*K_{21}^*+K_{12}K_{21}K_{11}^*K_{22}^*$ in Eq. (\ref{sub-4}), which is a result of two-photon interference for these two photons emitted by one source each. If one can eliminate all the other terms and keep only one of the last two terms in Eq. (\ref{sub-5}), one can observe subwavelength interference. The experiment reported in Ref. \cite{liu2010unified} employed random phases between these two lasers to eliminate all the other terms except the constant and the last term in Eq. (\ref{sub-5}). Subwavelength interference is observed when these two detectors are scanning in the opposite directions. Equations (\ref{sub-4}) and (\ref{sub-5}) shows that there is no intrinsic difference between the subwavelength interference when these two detectors are scanning in the same direction and opposite directions. They are results of two-photon interference with different two-photon paths.

\section{Third- and higher-order interference of light}\label{sec-third}

The third- and higher-order interference of light is a natural generalization of the second-order interference of light. The method introduced above can also be employed to calculate the third- and higher-order interference of light.

The history of the third- and higher-order interference dates back to 1963 when Glauber gave a unified definition of $N$th-order coherence function of light \cite{glauber1963coherent,glauber1963quantum}. Greenberg \textit{et al.} generalized entangled photon pair to three-photon entanglement \cite{greenberger1989going,greenberger1990bell}. It was difficult to generate entangled three-photon directly then. Bouwmeester \textit{et al.} employed two entangled photon pairs to generate entangled three-photon by post-selection \cite{bouwmeester1999observation}. Mitchell \textit{et al.} employed a laser photon and an entangled photon pair to generate entangled three-photon and observed three-photon subwavelength interference \cite{mitchell2004super}. In 2010, H{\"u}bel \textit{et al.} employed cascade periodically poled nonlinear crystals to generate entangled three-photon directly \cite{hubel2010direct}. Based on the work of Bouwmeester \textit{et al.} \cite{bouwmeester1999observation}, Walther \textit{et al.} observed four-photon subwavelength interference with two entangled photon pairs \cite{walther2004broglie} and Pan\rq{}s group observed as high as 12-photon subwavelength interference \cite{zhao2004experimental,pan2012multiphoton,zhong201812}.

Ghost imaging was originally realized with entangled photon pairs \cite{pittman1995optical} and later with thermal light \cite{gatti2004ghost,cheng2004incoherent,cai2005ghost,valencia2005two} via the second-order interference of light. Ghost imaging with the third- and higher-order interference of light was discussed for entangled three-photon \cite{wen2007transverse} and thermal light \cite{ou2007ghost,bai2007ghost,agafonov2008high,zhou2010third,cao2008enhancing,chan2009high,liu2009role}.

In the following, third-order interference of light is taken as an example to show how to employ Feyman\rq{}s path integral to calculate the third- and high-order interference of light.

\subsection{Third-order interference of one light beam}

\begin{figure}[htb]
\centering
\includegraphics[width=60mm]{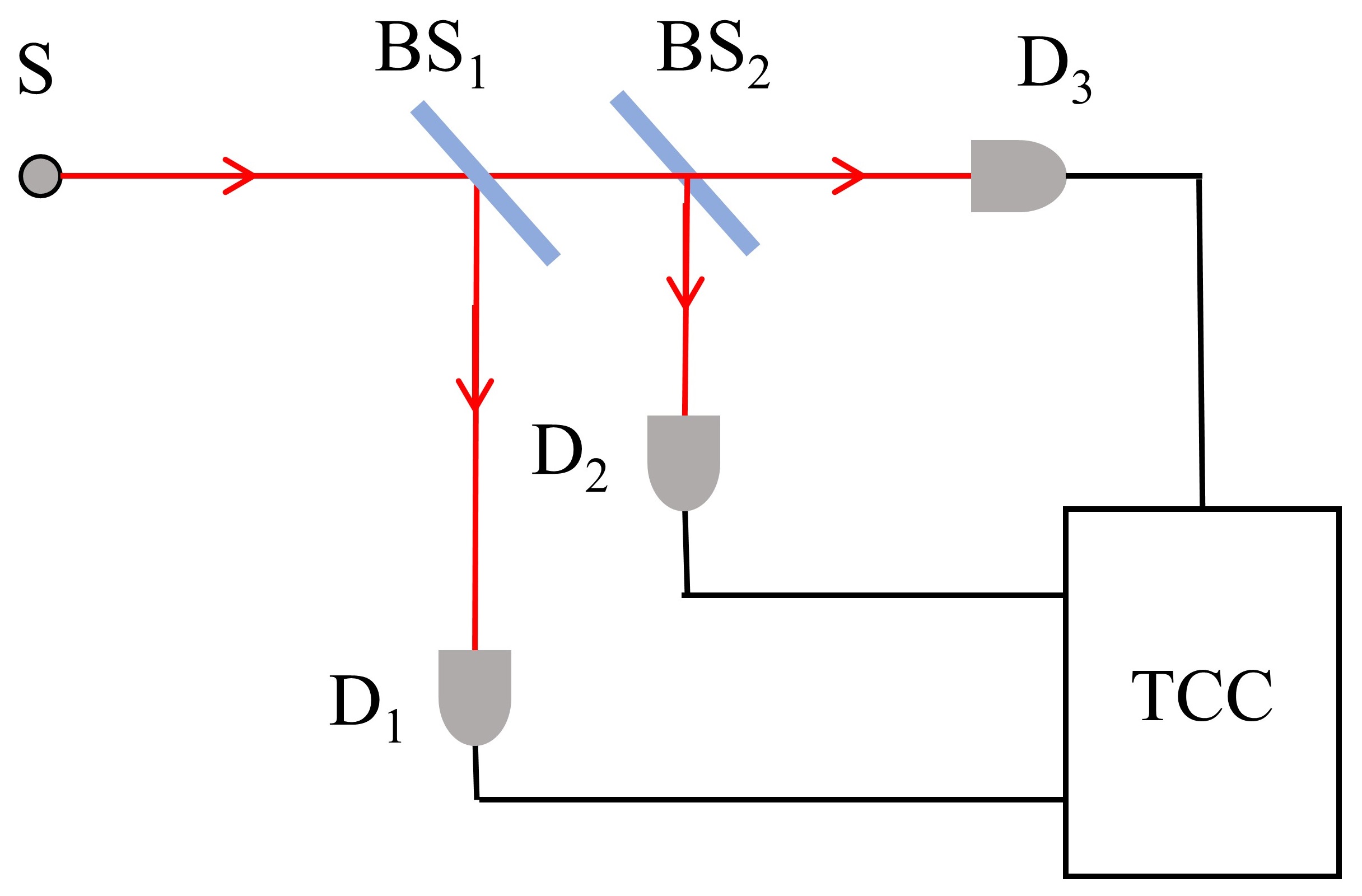}
\caption{Third-order HBT interferometer. TCC is short for three-photon coincidence count and is usually written as CC for convenience. BS$_1$ is a 1:2 non-polarizing beam splitter. The meanings of other symbols are the same as the ones in a usual HBT interferometer. }\label{27-3HBT}
\end{figure}

Figure \ref{27-3HBT} shows the scheme for the third-order HBT interferometer by analogy of the HBT interferometer shown in Fig. \ref{11-HBT-interferometer}. The light emitted by a light source, S, is split into two via a 1:2 non-polarizing beam splitter, BS$_1$. The transmitted light beam is split into two via a 1:1 non-polarizing beam splitter, BS$_2$. The intensities detected by D$_1$, D$_2$, and D$_3$ are equal. TCC is short for three-photon coincidence count detection system and usually written as CC for convenience.  The distance between the source and detector planes are equal.

\begin{figure}[htb]
\centering
\includegraphics[width=85mm]{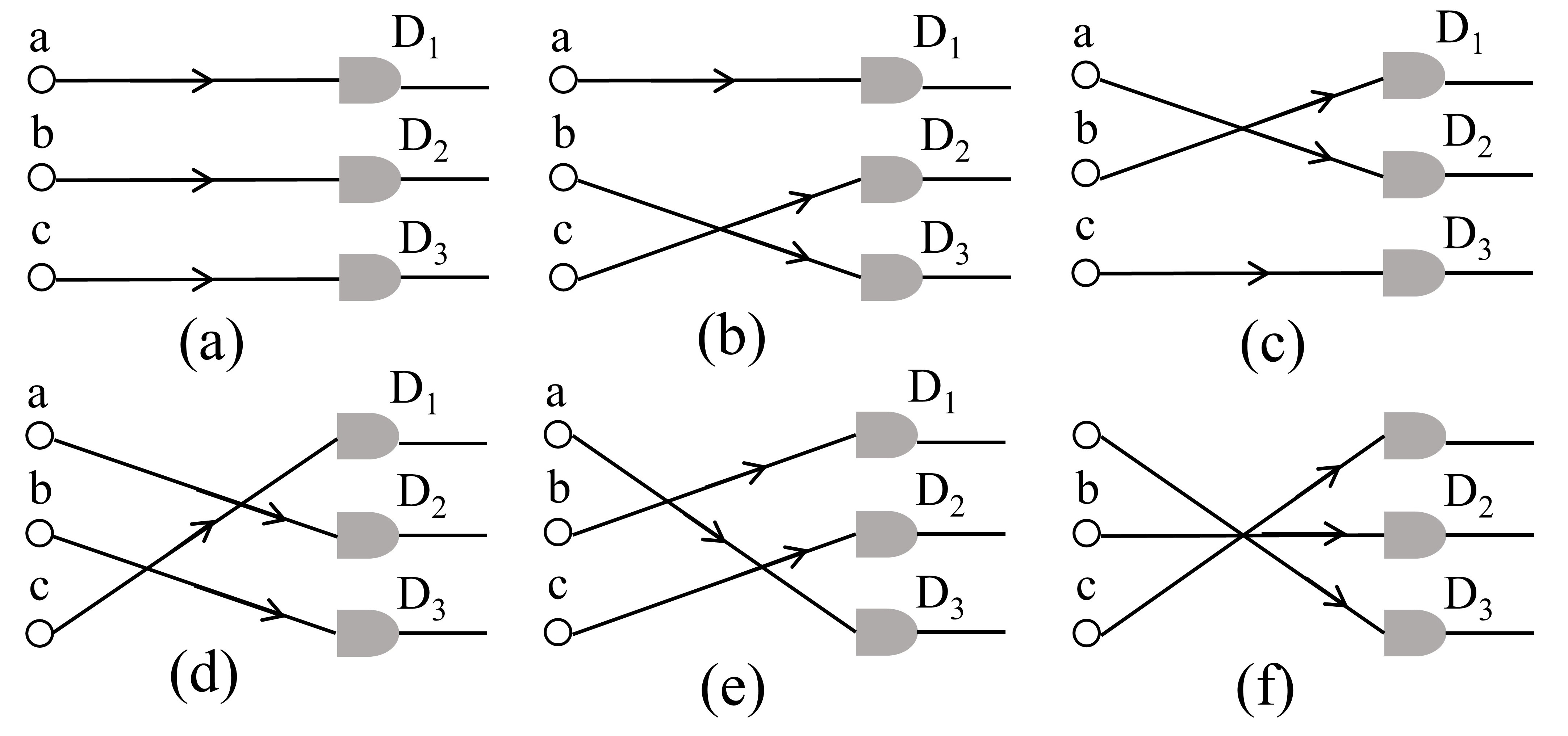}
\caption{Six different paths to trigger a three-photon coincidence count. (a) shows the path for photon a is detected by D$_1$, photon b is detected by D$_2$ and photon c is detected by D$_3$. Other five paths are defined similarly. }\label{28-six-ways}
\end{figure}

Assume S is a thermal light source in Fig. \ref{27-3HBT}. There are six different paths to trigger a three-photon coincidence count in  D$_1$, D$_2$, and D$_3$. Figure \ref{28-six-ways} shows the schematic diagram of the six different paths. Fig.  \ref{28-six-ways}(a) shows the path for that photon a is detected by D$_1$, photon b is detected by D$_2$ and photon c is detected by D$_3$. The corresponding three photon probability amplitude is $A_{a1,b2,c3}$. Other five paths are defined similarly. If all the six different paths are indistinguishable, the three-photon probability function is
\begin{eqnarray}\label{third-1}
&&P^{(3)}(\vec{r}_1,t_1;\vec{r}_2,t_2;\vec{r}_3,t_3)\nonumber\\
&=&\langle |\frac{1}{\sqrt{6}}(A_{a1,b2,c3}+A_{a1,b3,c2}+A_{a2,b1,c3}\nonumber\\
&&+A_{a2,b3,c1}+A_{a3,b1,c2}+A_{a3,b2,c1})|^2 \rangle.
\end{eqnarray}
The initial phases of photons in thermal light are random, Eq. (\ref{third-1}) can be simplified as
\begin{eqnarray}\label{third-2}
&&P^{(3)}(\vec{r}_1,t_1;\vec{r}_2,t_2;\vec{r}_3,t_3)\nonumber\\
&\propto&\langle |K_{a1}K_{b2}K_{c3} +K_{a1} K_{b3} K_{c2}+K_{a2} K_{b1} K_{c3} \nonumber\\
&&+K_{a2} K_{b3} K_{c1}+K_{a3} K_{b1} K_{c2}+K_{a3} K_{b2} K_{c1}|^2 \rangle,
\end{eqnarray}
where $K_{a1}$ is the photon\rq{}s Feynman propagator from S to D$_1$. 

\begin{figure}[htb]
\centering
\includegraphics[width=90mm]{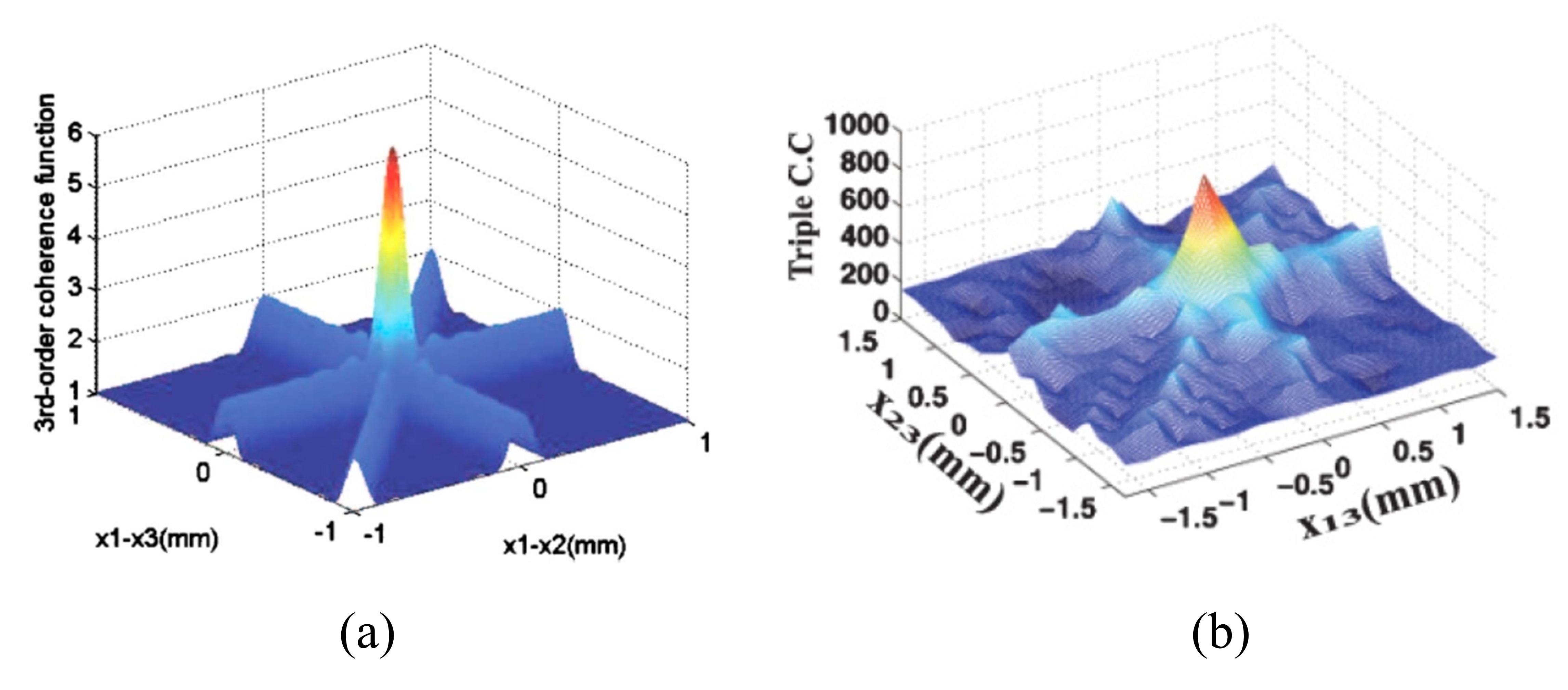}
\caption{Simulated  spatial third-order coherence function \cite{liu2009n} and measured three-photon coincidence counts \cite{zhou2010third} of pseudothermal light in a third-order HBT interferometer. $x1-x3$ in (a) and $x_{13}$ in (b) represent transverse position difference of D$_1$ and D$_3$. Similar definition holds for $x1-x2$ in (a) and $x_{23}$ in (b). Triple C.C. in (b) is three-photon coincidence counts, which is proportional to the normalized third-order coherence function in (a).}\label{29-spatial}
\end{figure}

With the same method as the one for thermal light in a HBT interferometer in Sect. \ref{second-thermal-one}, the spatial three-photon probability function is \cite{liu2009n}
\begin{eqnarray}\label{third-3}
&&P^{(3)}(x_1;x_2;x_3)\nonumber\\
&\propto& 1+\text{sinc}^2\frac{\pi d }{L \lambda}(x_1-x_2)\nonumber\\
&&+\text{sinc}^2\frac{\pi d }{L \lambda}(x_2-x_3)+\text{sinc}^2\frac{\pi d }{L \lambda}(x_3-x_1)\\
&&+2\text{sinc}\frac{\pi d }{L \lambda}(x_1-x_2)\text{sinc}\frac{\pi d }{L \lambda}(x_2-x_3)\text{sinc}\frac{\pi d }{L \lambda}(x_3-x_1),\nonumber
\end{eqnarray}
where paraxial approximation and one-dimension case were assumed, $\lambda$ is the wavelength of thermal light and single-frequency light is employed, $d$ is the length of the one-dimension thermal light source, $L$ is the distance between the source and detector planes. Figure \ref{29-spatial}(a) shows the simulated spatial third-order coherence function of thermal light in a third-order HBT interferometer. Figure \ref{29-spatial}(b) shows the measured three-photon coincidence counts vs. the transverse distances between these three detectors. The theoretical and experimental results are consistent with each other.

Temporal three-photon probability function can be calculated in the same way \cite{zhou2010third},
\begin{eqnarray}\label{third-4}
&&P^{(3)}(t_1;t_2;t_3)\nonumber\\
&\propto& 1+\text{sinc}^2\frac{\Delta \omega}{2}(t_1-t_2)\nonumber\\
&&+\text{sinc}^2\frac{\Delta \omega}{2}(t_2-t_3)+\text{sinc}^2\frac{\Delta \omega}{2}(t_3-t_1)\\&&+2\text{sinc}\frac{\Delta \omega}{2}(t_1-t_2)\text{sinc}\frac{\Delta \omega}{2}(t_2-t_3)\text{sinc}\frac{\Delta \omega}{2}(t_3-t_1),\nonumber
\end{eqnarray}
where $\Delta \omega$ is the frequency bandwidth of thermal light and point thermal light source is assumed.

The degree of third order coherence, $g^{(3)}(0)$, of thermal light equals 6, which means that the probability of detecting three photons when these three detectors are in the same space-time coordinate is 6 times of the one when these three detectors are in different coherence volumes. It is defined as three-photon bunching of thermal light. The degree of $N$th-oder coherence of thermal light equals $N!$  \cite{liu2009n}. Three-photon superbunching can be observed with superbunching pseudothermal light \cite{zhou2019experimental}. 

The third-order interference of single-mode continuous-wave laser light in a third-order HBT interferometer is simple. All the six different paths shown in Fig. \ref{28-six-ways} to trigger a three-photon coincidence count are identical. The superposition of these six probability amplitudes will not give three-photon interference. The reason is the same as the one for there is no two-photon bunching for single-mode continuous-wave laser light in a HBT interferometer. No three-photon bunching for single-mode continuous-wave laser light exists in the third-order HBT interferometer.

\subsection{Third-order interference of two independent light beams}

\begin{figure}[htb]
\centering
\includegraphics[width=50mm]{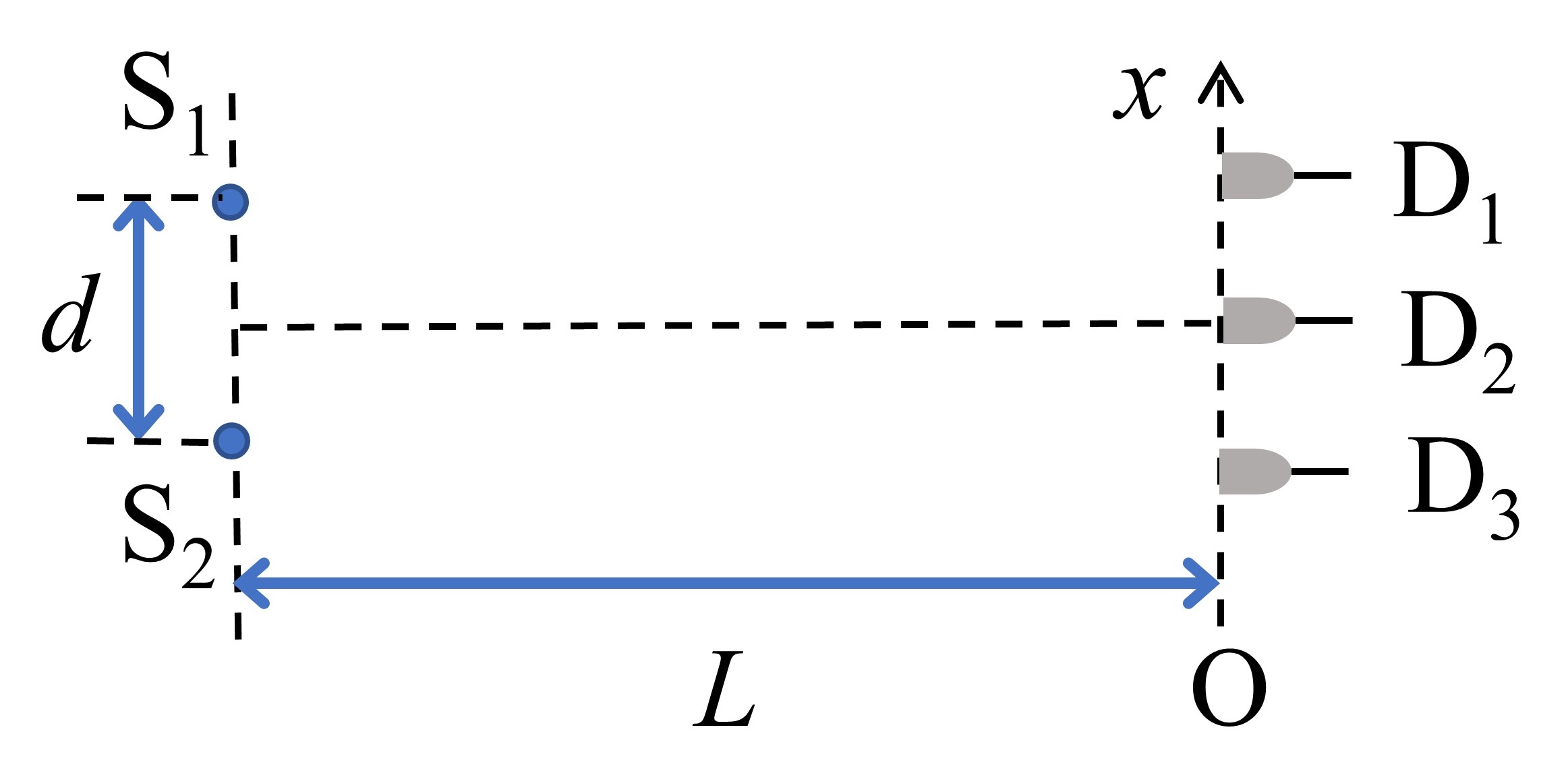}
\caption{Scheme for the third-order interference of two independent light beams. S$_1$ and  S$_2$ are two light sources. D$_1$, D$_2$ and D$_3$ are three single-photon detectors. The output signals of D$_1$, D$_2$ and D$_3$ are sent into a three-photon coincidence count detection system, which is not shown.}\label{30-two-sources}
\end{figure}

Figure \ref{30-two-sources} shows the scheme for the third-order interference of two independent light beams. Assume S$_1$ is a single-photon source and S$_2$ is a single-mode continuous-wave laser. There are two different ways to trigger a three-photon coincidence count in Fig. \ref{30-two-sources}. The first way is that one photon is emitted by S$_1$ and two photons are emitted by S$_2$. The probability is $3I_1I_2^2/(3I_1I_2^2+I_2^3)$, where $I_1$ and $I_2$ are the intensities of light beams emitted by S$_1$ and S$_2$, respectively. There are three different paths and the corresponding three-photon probability amplitudes are $A_{11,22,23}$, $A_{12,21,23}$, and $A_{13,21,22}$, respectively. The second way is these three photons are all emitted by S$_2$. The probability is $I_2^3/(I_1I_2^2+I_2^3)$. There is one path and the corresponding three-photon probability amplitude is $A_{21,22,23}$.

If these four different paths are indistinguishable, the three-photon probability function is
\begin{eqnarray}\label{third-two-1}
&&P^{(3)}(\vec{r}_1,t_1;\vec{r}_2,t_2;\vec{r}_3,t_3)\nonumber\\
&=& \langle |\sqrt{\frac{3I_1I_2^2}{3I_1I_2^2+I_2^3}}\frac{1}{\sqrt{3}}(A_{11,22,23}+A_{12,21,23}+A_{13,21,22})\nonumber\\
&&+\sqrt{\frac{I_2^3}{3I_1I_2^2+I_2^3}}A_{21,22,23}|^2 \rangle.
\end{eqnarray}
The initial phases of photons emitted by the single-photon source, S$_1$, are random. The initial phases of photons emitted by the single-mode continuous-wave laser, S$_2$, are identical within one coherence volume. Equation (\ref{third-two-1}) can be simplified as
\begin{eqnarray}\label{third-two-2}
&&P^{(3)}(\vec{r}_1,t_1;\vec{r}_2,t_2;\vec{r}_3,t_3)\nonumber\\
&=& \frac{3I_1}{3I_1+I_2} \langle |K_{11}K_{22}K_{23}+K_{12}K_{21}K_{23}+K_{13}K_{21}K_{22}|^2 \rangle\nonumber\\
&& + \frac{I_2}{3I_1+I_2}\langle |K_{21}K_{22}K_{23}|^2 \rangle.
\end{eqnarray}
Assume S$_1$ and S$_2$ are point light sources. With the same method and approximations above, one-dimension spatial three-photon probability function is
\begin{eqnarray}\label{third-two-3}
&&P^{(3)}(x_1;x_2;x_3)\nonumber\\
&\propto& 3I_1^2[3+  2\cos \frac{2\pi d}{L \lambda}(x_1-x_2)+ 2\cos \frac{2\pi d}{L \lambda}(x_2-x_3)\nonumber\\
&&+ 2\cos \frac{2\pi d}{L \lambda}(x_1-x_3)]+I_2^2.
\end{eqnarray}
There is third-order interference pattern. The visibility of the third-order interference pattern is dependent on the ratio between the intensities of light emitted by the single-photon source and single-mode continuous-wave laser.

Similar method can be employed to calculate the third-order interference of two other types of light beams.

\subsection{Third-order interference of multiple independent light beams}

\begin{figure}[htb]
\centering
\includegraphics[width=50mm]{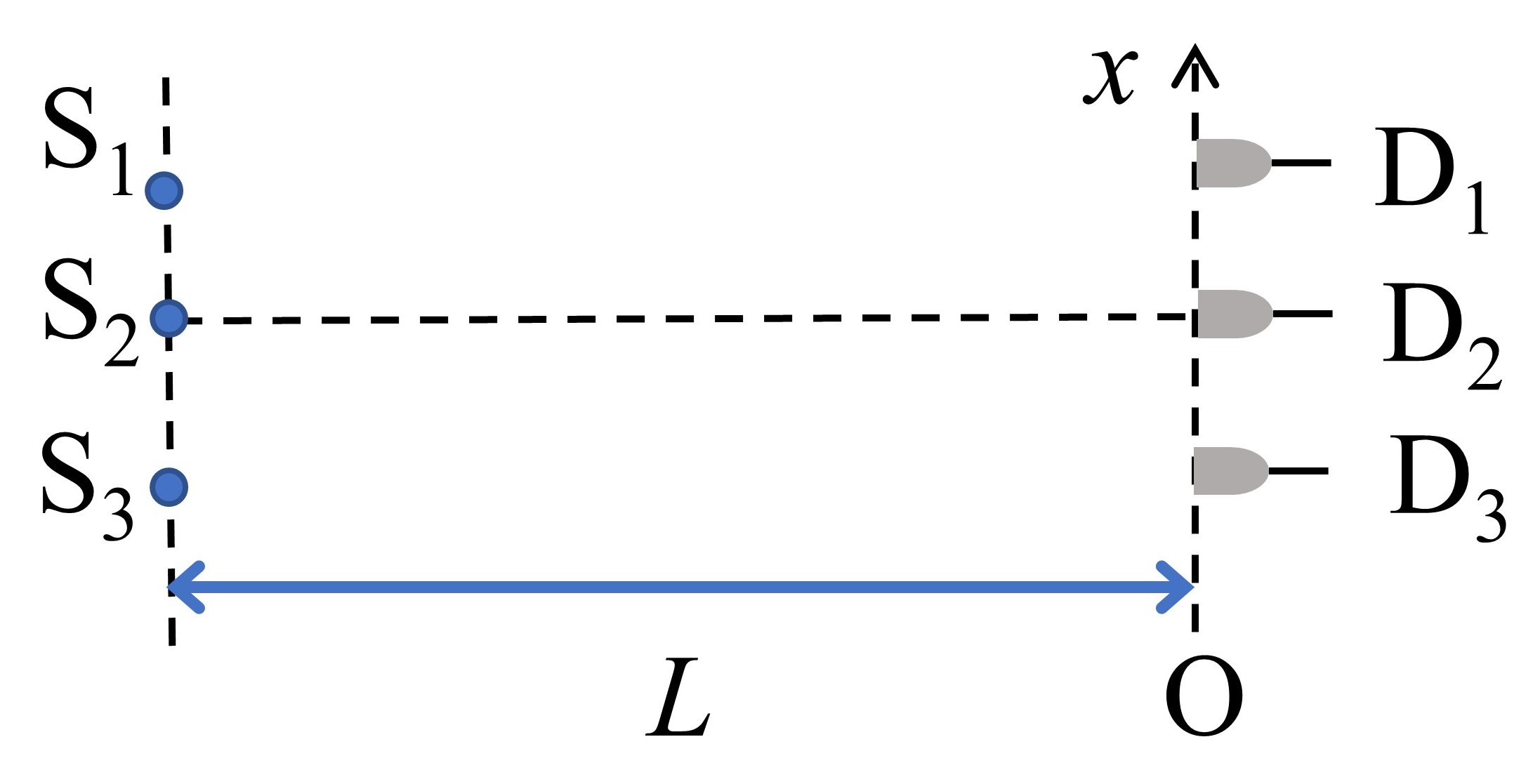}
\caption{Scheme for the third-order interference of three independent light beams. The meanings of the symbols are similar as the ones in Fig. \ref{30-two-sources}.}\label{31-three-sources}
\end{figure}

Figure \ref{31-three-sources} shows the scheme for the third-order interference of three independent light beams. Assume S$_1$,  S$_2$, and S$_3$ are three identical and independent single-photon sources. The intensities, polarizations, and frequency spectra of the photons emitted by these three sources are the same. There are six different paths to trigger a three-photon coincidence count. The corresponding three-photon probability amplitudes are $A_{11,22,33}$, $A_{11,23,31}$, $A_{12,21,33}$, $A_{12,23,31}$, $A_{13,22,31}$, and $A_{13,21,32}$, where $A_{11,22,33}$ is the probability amplitude for that the photon emitted by S$_1$ is detected by D$_1$, the photon emitted by S$_2$ is detected by D$_2$, and the photon emitted by S$_3$ is detected by D$_3$. The meanings of other symbols are defined similarly.

If all the six different paths are indistinguishable, three-photon probability function is
\begin{eqnarray}\label{third-three-1}
&&P^{(3)}(\vec{r}_1,t_1;\vec{r}_2,t_2;\vec{r}_3,t_3)\nonumber\\
&=& \langle|\frac{1}{\sqrt{6}}(A_{11,22,33}+A_{11,23,31}+A_{12,21,33}\nonumber\\
&&+A_{12,23,31}+A_{13,22,31}+A_{13,21,32})|^2\rangle\nonumber\\
&\propto& \langle |e^{i(\varphi_1+\varphi_2+\varphi_3)}(K_{11}K_{22}K_{33}+K_{11}K_{23}K_{31}\nonumber\\
&&+K_{12}K_{21}K_{33}+K_{12}K_{23}K_{31}\nonumber\\
&&+K_{13}K_{22}K_{31}+K_{13}K_{21}K_{32})|^2\rangle,
\end{eqnarray}
where $\varphi_1$, $\varphi_2$, and $\varphi_3$ are the initial phases of photons emitted by S$_1$,  S$_2$, and S$_3$, respectively. 

\begin{figure}[htb]
\centering
\includegraphics[width=60mm]{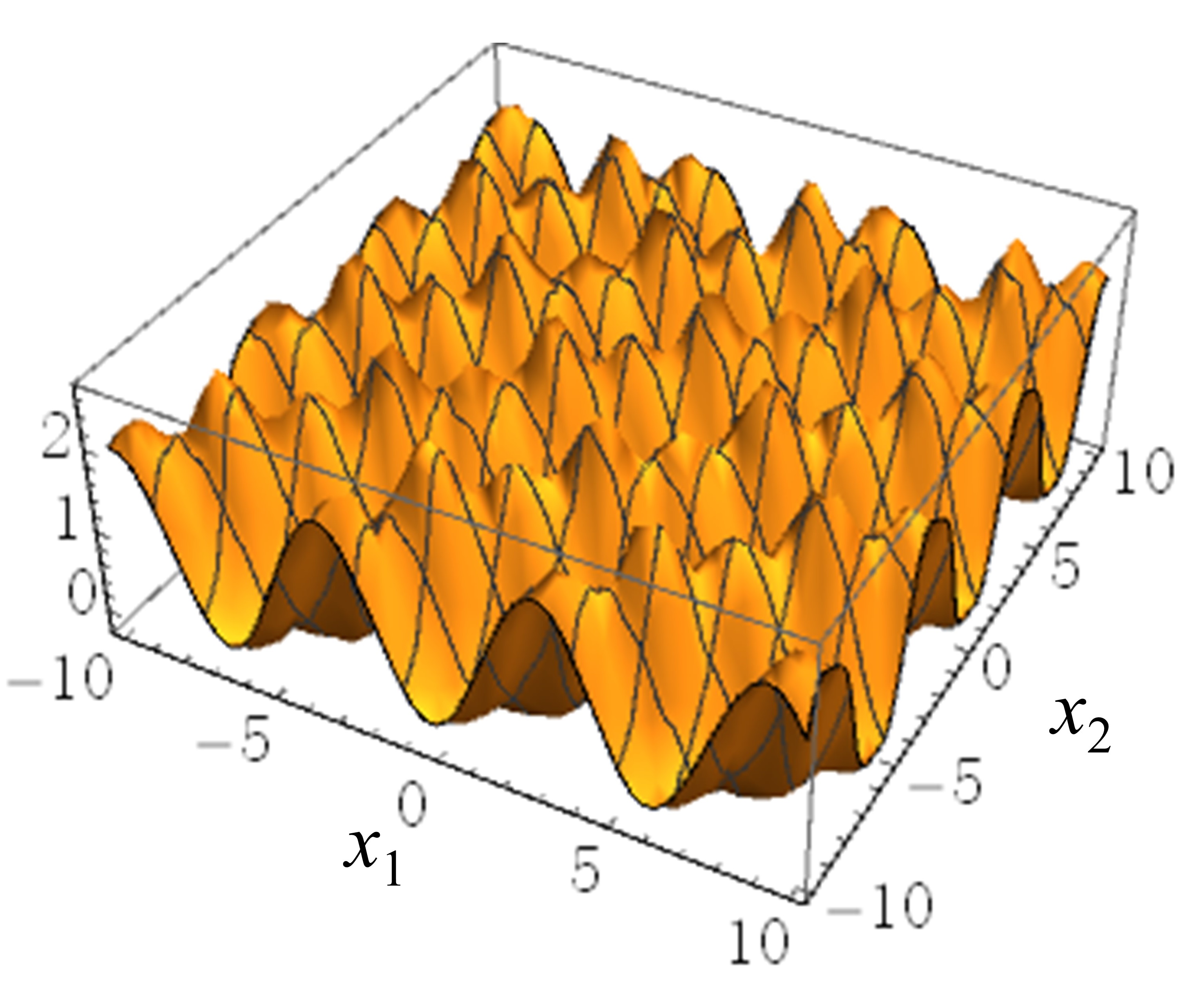}
\caption{Third-order interference pattern of three independent single-photon states. $x_1$ and $x_2$ are the transverse positions of D$_1$ and D$_2$, respectively.}\label{32-three-pattern-light}
\end{figure}

With the same method and approximations above, one-dimension spatial three-photon probability function can be simplified as
\begin{eqnarray}\label{third-three-2}
&&P^{(3)}(x_1;x_2;x_3)\nonumber\\
&=& 6+6\cos \frac{2\pi d_{12}}{L \lambda}(x_1-x_2)\\
&+&6 \cos \frac{2\pi d_{13}}{L \lambda}(x_1-x_3)+6 \cos \frac{2\pi d_{23}}{L \lambda}(x_2-x_3)\nonumber\\
&+&2\cos \frac{2\pi d_{13}}{L \lambda}x_1 \cos \frac{2\pi d_{12}}{L \lambda}(x_2-\frac{d_{23}}{2})\cos \frac{2\pi d_{23}}{L \lambda}(x_3+\frac{d_{12}}{2})\nonumber\\
&+&2\cos \frac{2\pi d_{13}}{L \lambda}x_1 \cos \frac{2\pi d_{12}}{L \lambda}(x_3-\frac{d_{23}}{2})\cos \frac{2\pi d_{23}}{L \lambda}(x_2+\frac{d_{12}}{2})\nonumber\\
&+&2\cos \frac{2\pi d_{13}}{L \lambda}x_2 \cos \frac{2\pi d_{12}}{L \lambda}(x_1-\frac{d_{23}}{2})\cos \frac{2\pi d_{23}}{L \lambda}(x_3+\frac{d_{12}}{2})\nonumber\\
&+&2\cos \frac{2\pi d_{13}}{L \lambda}x_2 \cos \frac{2\pi d_{12}}{L \lambda}(x_3-\frac{d_{23}}{2})\cos \frac{2\pi d_{23}}{L \lambda}(x_1+\frac{d_{12}}{2})\nonumber\\
&+&2\cos \frac{2\pi d_{13}}{L \lambda}x_3 \cos \frac{2\pi d_{12}}{L \lambda}(x_1-\frac{d_{23}}{2}) \cos \frac{2\pi d_{23}}{L \lambda}(x_2+\frac{d_{12}}{2})\nonumber\\
&+&2\cos \frac{2\pi d_{13}}{L \lambda}x_3 \cos \frac{2\pi d_{12}}{L \lambda}(x_2-\frac{d_{23}}{2}) \cos \frac{2\pi d_{23}}{L \lambda}(x_1+\frac{d_{12}}{2}).\nonumber
\end{eqnarray}
In order to see the third-order interference pattern more clearly, assuming $d_{12}=d_{23}=d$ and $x_2=x_3$, Eq. (\ref{third-three-2}) can be simplified as
\begin{eqnarray}\label{third-three-3}
&&P^{(3)}(x_1;x_2;x_3=x_2)\nonumber\\
&\propto& 1+\cos \frac{2\pi d}{L \lambda}(x_1-x_2)\\
&&+\frac{1}{3}\cos \frac{4\pi d}{L \lambda}x_1 \cos \frac{2\pi d}{L \lambda}(x_2-\frac{d}{2})\cos \frac{2\pi d}{L \lambda}(x_2+\frac{d}{2})\nonumber\\
&&+\frac{1}{3}\cos \frac{4\pi d}{L \lambda}x_2 \cos \frac{2\pi d}{L \lambda}(x_1-\frac{d}{2})\cos \frac{2\pi d}{L \lambda}(x_2+\frac{d}{2})\nonumber\\
&&+\frac{1}{3}\cos \frac{4\pi d}{L \lambda}x_2 \cos \frac{2\pi d}{L \lambda}(x_2-\frac{d}{2})\cos \frac{2\pi d}{L \lambda}(x_1+\frac{d}{2}).\nonumber
\end{eqnarray}
Figure \ref{32-three-pattern-light} shows the third-order interference pattern given by Eq. (\ref{third-three-3}). The periodic structures can be observed in $x_1$, $x_2$, and $x_1-x_2$ directions.

Similar method can be employed to calculate the third-order interference other types of light beams and higher-order interference of light. In fact, all optical coherence problems can be calculated in the same method introduced above. 

\section{Quantum atomic coherence theory based on Feynman\rq{}s path integral}\label{sec-atom}

It is straightforward to generalize quantum optical coherence theory based on Feynman\rq{}s path integral to discuss the coherence of massive particles. The history of atomic coherence traces back to de Broglie in 1924 when he proposed matter wave assumption \cite{de1924recherches} by analogy of Planck\rq{}s \cite{planck1901law} and Einstein\rq{}s \cite{einstein1905heuristic} quanta assumption of light. The de Broglie wavelength of matter wave equals
\begin{equation}\label{mass-1}
\lambda_D=h/p.
\end{equation}
De Broglie\rq{}s assumption was soon verified by Davisson \textit{et al.} and Thomson \textit{et al.} with diffraction pattern of electrons \cite{davisson1927diffraction,thomson1927diffraction}. Due to the very short de Broglie wavelength of matter wave, it was difficult to observe the first-order interference pattern like the one of light then. In 1953, Marton \textit{et al.} observed first-order interference pattern of electrons in a Mach-Zehnder interferometer \cite{marton1953electron}. Then the first-order interference pattern of neutron \cite{maier1962interferometer}, atom \cite{keith1991interferometer}, molecule \cite{arndt1999wave} were also reported. The first-order interference pattern of two independent Bose Einstein Condensations (BEC) was observed \cite{andrews1997observation}.

The first experiment of the second-order interference with matter wave was done by Yasuda and Shimizu, in which they observed two-atom bunching with cold atoms \cite{yasuda1996observation}. Burt \textit{et al.} found that the ratio between the degrees of third-order coherence of cold atoms and BEC equals $7.4\pm 2.6$ \cite{burt1997coherence}. The second-order interference of fermions was reported by several groups \cite{henny1999fermionic,oliver1999hanbury,iannuzzi2006direct,rom2006free,bocquillon2013coherence}.

 Quantum atomic coherence theory is mainly based on Glauber\rq{}s quantum optical coherence theory \cite{adams1994atom,meystre2001atom,cronin2009optics}. Feynman himself introduced how to employ Fenman\rq{}s path integral to calculate the first-order interference of electrons in a Young\rq{}s double-slit interferometer \cite{feynman2010quantum,feynman2011feynman}. Storey and Cohen-Tannoudji provided an introduction of how to employ Feynman\rq{}s path integral to calculate the first-order interference of atoms \cite{storey1994feynman}. Employing Feynman\rq{}s path integral to calculate the first-order interference of matter wave can be found in Refs. \cite{deng2006theory,deng2008theory,gondran2005numerical}, too.

This section will focus on how to employ Feynman\rq{}s path integral to calculate the first-, second, and higher-order interference of matter waves, with special attentions on the analogy between cold atoms (BEC) and thermal light (laser light).

\subsection{First-order interference of matter wave}

\subsubsection{First-order interference of one matter wave beam}

\begin{figure}[htb]
\centering
\includegraphics[width=40mm]{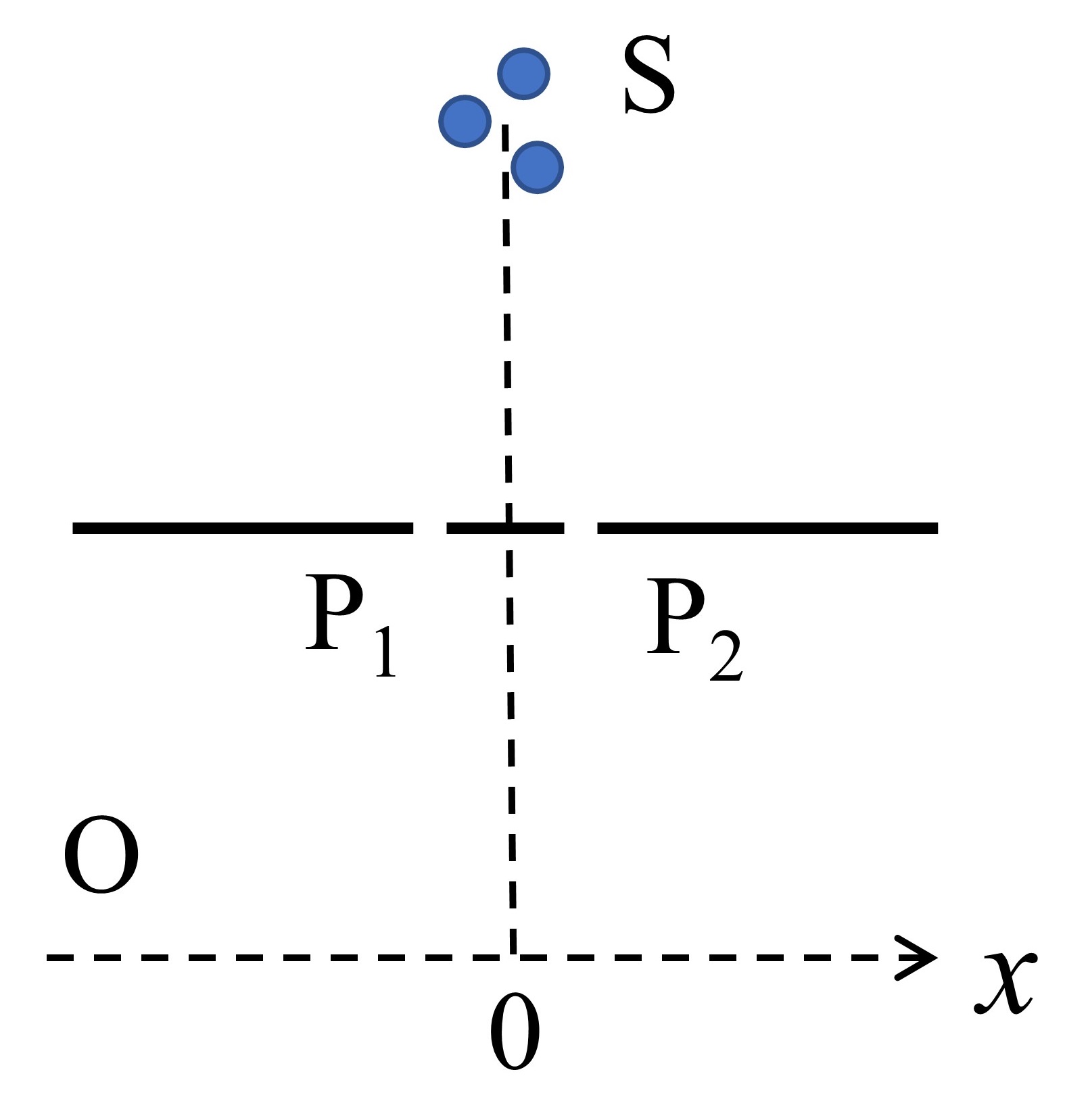}
\caption{Scheme for the first-order interference of one matter wave beam. S is a particle source. P$_1$ and P$_2$ are two pinholes. O is the observation plane.}\label{33-atom-interferometer}
\end{figure}

Figure \ref{33-atom-interferometer} shows the scheme for the first-order interference of one matter wave beam. S is a particle source. P$_1$ and P$_2$ are two pinholes with equal size. O is the observation plane. For simplicity, free atoms are assumed here and in the following discussions.

There are two different paths to trigger a particle detection event in the observation plane. The first path is the particle going through P$_1$. The probability is 1/2 and the probability amplitude is $A_{S1x}$. The second path is the particle going through P$_2$. The probability is 1/2 and the probability amplitude is $A_{S2x}$. If these two different paths are indistinguishable, the probability  function of particles in the observation plane is
\begin{eqnarray}\label{atom-1}
P^{(1)}(x,t)&=&\langle |\frac{1}{\sqrt{2}}(A_{S1x}+A_{S2x})|^2 \rangle \nonumber\\
&\propto& \langle |e^{i\varphi_j}K_{S1x}+e^{i\varphi_j}K_{S2x}|^2 \rangle \nonumber\\
&=& \langle |K_{S1}K_{1x}+K_{S2}K_{2x}|^2 \rangle,
\end{eqnarray}
where $\varphi_j$ is the initial phase of the $j$th detected particle, $K_{S1x}$ is the particle\rq{}s Feynman propagator from S to $x$ via P$_1$, the meanings of other symbols are defined similarly. 

Massive particle\rq{}s Feynman propagator can be calculated in the same method as the one for photon and has a simple form for free particles  \cite{feynman2010quantum}
\begin{equation}\label{atom-2}
K(x_a,t_a;x_b,t_b)=\sqrt{\frac{m}{2\pi i \hbar (t_b-t_a)}}\exp\frac{im(x_b-x_a)^2}{2\hbar(t_b-t_a)},
\end{equation}
where $m$ is the mass of particle, $\hbar=h/(2\pi)$, $(x_a,t_a)$ and $(x_b,t_b)$ are the starting and ending space-time coordinates of the particle, respectively. For simplicity, one-dimension case is assumed.

Assuming S is at the symmetrical position of P$_1$ and P$_2$, there is no path difference from S to two pinholes. The path difference is introduced by the distances from the pinholes to the observation plane. Substituting Eq. (\ref{atom-2}) into Eq. (\ref{atom-1}), it is straightforward to have one-dimension spatial first-order interference pattern in the Fraunhofer range,
\begin{eqnarray}\label{atom-3}
P^{(1)}(x)\propto 1+ \cos\frac{2\pi d}{L \lambda_D}x,
\end{eqnarray}
in which paraxial approximation has been employed, $\lambda_D=h/(m v)$ is the de Broglie wavelength of the particle, $v$ is the speed of the particle and the speed of all the particles are assumed to be the same, $d$ is the distance between these two pinholes, and $L$ is the distance between the pinhole and observation planes. 

There is first-order interference pattern when these two different paths are indistinguishable. The patterns are the same for fermions and bosons. Just like there is no difference for the first-order interference patterns of thermal and laser light in a Young\rq{}s double-slit interferometer, the first-order interference patterns of bosons in thermal state or BEC state in a Young\rq{}s double-slit interferometer are the same. The initial phases of the particles do not influence the first-order interference pattern of one beam.

\subsubsection{First-order interference of two independent matter wave beams}

\begin{figure}[htb]
\centering
\includegraphics[width=40mm]{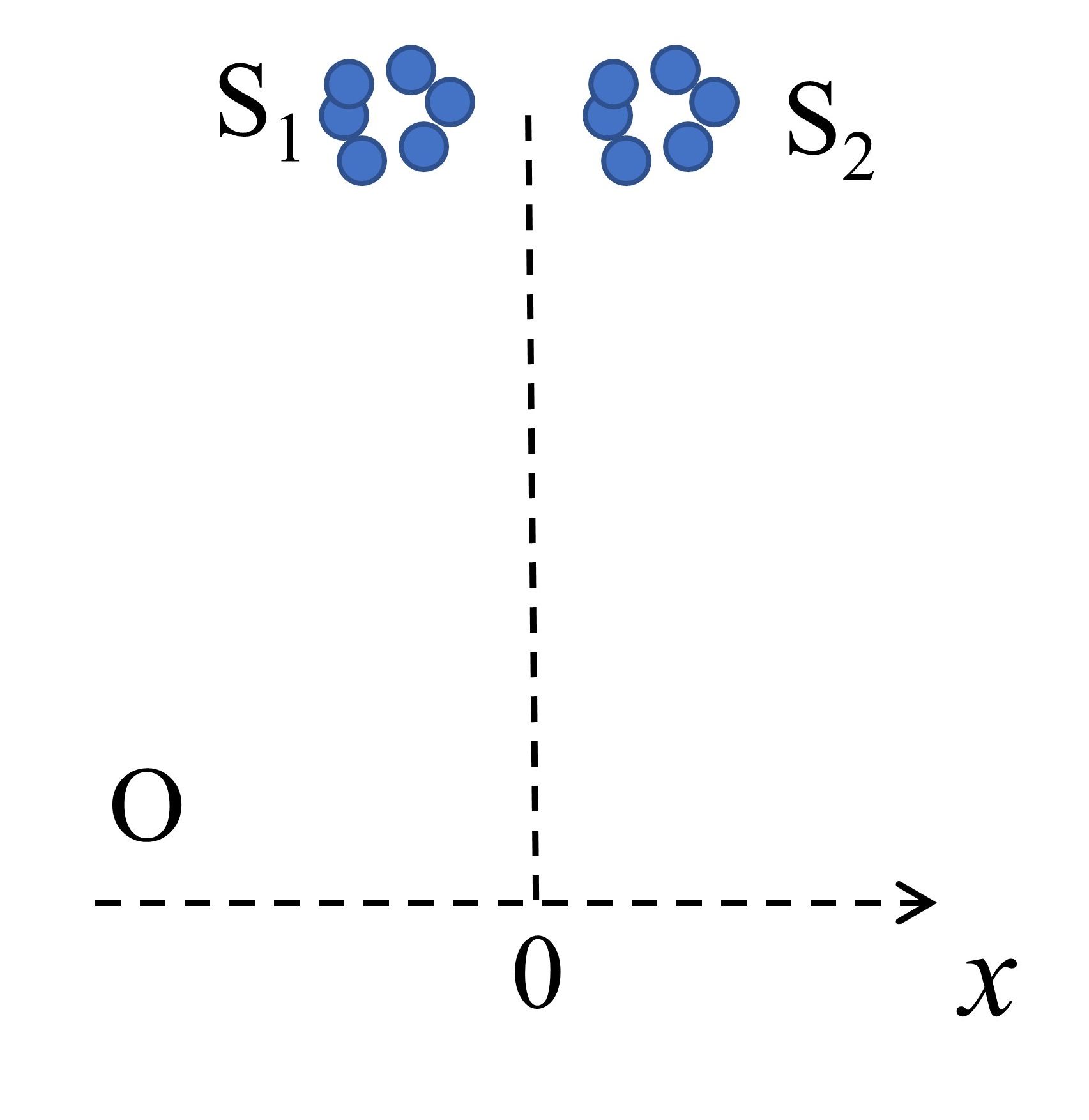}
\caption{Scheme for the first-order interference of two independent matter wave beams. S$_1$ and S$_2$ are two particle sources.}\label{34-two-atoms}
\end{figure}

Figure \ref{34-two-atoms} shows the scheme for the first-order interference of two independent matter wave beams. S$_1$ and S$_2$ are two independent particle sources. O is the observation plane.

There are two different paths to trigger a particle detection event in the observation plane. The first one is that the detected particle is emitted S$_1$. The probability is 1/2 and the probability amplitude is $A_{1x}$.  The second one is that the detected particle is emitted S$_2$. The probability is 1/2 and the probability amplitude is $A_{2x}$. If these two different paths are indistinguishable, the probability function of detecting a particle in the observation plane is
\begin{eqnarray}\label{atom-4}
P^{(1)}(x,t)&=& \langle |\frac{1}{\sqrt{2}}(A_{1x}+A_{2x})|^2 \rangle \nonumber\\
&\propto& \langle |e^{i\varphi_{1j}}K_{1x}+e^{i\varphi_{2j}}K_{2x}|^2 \rangle,
\end{eqnarray}
where $\varphi_{1j}$ and $\varphi_{2j}$ are the initial phases of particles emitted by S$_1$ and S$_2$, respectively. With the same approximations above, the spatial probability function can be obtained via Eq. (\ref{atom-4}),
\begin{eqnarray}\label{atom-5}
P^{(1)}(x)&=& \langle 1+ \cos[\frac{2\pi d}{L \lambda_D}x+(\varphi_{1j}-\varphi_{2j})] \rangle.
\end{eqnarray}

The interference pattern is depend on the initial phases of particles. Comparing the experimental results of the degrees of second- and third-order coherence of cold atoms and BEC to the ones of thermal and laser light \cite{yasuda1996observation,burt1997coherence,lopes2015atomic}, it is reasonable to assume that the coherence properties of cold atoms are similar as the ones of thermal light and the coherence properties of BEC  are similar as the ones of laser light. The initial phases of particles in cold atoms are random. The initial phases of particles in one BEC are identical. 

For the superposition of two independent cold atom sources, the initial phases of particles are random. Equation (\ref{atom-5}) is simplified as
\begin{eqnarray}\label{atom-6}
P^{(1)}(x)&\propto& 1+\frac{1}{\sqrt{N}}\cos[\frac{2\pi d}{L \lambda_D}x+\varphi] ,
\end{eqnarray}
where $N$ is the number of detected particles and $\varphi$ is a random phase due to the sum of $\varphi_{1j}-\varphi_{2j}$ for $N$ detected particles. The first-order interference pattern of two independent cold atom sources can not be observed.

For the superposition of of two independent BECs, the initial phases of particles in one BEC are identical. Equation (\ref{atom-5}) is simplified as
\begin{eqnarray}\label{atom-7}
P^{(1)}(x)&\propto &  1+ \cos[\frac{2\pi d}{L \lambda_D}x+(\varphi_{1j}-\varphi_{2j})].
\end{eqnarray}
$\varphi_{1j}-\varphi_{2j}$ is a random but fixed number for two independent BECs. The first-order interference pattern can be observed \cite{andrews1997observation}, just like there is transient first-order interference pattern by superposing two independent laser light beams \cite{magyar1963interference}.

Similar method can be employed to calculate the first-order interference of multiple independent matter wave beams.

\subsection{Second-order interference of matter wave}

\subsubsection{Second-order interference of one matter wave beam}

The scheme for the second-order interference of one matter wave beam is similar as the one shown in Fig. \ref{11-HBT-interferometer} except that the source, S, is a particle source, BS is a beam splitter for particles and these two detectors are particle detectors.

Let us first discuss the second-order interference of one boson beam. Assuming S is a cold atom source, there are two different paths to trigger a two-particle coincidence count. The first path is that particle a is detected by D$_1$ and particle b is detected by D$_2$. The probability is 1/2 and the corresponding two-particle amplitude is $A_{a1,b2}$. The second path is that particle a is detected by D$_2$ and particle b is detected by D$_1$. The probability is 1/2 and the corresponding two-particle amplitude is $A_{a2,b1}$. If these two different paths are indistinguishable, the two-particle probability function is
\begin{eqnarray}\label{two-atom-1}
&&P^{(2)}(\vec{r}_1,t_1;\vec{r}_2,t_2)\nonumber\\
&=& \langle |\frac{1}{\sqrt{2}}(A_{a1,b2}+A_{a2,b1})|^2\rangle\nonumber\\
&\propto& \langle |e^{i\varphi_a}K_{a1}e^{i\varphi_b}K_{b2}+e^{i\varphi_a}K_{a2}e^{i\varphi_b}K_{b1}|^2\rangle\nonumber\\
&=& \langle |K_{a1}K_{b2}+K_{a2}K_{b1}|^2 \rangle.
\end{eqnarray}
Substituting particle\rq{}s Feynman propagator into Eq. (\ref{two-atom-1}), the spatial two-particle probability function is
\begin{eqnarray}\label{two-atom-2}
P^{(2)}(x_1-x_2)\propto 1+ \text{sinc}^2 \frac{\pi D}{L \lambda_D}(x_1-x_2),
\end{eqnarray}
where $D$ is the length of S and single-speed particles are assumed. 

The temporal two-particle probability function is 
\begin{eqnarray}\label{two-atom-3}
P^{(2)}(t_1-t_2)\propto 1+ \text{sinc}^2 \frac{\Delta \omega}{2}(t_1-t_2),
\end{eqnarray}
where $\Delta \omega=2mv_0\Delta v/\hbar$ is the angular frequency bandwidth, $v_0$ is the central speed, $\Delta v$ is the speed bandwidth. Point source is assumed to simplify the calculation.

When these two detectors are at the same space-time coordinate, the probability to detect two particles is two times of the one when these two detectors are in different coherence volumes. There is two-particle bunching for cold boson particles in a HBT interferometer \cite{yasuda1996observation}, just like there is two-photon bunching for thermal light in a HBT interferometer.

Assuming S is a BEC in Fig. \ref{11-HBT-interferometer}, there is only one path to trigger a two-particle coincidence count. The initial phases of the particles in a BEC are identical, just like the one for single-mode continuous-wave laser light. No two-particle bunching can be observed, which is consistent with Burt \textit{et al.}\rq{}s experiments \cite{burt1997coherence}.

Then discuss the second-order interference of one fermion beam. There are also two different paths to trigger a two-particle coincidence count and the corresponding two-particle probability amplitudes are $A_{a1,b2}$ and $A_{a2,b1}$. If these two different ways are indistinguishable, the two-particle probability function is
\begin{eqnarray}\label{two-atom-4}
&&P^{(2)}(\vec{r}_1,t_1;\vec{r}_2,t_2)\nonumber\\
&=& \langle |\frac{1}{\sqrt{2}}(A_{a1,b2}-A_{a2,b1})|^2\rangle\nonumber\\
&\propto& \langle |e^{i\varphi_a}K_{a1}e^{i\varphi_b}K_{b2}-e^{i\varphi_a}K_{a2}e^{i\varphi_b}K_{b1}|^2\rangle\nonumber\\
&=& \langle |K_{a1}K_{b2}-K_{a2}K_{b1}|^2 \rangle,
\end{eqnarray}
where the minus sign is due to the exchange antisymmetry of fermions \cite{griffiths2018introduction}. By analogy of cold boson particles in a HBT interferometer, one-dimension spatial two-particle probability function for fermions in a HBT interferometer is
\begin{eqnarray}\label{two-atom-5}
P^{(2)}(x_1-x_2)\propto 1- \text{sinc}^2 \frac{\pi D}{L \lambda_D}(x_1-x_2),
\end{eqnarray}
where $D$ is the length of the source and single-speed particles are assumed. 

The temporal two-particle probability function of fermions in a HBT interferometer is
\begin{eqnarray}\label{two-atom-6}
P^{(2)}(t_1-t_2)\propto 1- \text{sinc}^2 \frac{\Delta \omega}{2}(t_1-t_2),
\end{eqnarray}
where the meanings of symbols are similar as the ones in Eq. (\ref{two-atom-3}).

When these two detectors are at the same space-time coordinates, the probability of detecting two fermions equals 0. As the distance between these two detectors increases, the probability of detecting two fermions increases to 1. There is two-particle antibunching effect for fermions in a HBT interferometer \cite{henny1999fermionic,jeltes2007comparison}, which is in contrast to the two-particle bunching effect for bosons in a HBT interferometer \cite{jeltes2007comparison,hanbury1956test,brown1956correlation}.

\subsubsection{Second-order interference of two independent matter wave beams}

Figure \ref{18-HOM-interferomter} shows the scheme for the second-order interference of two independent matter wave beams. S$_1$ and S$_2$ are two independent particle sources. BS is a 1:1 particle beam splitter.  D$_1$ and D$_2$ are two single-particle detectors. CC is two-particle coincidence count detection system.

First discuss the second-order interference of two single-boson sources. Assuming S$_1$ and S$_2$ are two identical and independent single-particle sources, there are two different paths to trigger a two-particle coincidence count. The first one is that the particle emitted by S$_1$ is detected by D$_1$ and the particle emitted by S$_2$ is detected by D$_2$. The probability is 1/2 and the two-particle probability amplitude is $A_{11,22}$. The second path is that the particle emitted by S$_1$ is detected by D$_2$ and the particle emitted by S$_2$ is detected by D$_1$. The probability is 1/2 and the two-particle probability amplitude is $A_{12,21}$. If these two different paths are indistinguishable, the two-particle probability function for bosons is
\begin{eqnarray}\label{hom-1}
&&P^{(2)}(\vec{r}_1,t_1;\vec{r}_2,t_2)\nonumber\\
&=& \langle \frac{1}{\sqrt{2}}(|A_{11,22}+A_{12,21})|^2\rangle\nonumber\\
&\propto& \langle |e^{i\varphi_1}K_{11}e^{i\varphi_2}K_{22}+e^{i(\varphi_1+\frac{\pi}{2})}K_{12}e^{i(\varphi_2+\frac{\pi}{2})}K_{21}|^2\rangle\nonumber\\
&=& \langle |K_{11}K_{22}-K_{12}K_{21}|^2 \rangle,
\end{eqnarray}
where the extra $\pi/2$ phase is due to the reflection of BS comparing to the transmission like the one for photons. Substituting Eq. (\ref{atom-2}) into Eq. (\ref{hom-1}), temporal two-particle probability function is
\begin{eqnarray}\label{hom-2}
P^{(2)}(t_1-t_2)\propto 1- \text{sinc}^2 \frac{\Delta \omega}{2}(t_1-t_2),
\end{eqnarray}
where point particle sources are assumed. It is impossible to detect two boson particles when these two detectors are at symmetrical positions of BS. Particle HOM dip with visibility higher than 50\% was observed \cite{lopes2015atomic}.

With the same method, temporal two-particle probability function in Fig. \ref{18-HOM-interferomter} for fermions is
\begin{eqnarray}\label{hom-21}
P^{(2)}(t_1-t_2)\propto 1+ \text{sinc}^2 \frac{\Delta \omega}{2}(t_1-t_2),
\end{eqnarray}
in which two particles tends to leave BS from different output ports \cite{bocquillon2013coherence}.

Then calculate the second-order interference of two independent and identical BECs in Fig. \ref{18-HOM-interferomter}. There are three different ways to trigger a two-particle coincidence count. The first way is that these two particles are both emitted by S$_1$. The probability is 1/4 and there is one path, $A_{11,12}$. The second one is that these two particles are both emitted by S$_2$. The probability is 1/4 and there is one path, $A_{21,22}$. The third way is that these two particles are emitted by S$_1$ and S$_2$ one each. The probability is 1/2 and there are two different paths, $A_{11,22}$ and $A_{12,21}$. When these four different paths are indistinguishable, the two-particle probability function is 
\begin{eqnarray}\label{hom-3}
&&P^{(2)}(\vec{r}_1,t_1;\vec{r}_2,t_2)\nonumber\\
&=& \langle |\frac{1}{\sqrt{4}}A_{11,12}+\frac{1}{\sqrt{4}}A_{21,22}+\frac{1}{\sqrt{4}}(A_{11,22}+A_{12,21})|^2\rangle\nonumber\\
&\propto& \langle |e^{i(2\varphi_1+\frac{\pi}{2})}K_{11}K_{12}+e^{i((2\varphi_2+\frac{\pi}{2})}K_{21}K_{22}\nonumber\\
&&+e^{i(\varphi_1+\varphi_2)}(K_{11}K_{22}-K_{12}K_{21})|^2\rangle,
\end{eqnarray}
where $\varphi_1$ and $\varphi_2$ are initial phases of particles emitted by S$_1$ and S$_2$, respectively. Since S$_1$ and S$_2$ are independent, Eq. (\ref{hom-3}) can be simplified as
\begin{eqnarray}\label{hom-4}
&&P^{(2)}(\vec{r}_1,t_1;\vec{r}_2,t_2)\\
&\propto& \langle |K_{11}K_{12}|^2\rangle +\langle |K_{21}K_{22}|^2 \rangle +\langle |(K_{11}K_{22}-K_{12}K_{21})|^2\rangle.\nonumber
\end{eqnarray}
Substituting Eq. (\ref{atom-2}) into Eq. (\ref{hom-4}), it is straightforward to have temporal two-particle probability function,
\begin{eqnarray}\label{hom-5}
P^{(2)}(t_1-t_2) \propto   1- \frac{1}{2} \text{sinc}^2 \frac{\Delta \omega}{2}(t_1-t_2).
\end{eqnarray}
Comparing to the second-order interference of two single boson sources, the visibility of the second-order interference of two independent BECs can not exceed 50\%, which is similar as the conclusion for the visibilities of the second-order interference of two independent classical and non-classical light beams \cite{mandel1983photon}.

\subsection{Third-order interference of matter wave}

Based on the results above, it is easy to predict that there is three-particle bunching for a cold boson  beam and no three-particle bunching for a BEC in a third-order HBT interferometer \cite{burt1997coherence}. This section will focus on the third-order interference of fermions. The scheme for the third-order interference of one fermion beam is the same as the one shown in Fig. \ref{27-3HBT}, in which the meanings of all the symbols are similar as the one for photons. There are six different ways to trigger a three-particle coincidence count and the corresponding three-particle probability amplitudes are $A_{a1,b2,c3}$, $A_{a1,b3,c2}$, $A_{a2,b1,c3}$, $A_{a2,b3,c1}$, $A_{a3,b1,c2}$ and $A_{a3,b2,c1}$. When these six different paths are indistinguishable, the three-particle probability function is
\begin{eqnarray}\label{three-f-1}
&&P^{(3)}(\vec{r}_1,t_1;\vec{r}_2,t_2;\vec{r}_3,t_3)\nonumber\\
&=& \langle |\frac{1}{\sqrt{6}}(A_{a1,b2,c3}-A_{a1,b3,c2}-A_{a2,b1,c3}\nonumber\\
&&+A_{a2,b3,c1}+A_{a3,b1,c2}-A_{a3,b2,c1})|^2 \rangle,
\end{eqnarray}
where the minus signs are due to exchange antisymmetry of fermions. With similar method above, temporal three-particle probability function is
\begin{eqnarray}\label{three-f-2}
&&P^{(3)}(t_1;t_2;t_3)\nonumber\\
&\propto& 1-\text{sinc}^2\frac{\Delta \omega}{2}(t_1-t_2)\nonumber\\
&&-\text{sinc}^2\frac{\Delta \omega}{2}(t_2-t_3)-\text{sinc}^2\frac{\Delta \omega}{2}(t_3-t_1)\\
&&+2\text{sinc}\frac{\Delta \omega}{2}(t_1-t_2)\text{sinc}\frac{\Delta \omega}{2}(t_2-t_3)\text{sinc}\frac{\Delta \omega}{2}(t_3-t_1). \nonumber
\end{eqnarray}

\begin{figure}[htb]
\centering
\includegraphics[width=70mm]{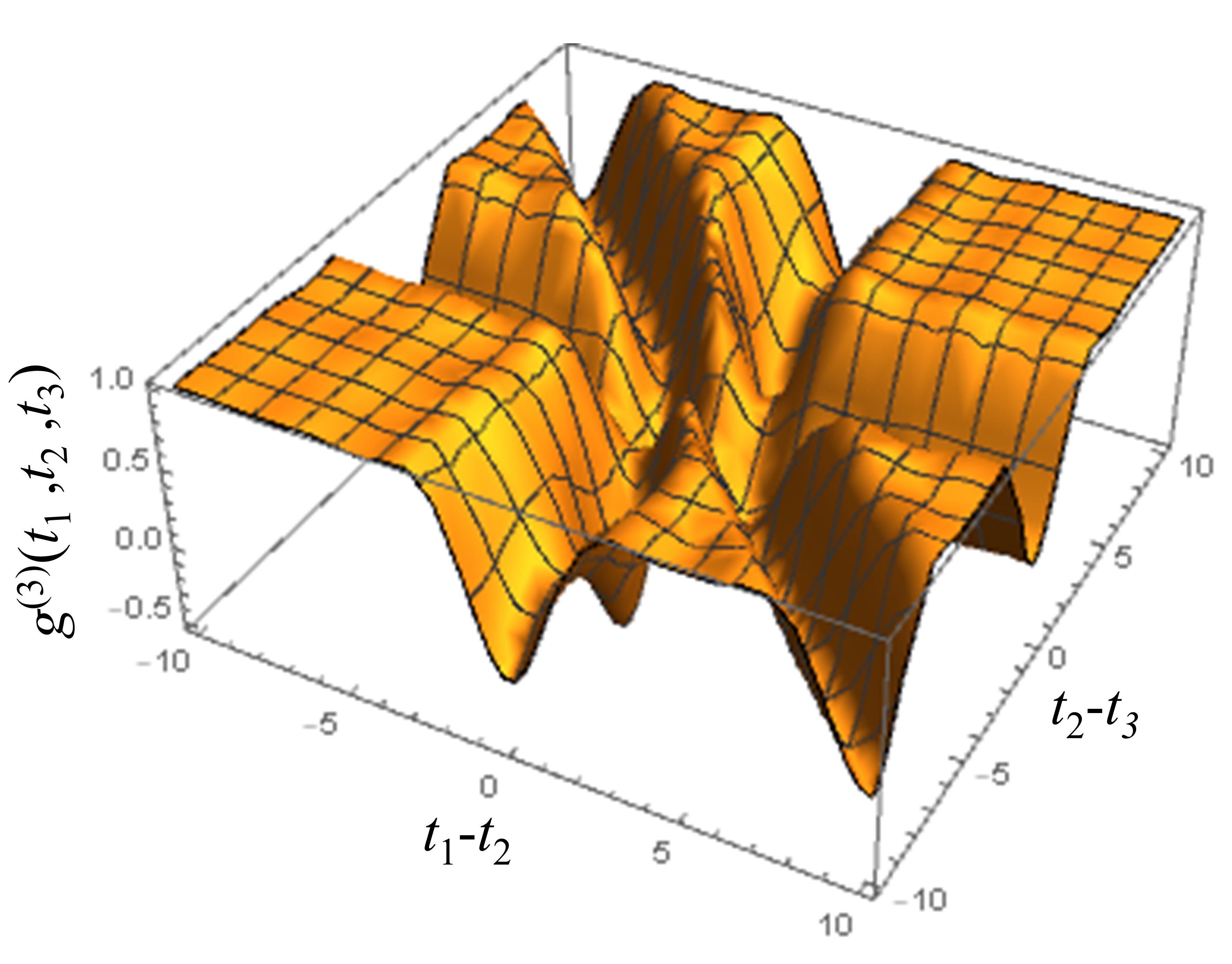}
\caption{Third-order temporal coherence function of fermions in a third-order HBT interferometer. $g^{(3)}(t_1,t_2,t_3)$ is the normalized third-order temporal coherence function. $t_1-t_2$ is the time difference between photon detection event at D$_1$ and D$_2$. Similar definition holds for $t_2-t_3$.}\label{35-three-fermion}
\end{figure}

Figure \ref{35-three-fermion} shows the normalized third-order temporal coherence function of fermions in a third-order HBT interferometer, which is proportional to Eq. (\ref{three-f-2}). The probability of detecting three particles when these three detectors are at the symmetrical positions equals 0, which is consistent with the results obtained by Liu \textit{et al.} \cite{liu2016high}.

Similar method can be employed to calculate the third- and higher-order interference of multiple matter wave beams.

\section{Conclusions}\label{sec-summary}

It is interesting to notice that there is some difference between the points of view about the interference of light in classical optical coherence theory and quantum optical coherence theory based on Feynman\rq{}s path integral. In classical optical theory, the first-order interference of light is the foundation of the second- and higher-order interference of light. All the second- and higher-order interference of light is interpreted as the results of first-order interference of light. For instance, two-photon bunching of thermal light is interpreted as the intensity fluctuations correlation in classical optical coherence theory.  In quantum optical coherence theory based on Feynman\rq{}s path integral, the first-, second-, and higher-order interference of light are interpreted by the same superposition principle. The first-order interference of light is not the foundation of second- and higher-order interference of light. There is no right and wrong about these two different points of view. It is just the interpretations about optical coherence are different in these two different theories.

In summary, we have presented a novel quantum optical coherence theory based on Feynman\rq{}s path integral, which provides a useful tool for studying the physics of optical coherence besides classical optical coherence theory based on Maxwell\rq{}s electromagnetic theory and Glauber\rq{}s quantum optical coherence theory based on matrix mechanics formalism of quantum mechanics. Some interesting conclusions are drawn. For instance, unlike the well-accepted conclusion that the electric field models are the same for thermal and laser light within the coherence time, it is predicted by quantum optical coherence theory based on Feynman\rq{}s path integral that there may be some difference between these two. The physics behind two-photon bunching of thermal light in a HBT interferometer and HOM dip of entangled photon pairs in a HOM interferometer is the same, both of which can be interpreted by two-photon interference. In order to show how quantum optical coherence theory based on Feynman\rq{}s path integral can be helpful to understand the physics of optical coherence, subwavelength interference is taken as an example and it is shown that there is no intrinsic difference between the observed two-photon subwavelength interference for two detectors scanning in the same direction and  opposite directions. Finally, quantum particle coherence theory based on Feynman\rq{}s path integral is also given by analogy of the developed quantum optical coherence theory based on Feynman\rq{}s path integral.

We believe that quantum optical coherence theory based on Feynman\rq{}s path integral is helpful to understand the physics of optical and atomic coherence, and may eventually lead us to the answer to the question puzzling us for a very long time: what is light?

\section*{Acknowledgments}

Jianbin Liu would like to thank the help from Guoquan Zhang and Yanhua Shih. This project is supported by National Science Foundation of China (No.11404255), Doctoral Fund of Ministry of Education of China (No.20130201120013), Shanxi Key Research and Development Project (No. 2019ZDLGY09-08), Xi'an Science and Technology Program Project (No. GX2331), Open fund of MOE Key Laboratory of Weak-Light Nonlinear Photonics (No. OS19-2), and Fundamental Research Funds for the Central Universities.

\bibliography{bibliography}

\end{document}